%% file: GAIA_DR3_Crossmatch.tex
\documentclass{aa}  

\usepackage{graphicx}
\usepackage{numprint}
\usepackage{txfonts}
\usepackage{lscape}

\usepackage{hyperref}
\usepackage{url}
\hypersetup{
    colorlinks=true,
    linkcolor=blue,
    urlcolor=magenta,
    citecolor=blue
}
\newcommand{\orcit}[1]{\protect\href{https://orcid.org/#1}{\protect\includegraphics[width=8pt]{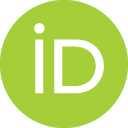}}}

\titlerunning{\textit{Gaia}~DR3: Cross-match of \textit{Gaia} sources with variable objects} 

\begin{document} 

\title{\textit{Gaia} Data Release 3}
\subtitle{Cross-match of \textit{Gaia} sources with variable objects from the literature}

\author{Panagiotis Gavras\orcit{0000-0002-4383-4836}\inst{\ref{RheaESA}}\fnmsep\thanks{\email{panagiotis.gavras@esa.int}}
\and
Lorenzo Rimoldini\orcit{0000-0002-0306-585X}\inst{\ref{inst4}}
\and
Krzysztof Nienartowicz\orcit{0000-0001-5415-0547}\inst{\ref{inst3}}
\and
Gr{\'e}gory Jevardat~de~Fombelle\inst{\ref{inst4}}
\and 
Berry Holl\orcit{0000-0001-6220-3266} \inst{\ref{inst2},\ref{inst4}}
\and 
Péter Ábrahám\orcit{0000-0001-6015-646X} \inst{\ref{budapest_1},\ref{budapest_2}}
\and 
Marc Audard\orcit{0000-0003-4721-034X}\inst{\ref{inst2},\ref{inst4}}
\and 
Maria I. Carnerero\orcit{0000-0001-5843-5515}\inst{\ref{inaf_torino}}
\and 
Gisella Clementini\orcit{0000-0001-9206-9723}\inst{\ref{inaf_bologna}}
\and 
Joris De Ridder\orcit{0000-0001-6726-2863}\inst{\ref{leuven}}
\and 
Elisa Distefano\orcit{0000-0002-2448-2513}\inst{\ref{inaf_catania}}
\and 
Pedro Garcia-Lario\orcit{0000-0003-4039-8212}\inst{\ref{esac}}
\and 
Alessia Garofalo\orcit{0000-0002-5907-0375}\inst{\ref{inaf_bologna}}
\and 
Ágnes Kóspál \inst{\ref{budapest_1},\ref{max_planck},\ref{budapest_2}}
\and 
Katarzyna Kruszy{\'n}ska\orcit{0000-0002-2729-5369}\inst{\ref{warsaw}}
\and 
Mária Kun\orcit{0000-0002-7538-5166}\inst{\ref{budapest_1}}
\and 
Isabelle Lecoeur-Ta{\"i}bi\orcit{0000-0003-0029-8575}\inst{\ref{inst4}}
\and 
Gábor Marton\orcit{0000-0002-1326-1686}\inst{\ref{budapest_1}}
\and 
Tsevi Mazeh\orcit{0000-0002-3569-3391}\inst{\ref{tel_aviv_1}}
\and 
Nami Mowlavi\orcit{0000-0003-1578-6993}\inst{\ref{inst2},\ref{inst4}}
\and 
Claudia M. Raiteri\orcit{0000-0003-1784-2784}\inst{\ref{inaf_torino}}
\and 
Vincenzo Ripepi\orcit{0000-0003-1801-426X}\inst{\ref{inaf_napoli}}
\and 
L\'aszl\'o Szabados\orcit{0000-0002-2046-4131}\inst{\ref{budapest_1}}
\and 
Shay Zucker\orcit{0000-0003-3173-3138}\inst{\ref{tel_aviv_2}}
\and 
Laurent Eyer\orcit{0000-0002-0182-8040}\inst{\ref{inst2}}
}

\authorrunning{Gavras et al.}
\institute{RHEA for European Space Agency (ESA), Camino bajo del Castillo, s/n, Urbanizacion Villafranca del Castillo, Villanueva de la Ca{\~n}ada, 28692 Madrid, Spain
\label{RheaESA}
\and Department of Astronomy, University of Geneva, Chemin d'Ecogia 16, 1290 Versoix, Switzerland\label{inst4}
\and
Sednai Sàrl, Geneva, Switzerland\label{inst3}
\and
Department of Astronomy, University of Geneva, Chemin Pegasi 51, 1290 Versoix, Switzerland\label{inst2}
\and
Konkoly Observatory, Research Centre for Astronomy and Earth Sciences, E{\"o}tv{\"o}s Lor{\'a}nd Research Network, Konkoly Thege 15-17, 1121, Budapest, Hungary\label{budapest_1}
\and
ELTE E{\"o}tv{\"o}s Lor{\'a}nd University, Institute of Physics, P{\'a}zm{\'a}ny P{\'e}ter s{\'e}t{\'a}ny 1/A, 1117 Budapest, Hungary \label{budapest_2}
\and 
INAF – Osservatorio Astrofisico di Torino, Via Osservatorio 20, 10025 Pino Torinese, Italy\label{inaf_torino}
\and
INAF – Osservatorio di Astrofisica e Scienza dello Spazio di Bologna, Via Gobetti 93/3, 40129 Bologna, Italy \label{inaf_bologna}
\and
Instituut voor Sterrenkunde, KU Leuven, Celestijnenlaan 200D, 3001 Leuven, Belgium \label{leuven}
\and
INAF – Osservatorio Astrofisico di Catania, Via S.\ Sofia 78, 95123 Catania, Italy \label{inaf_catania}
\and
European Space Agency (ESA), European Space Astronomy Centre (ESAC), 
 Camino Bajo del Castillo s/n, Urb.\ Villafranca del Castillo, 28692 Villanueva de la Cañada, Spain \label{esac}
\and
Max Planck Institute for Astronomy, K{\"o}nigstuhl17, 69117 Heidelberg, Germany \label{max_planck}
\and
Warsaw University, Astronomical Observatory, Department of Physics, Al. Ujazdowskie 4, 00-478, Warszawa, Poland \label{warsaw}
\and 
School of Physics and Astronomy, Tel Aviv University, Tel Aviv 6997801, Israel \label{tel_aviv_1}
\and
INAF – Osservatorio Astronomico di Capodimonte, Via Moiariello 16, 80131 Napoli, Italy \label{inaf_napoli}
\and 
Porter School of the Environment and Earth Sciences, Tel Aviv University, Tel Aviv 6997801, Israel\relax  \label{tel_aviv_2}
}

\date{Received -, -; accepted - -, -}

 
  \abstract
  {In the current ever increasing data volumes of astronomical surveys, automated methods are essential. Objects of known classes from the literature are necessary for training supervised machine learning algorithms, as well as for verification/validation of their results. }
  {The primary goal of this work is to provide a comprehensive data set of known variable objects from the literature  cross-matched with \textit{Gaia}~DR3 sources, including a large number of both variability types and representatives, in order to cover as much as possible sky regions and magnitude ranges relevant to each class. In addition, non-variable objects from selected surveys are targeted to probe their variability in \textit{Gaia} and possible use as standards. This data set can be the base for a training set applicable in variability detection, classification, and validation.}
   {A statistical method that employed both astrometry (position and proper motion) and photometry (mean magnitude) was applied to selected  literature catalogues in order to identify the correct counterparts of the known objects in the \textit{Gaia} data. The cross-match strategy was adapted to the properties of each catalogue and the verification of results excluded dubious matches.}
  {Our catalogue gathers  7\,841\,723 \textit{Gaia} sources among which 1.2~million non-variable objects and 1.7~million galaxies, in addition to 4.9~million variable sources representing over 100~variability (sub)types.}
   {This data set served the requirements of \textit{Gaia}'s variability pipeline for its third data release (DR3), from classifier training to result validation, and it is expected to be a useful resource for the scientific community that is interested in the analysis of variability in the \textit{Gaia} data and other surveys.}


   \maketitle
%

\section{Introduction}
Variable stars have been proven extremely useful tool to investigate a diverse set of 
astronomical problems. Their variability properties allowed us to measure physical 
quantities such as distances using the luminosity-period relations of stars as Cepheids \citep{1926ApJ....64..321H} 
and RR\,Lyrae stars \citep{1978ApJ...223..351D} or using their pulsation velocities in Baade-Wesselink method \citep{1926AN....228..359B,1946BAN....10...91W}, while
eclipsing binaries enabled us to have a measurement of masses and radii of stars \citep{1967ARA&A...5...85P}.
Other types of variable sources like AGNs are useful to enrich our
knowledge on the early universe. Thus, since the early days scientists have started to register and classify
sources that appear to be variable. Over the years the number of 
known variables and the number of (sub)types of variability have increased significantly.
GCVS \citep{2017ARep...61...80S} has been one of the first catalogues of variable stars started in 1946. The American Association of Variable Star Observers (AAVSO) maintain 
the international variable star index \citep[VSX;][]{2006SASS...25...47W} that in its latest version
contains more than 2.1~million objects. The advance of modern astronomy allowed the identification of 
variable sources by large-scale surveys. All-Sky Automated Survey
\citep[ASAS;][]{2002AcA....52..397P}, All-Sky Automated Survey for Supernovae \citep[ASAS-SN;][]{2014ApJ...788...48S,2018MNRAS.477.3145J,2019MNRAS.486.1907J,2019MNRAS.485..961J}, the Optical Gravitational Lensing Experiment 
\citep[OGLE;][] {2015AcA....65....1U},
the Catalina Real-Time Transient Survey \citep{2014ApJS..213....9D}, Zwicky Transient Facility \citep[ZTF;][] {2019PASP..131g8001G} and \textit{Gaia} \citep{2016A&A...595A.133C,2019A&A...625A..97R,2019A&A...622A..60C} are only some projects that have increased significantly the number of known variables. 

The \textit{Gaia} consortium released 3194 variable stars of 2 variability types in its first data release \citep[DR1;][]{2017arXiv170203295E}, which increased to 550\,737 variables and 6 types in DR2 \citep{2018A&A...618A..30H}, and to 13 million and 30 (sub)types including galaxies in DR3 \citep{DR3-DPACP-162}.
Moreover, it is foreseen that the increase in the number of variables will continue in DR4 by an order of magnitude. 
This abundance of data has made the need of automated methods of detection and classification of sources imperative. Thus, most of the modern all-sky surveys use some type of machine learning method for the identification
of variables. Supervised machine learning methods use a labelled set of known variables (usually from the literature) 
in order to train classifiers. The creation of an unbiased training set is a challenging task. It needs to have a large number of sources adequately covering all variability classes aimed by the project, in order to be able to select training sources that do not suffer from selection biases, e.g., in the distribution in the sky or by incomplete coverage of magnitudes. It may also include contaminants, where in the case of variable sources can be non variable or other types of objects that exhibit artificial variability. Details on artificial variability in \textit{Gaia} can be found in \cite{DR3-DPACP-164}. 

Producing an optical catalogue by cross-matching many input catalogues, with data in the radio, mid and near-infrared, optical, and X-ray bands, is not a straightforward task. 
Each catalogue has its own unique properties, such as astrometric and photometric qualities, 
observational bands, and with different needs of propagation of proper motion (when available), depending on object distance and observational time difference (i.e. different survey epoch),  
which need to be fine tuned, and some fraction of mismatches becomes  inevitable. 
In the case of \textit{Gaia}, a cross-match with external catalogues was provided in all data releases (e.g., see \citealt{2019A&A...621A.144M} 
and  \href{https://gea.esac.esa.int/archive/documentation/GEDR3/Catalogue\_consolidation/chap\_crossmatch/}{online documentation}), 
but their focus were not variable objects, leaving the vast majority of the known variables unmatched.

The variability processing of \textit{Gaia} employed data sets from literature to train its classifiers. Cross-match techniques varied in each data release but their results were not published before. 
DR1 was limited to two variability types and a specific region in the sky \citep{2017arXiv170203295E}, for which 7 literature catalogues were cross-matched with \textit{Gaia} using a random forest classifier \citep{2019ASPC..521..307R}. In DR2, astrometry was combined with transformed photometry and time series features to create a multi-dimensional distance, which was used to match 70 catalogues from the literature (\citealt{2019A&A...625A..97R}, \href{https://gea.esac.esa.int/archive/documentation/GDR2/Data_analysis/chap_cu7var/ssec_cu7var_sos_allsky/ssec_cu7var_allsky_proc.html#SSS2}{online documentation}) with \textit{Gaia} sources. Machine learning supervised classification and special variability detection in the third data release of \textit{Gaia} contains $\sim$10.5 million variables sources and 24 different classes, 
which required a larger and more diverse training data set. The base of this training set is our cross-match catalogue.
In this first publication of the cross-match catalogue we cross-matched the sources found in a selection of 152 catalogues with \textit{Gaia} results. Our catalogue contains  7.8 million unique objects. 

This paper presents the method, the results and the caveats of the cross-match between the 152 catalogues and \textit{Gaia}~DR3 sources.
We describe the creation of this data set in Sect.~\ref{Sec:Creation}. Section~\ref{Sec:Properties} presents the properties of the produced catalogue. We discuss the properties of selected variability types in Sect.~\ref{Sec:Quality} indicating the overall quality of the catalogue. Section~\ref{sec:leastvariables} shows an effort to identify stars that are the least variable and conclusions are in Sect.~\ref{Sec:Conclusion}.
The cross-match catalogue is made available exclusively online through the Centre de Donn\'ees astronomiques de Strasbourg website\footnote{\url{http://cdsarc.u-strasbg.fr/}}.

\section{Creation of the cross-match catalogue}\label{Sec:Creation}
\subsection{Input catalogues selection}
There are many interesting catalogues that we could select for this work. However as the idea was to create a large data set with many variability types, we used well-known diverse catalogues that contain various variability types. Also we selected smaller catalogues of objects of particular interest or of rare variability types. Finally we assembled a list of 152 different input catalogues.
Some of these were compiled and used internally by \textit{Gaia} Data Processing and Analysis Consortium (DPAC) members.

In order to facilitate the identification and basic properties of each catalogue, we constructed and used an informative catalogue label. This label is derived from the mission, survey or compilation name, the type of targets that the catalogue contains, the name of first author (or the person who compiled it), and the date of publication. We use this label throughout the rest of the paper.

All input catalogues are listed alphabetically in Table~\ref{table:catalogList}: the first column provides the catalogue label, the second column presents the number of stars finally cross-matched with \textit{Gaia} sources, and the last column lists the references for each catalogue. The selection of the literature catalogues is limited to those published before 2021 with only exception EROSITA\_AGN\_LIU\_2021 \citep{2021arXiv210614522L}.

In addition to variable sources, the cross-match catalogue includes a limited number of non-varying sources according to surveys with similar precision to \textit{Gaia} (named constants hereafter), for use e.g.\ in variability detection or to capture objects with insufficient or corrupt variability.
The HIPPARCOS\_VAR\_ESA\_1997 \citep{1997ESASP1200.....E} and SDSS\_CST\_IVEZIC\_2007  \citep{2007AJ....134..973I} catalogues are the main providers of non-varying objects, but the former lacks faint objects and the latter misses bright sources and is limited to the SDSS Stripe~82 footprint.
Given the gap in magnitude ($12<G<14$) from these two catalogues and the non-representative distribution in the sky for faint objects, two new catalogues of constant stars were created using data from TESS \citep{2015JATIS...1a4003R}, to fill the magnitude gap, and ZTF \citep{2019PASP..131a8003M}, for improved sky distribution.
This effort it is not the main focus of this paper and it is  is described in Sect.~\ref{sec:leastvariables}.
\subsection{The pipeline}
The pipeline we built to identify the correct counterpart of an input source is divided in two major parts. The fist part is performing a positional cross-match of each source in a literature catalogue with the \textit{Gaia}~DR3 sources and the second is the cleaning of the results of the first part from false identifications.

\subsubsection{Positional cross-match}
The cross-match of an input catalogue with the sources of \textit{Gaia}~DR3 was performed in the database deployed at the  data processing centre of Geneva (DPCG) at the homonym observatory. This process was divided in two steps to facilitate processing. The first step was to make a simple cone search with a radius typically of 1\arcmin ~ around the coordinates of each source existing in an input catalogue to Gaia sources. The large radius was used to cover the positional uncertainties and most of the proper motion effects while keeping the computational load low and speed up the cross-match. The second step was to make a cone search with a radius of 5\arcsec ~applying epoch propagation of the positions using the relevant function of the Q3C library \citep{2006ASPC..351..735K} and \textit{Gaia} proper motions. This way the cross-match was fine-tuned  in a fraction of sources instead of the $\sim$1.8 billion sources  in  \textit{Gaia}~DR3. 
 
The radius of the cross-match was adjusted in some catalogues to larger or smaller values, e.g., HIPPARCOS\_VAR\_ESA\_1997 \citep{1997ESASP1200.....E} to a larger value to take into account the proper motion effect that is more evident due to its bright magnitude limit (including mostly nearby stars) and the large difference in time of observations.
For the majority of catalogues, we were able to find the date of observations and perform epoch propagation of the positions. The epoch used was the mean epoch of observations. However, epoch propagation was not applied to catalogues that were compilations of papers or few others for which we were unable to identify the date of observations.

\subsubsection{Cleaning of false identifications}
The results of the positional cross-match may  return a large number of candidate counterparts, depending on the properties of the input catalogue. For the second part of the pipeline, to further refine from the many candidates, we selected matches using a synthetic distance metric $\rho_{\rm synth}$  that combines angular sky separation and photometric differences:
    \begin{equation}
        \label{eq:rsynth}
        \rho_{\rm synth}\!=\!\!\sqrt{\left[ \frac{\Delta\theta-{\rm median}({\Delta\theta})}{{\rm MAD}({\Delta\theta})} \right]^2\!\!+\!
        \left[ \frac{\Delta{\rm mag}-{\rm median}({\Delta{\rm mag}})}{{\rm MAD}({\Delta{\rm mag}})} \right]^2},
    \end{equation}    
where $\Delta\theta$ denotes the angular distance and $\Delta{\rm mag}$ is the magnitude difference between \textit{Gaia} and a given survey. Medians and median absolute deviations (MAD) are computed on all neighbours within 5\arcsec (or the adapted value used) radius of the targeted sources. The \textit{Gaia} magnitudes used in this process come directly from the photometry and have not passed through the data cleaning process of the variability detection described in \citep{DR3-DPACP-162}.

With respect to the DR2 approach, we reduced the complexity  for DR3 (excluding time series features and photometric transformations) to favour the use of a robust and consistent $\rho_{\rm synth}$ for most catalogues, at the cost of a loss in precision of $\Delta$mag when comparing stars of multiple spectral types in $G$ versus other bands. For example, the different wavelength coverage of the OGLE $I$ band with respect to \textit{Gaia} $G$ causes redder objects to be brighter in $I$ than in $G$. When this is combined with a catalogue that includes both blue and red objects, as for eclipsing binaries, the uncorrected photometric comparison of main sequence stars and red giants may form even separate $\Delta$mag clumps of valid counterparts.  Although in general the selection of matches was conservatively biased towards the clump associated with the smallest $\rho_{\rm synth}$, correct matches from secondary clumps could be recovered by sources overlapping with other catalogues (more specific to the group of missed matches, or more generic and thus with a larger MAD of $\Delta$mag).  Consequently, the completeness of the cross-match for a given catalogue can be larger than it appears from the simple association of sources with a catalogue.

The distance $\rho_{\rm synth}$ takes into account the angular distance of all sources in the table and the difference in magnitude between the input source and the \textit{Gaia} source in the $G$ band. 
Of course, most of the catalogues contain photometry in different filters than $G$ and some are in multiple bands (in which cases only one of the bands was used, typically the most similar to $G$ or the most sampled one).
For catalogues that are compilations of often many data sets, like the VSX, each source may have different positional precision or use different filters, so the quality of the cross-match cleaning is degraded as the efficiency of a common $\rho_{\rm synth}$ is reduced. Moreover, a number of catalogues was cross-matched when the \textit{Gaia}~DR3 photometry was not available yet, so the DR2 photometry was used instead. 
With the above constraints it is clear that the values of $\rho_{\rm synth}$ depend on each individual input catalogue and are not comparable between catalogues, therefore, a universal constraint in $\rho_{\rm synth}$ cannot be set.

Applying a threshold on $\rho_{\rm synth}$ reduces the number of multiple identifications however there were some left so a final cleaning was applied. The selection of the best match among other matches for the same target used the lowest value of $\rho_{\rm synth} $  or angular distance,
depending on the catalogue. Finally, the  sources flagged as astrometric duplicated source were rejected (these sources are not published in the \textit{Gaia} archive either).

\begin{figure}
  \centering
  \includegraphics[width=\hsize]{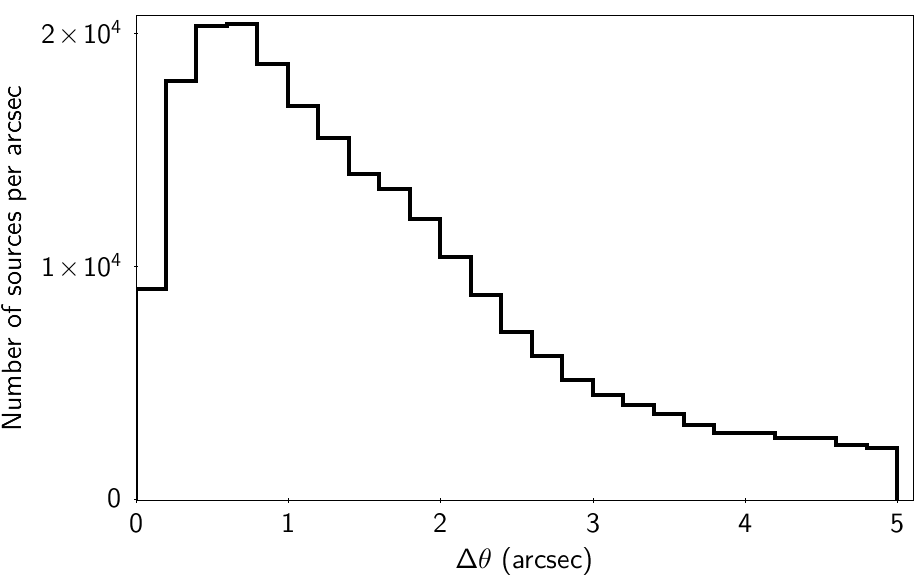}
  \caption{Distribution of angular distance between Catalina CSS South targets and their \textit{Gaia} candidate counterparts found within 5\arcsec ~ radius.}
  \label{Fig:catalinaDist}
\end{figure}

\begin{figure}
  \centering
  \includegraphics[width=\hsize]{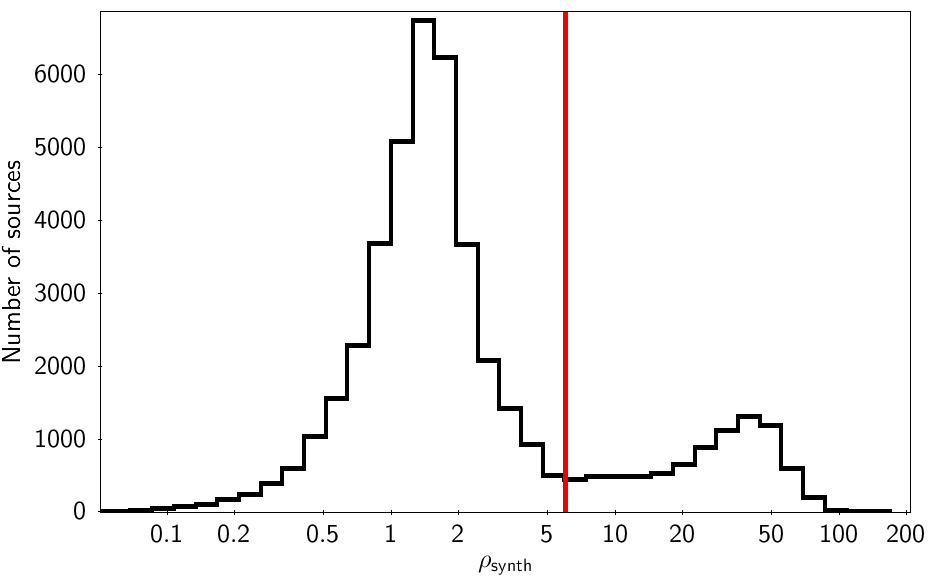}
  \caption{Distribution of the synthetic distance of candidate counterparts obtained by the cross-match between \textit{Gaia} and Catalina CSS South. An upper limit of $\rho_{\rm synth} = 6$ was applied to filter mismatches out.}
  \label{Fig:catalinaSdist}
\end{figure}

\begin{figure}
  \centering
  \includegraphics[width=\hsize]{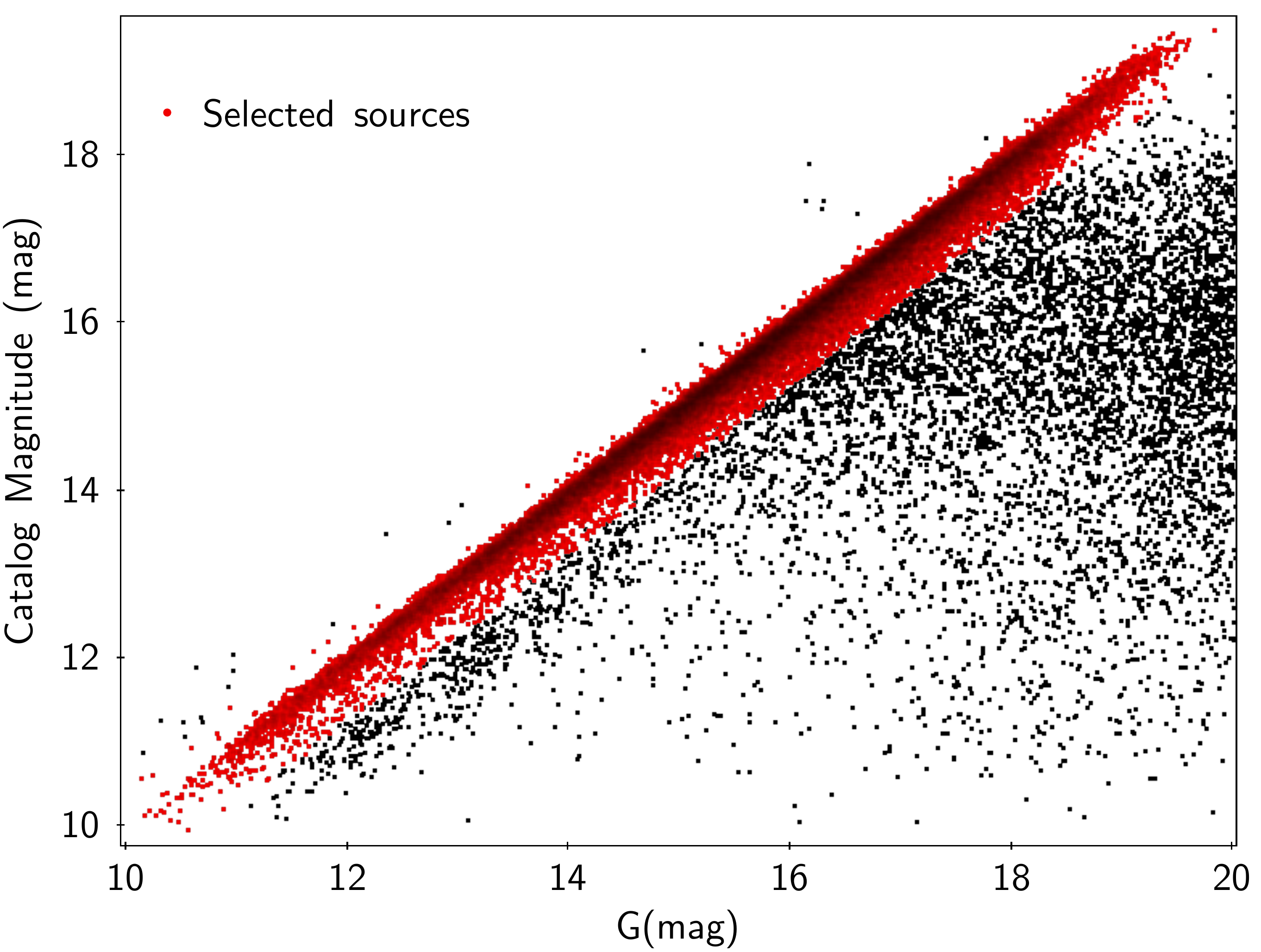}
  \caption{\textit{Gaia} magnitude versus Catalina magnitude for candidates. The counterpart sources with $\rho_{\rm synth}<6$ have red colour.}
   \label{Fig:catalinaCvsGmag}
\end{figure}

Figures \ref{Fig:catalinaDist}--\ref{Fig:catalinaCvsGmag} present examples with data taken from the processing step 2 of the CATALINA\_VAR\_DRAKE\_2017  \citep[Catalina Surveys Southern periodic variable star catalogue,][]{2017MNRAS.469.3688D}.
This catalogue contains 37\,745 variable sources. After the end of positional cross-match, we obtained $\sim$45\,000 candidate counterparts. Figure~\ref{Fig:catalinaDist} shows the distribution of the angular distance between all targets and their potential counterpart. The output of part~1 with a maximum angular distance of 5$\arcsec$ includes obvious mismatches, which are removed by setting an upper limit to $\rho_{\rm synth}$.  
Figure~\ref{Fig:catalinaSdist} shows the distribution of the $\rho_{\rm synth}$. The abscissa is in log-scale to facilitate the clarity of the plot. This plot led to the selection of the cut off value $\rho_{\rm synth} = 6$. The difference in photometric magnitudes between \textit{Gaia} and Catalina is shown in Fig.~\ref{Fig:catalinaCvsGmag}, where the \textit{Gaia} magnitude of the candidate \textit{Gaia} matches are in the horizontal axis and the corresponding Catalina magnitude is given in the vertical axis. The \textit{Gaia} sources selected by the $\rho_{\rm synth}<6$ constraint are coloured red and $\sim$36\,900 sources left. The next step is to reduce any remaining multiple matches to a single one. In this example, less than 200 sources are multiple matches and we kept those with the lowest value of $\rho_{\rm synth}$.  The final cross-matched catalogue for Catalina Surveys Southern periodic variable star catalogue contains 36\,584 sources.
\subsection{Assembly of the final catalogue}
\label{Sec:Assembly}
After cross-matching of all of the individual catalogues, we merged the per-catalogue results to form a single cross-match data set.  It is expected that many of the input catalogues overlap and some of the these sources may appear in several input catalogues with different information, as name, variability type, or their variability period. 
In order to guide users towards the most likely class and period, we defined an approximate catalogue ranking list.
This was not a perfect solution as  catalogue classifications may be more accurate for some types of objects rather than other ones. 
Catalogues that did not overlap with others had no reason to compete in ranking so their relative position is not meaningful and could go to any place. Table~\ref{table:rank} shows the rank-ordered list of literature catalogues, where generally the higher a catalogue is in the list, the more accurate it is.

Source matches of multiple catalogues to the same \textit{Gaia} source identifiers were merged. The resulting cross-match catalogue contains one \textit{Gaia} source per row and information from all relevant catalogues, sorted according to their rank. 
For convenience, information from the highest ranked catalogue, for a given source, are replicated in single-element `primary' fields (like {\tt{primary\_var\_type}} and {\tt{primary\_period}}, see Table~\ref{table:catalog_labels}). 

During the assembly of the catalogue, we tried to homogenize the labels of the variability classes used in the literature.So the often different literature class labels for the same types were made homogeneous following the nomenclature used by the AAVSO,\footnote{\url{https://www.aavso.org/vsx/index.php?view=about.vartypessort}} except for a few exceptions (e.g.,  SARG, OSARG, GTTS, IMTTS).
Some type labels were relabelled as `OMIT' as primary class, because they were too generic, uncertain, or with insufficient variability characterization, and thus should be omitted from training or completeness and purity estimates. 
There is a  large number of literature catalogues for eclipsing binaries and different authors use different labels in their works. In order to homogenise the naming of eclipsing binaries, we grouped them into four subclasses: EA, EB, EW, and ECL, with the latter denoting the generic class when there is no further information or the subclass is uncertain. Table~\ref{table:eb_naming_conv} shows the grouping of labels as defined in our catalogue. 
Information on the original labels from literature was however preserved.
As a special case, sources from the \textit{Gaia} alerts\footnote{\url{http://gsaweb.ast.cam.ac.uk/alerts/home}} have class labels set to OMIT if they were recorded after 28~May~2017 (\textit{Gaia}~DR3 observation time limit).  
There are 5676 such sources and 49\% of them were reported by \textit{Gaia} alerts, the rest were also included in other input catalogues. 
In some catalogues, sources could be associated with multiple types, in which cases \emph{OR} as $|$ and \emph{AND} as $+$ were used. 
We respected the source classification given in the original catalogues, thus class labels may refer to any level of a possible hierarchy. For example, a source may be classified as AGN, QSO, BLAZAR, or BLLAC, without implying that a subtype (e.g., BLLAC) does not belong to its superclass (like BLAZAR or AGN).

After merging information of overlapping catalogues, the final cross-match catalogue contains 7\,841\,723 unique \textit{Gaia}~DR3 source ids.
A subset of catalogues with particularly low contamination rates is indicated by a boolean column {\tt{selection}} and includes  6\,697\,530 sources. Sources with class 'OMIT' are filtered out from the {\tt{selection}}.
The properties of the final catalogue are discussed in the next section.

\subsection{Caveats and exceptions in the pipeline}
\begin{figure}
   \centering
   \includegraphics[width=\hsize]{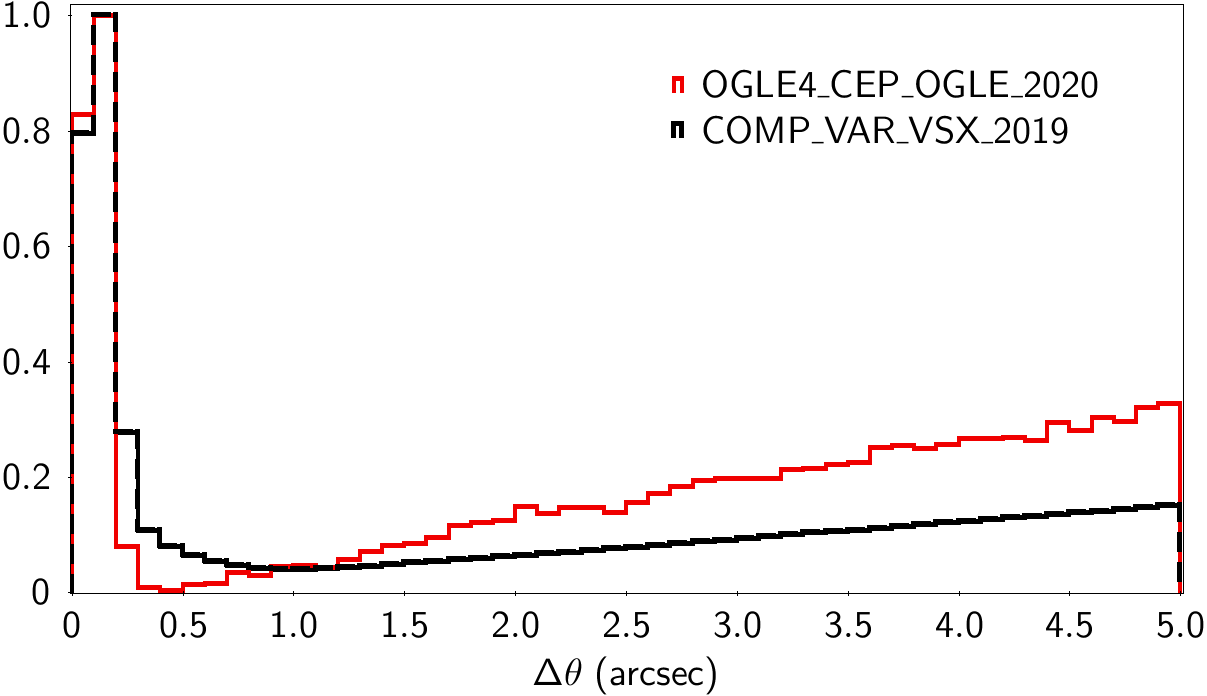}
      \caption{Angular distance distribution (normalised to maximum) of \textit{Gaia} cross-match candidates for OGLE4\_CEP\_OGLE\_2020 and COMP\_VAR\_VSX\_2019.}
     \label{FigAngDist}
  \end{figure}

The method described above, using the statistics of each catalogue, has the advantage of automatically eliminating large numbers of outliers and provides a clean data set.
However, as a statistical process, it may sometimes reject perfectly good candidates.
For example, \textit{Gaia} source\_id  4040728046945051264 exists  in both OGLE4\_CEP\_OGLE\_2020 \citep{2020AcA....70..101S}, as OGLE-BLG-T2CEP-0346, and COMP\_VAR\_VSX\_2019 \citep{2006SASS...25...47W}, with OID=33239.  
The angular distance of this source with respect to its \textit{Gaia} counterpart is $\Delta\theta=0.89$\arcsec in both catalogues (VSX includes OGLE stars). Figure~\ref{FigAngDist}  shows that for OGLE4\_CEP\_OGLE\_2020 the bulk of the counterpart sources exist within 0.3\arcsec, while in COMP\_VAR\_VSX\_2019, which is a compilation of sources from various catalogues, it is close to 1\arcsec. Thus, due to different cuts, this source is eliminated from OGLE4\_CEP\_OGLE\_2020 but not from COMP\_VAR\_VSX\_2019.
  
The cross-match was purely astrometric (based only on the smallest $\Delta\theta$) in the following special cases:  catalogues with highly non-uniform photometry (bands, methods, etc.), whose distribution of $\rho_{\rm synth}$ was not adequate to split matches from mismatches,  
catalogues whose photometry was biased by extreme variability (e.g., sampling only the peak brightness of cataclysmic variables), and very small catalogues for which a statistical procedure
 was not applicable. 
 
 Exceptionally, some catalogues that required no cross-match were included, such as DPAC internal catalogues with pre-assigned \textit{Gaia} source\_id 
 and EROSITA\_AGN\_LIU\_2021 
 for which the authors had already performed cross-match with \textit{Gaia} in \cite{2021arXiv210614520S} using methods optimised for X-ray data sets, therefore their results were used. 

\section{The cross-match catalogue} \label{Sec:Properties}
In this section, a description of the catalogue and its general properties are discussed. 
It is published online through the Centre de Donn\'ees astronomiques de Strasbourg website.

\subsection{Description of the catalogue}
The cross-match catalogue contains in total  7\,841\,723~ sources of various types (6\,697\,530  of them are flagged as {\tt{selection}}=true).

Table \ref{table:catalog_labels} shows the available fields in the published catalogue and provides a short description. Columns in plural may contain multiple values, separated by a semicolon, as a source may exist in multiple literature catalogues. Their order follows the ranking list. The fields start with \textit{primary} contain the information from the highest ranking catalogue that a specific source exists.
The {\tt{primary\_superclass}} field has been introduced in order to group smaller classes and facilitate the selection of generic types.
Table~\ref{table:sourcesPerSuperclass} presents the available types in {\tt{primary\_superclass}}, the number of sources, and classes assigned to each superclass. The assignment was performed based on the class in the highest ranking catalogue which is given in {\tt{primary\_var\_type}}. {\tt{var\_types}} contains the homogenised variability class, {\tt{original\_var\_types}} the original variability type from the literature (i.e. not homogenized), and {\tt{original\_alt\_var\_types}} any alternative variability types provided in literature.
Despite our best effort at minimising mismatches, the cross-match catalogue may still associate sources with incorrect classifications, because of remaining mismatched sources or inaccurate classifications in the literature.
No cleaning or corrections were performed with respect to the information from literature. 
Thus, depending on purpose, users might need to verify or clean some objects of interest, especially if not using the {\tt{selection}} flag. 

The final product contains 112 different types of objects. Some of them are not variable, like constants (CST), generic white dwarfs (WD), non variable DQ dwarfs (DQ, HOT DQ, WARM DQ), or galaxies which appear artificially as variable in \textit{Gaia} \citep{DR3-DPACP-164}, as they might be relevant (depending on purpose) to differentiate genuine vs spurious light variations.
The full list of the 112 different types alphabetically ordered is presented in Table~\ref{table:sourcesPerclass}, together with the number of objects: the first column shows the variability class, the following two columns present the number of objects of this type in {\tt{primary\_var\_type}} and the last two columns refer to the number of unique sources that were classified in any input catalogue as the specific class (in {\tt{var\_types}}). 

\subsection{Properties of the catalogue}\label{cat_properties}

\begin{figure}
  \centering
  \includegraphics[width=\hsize]{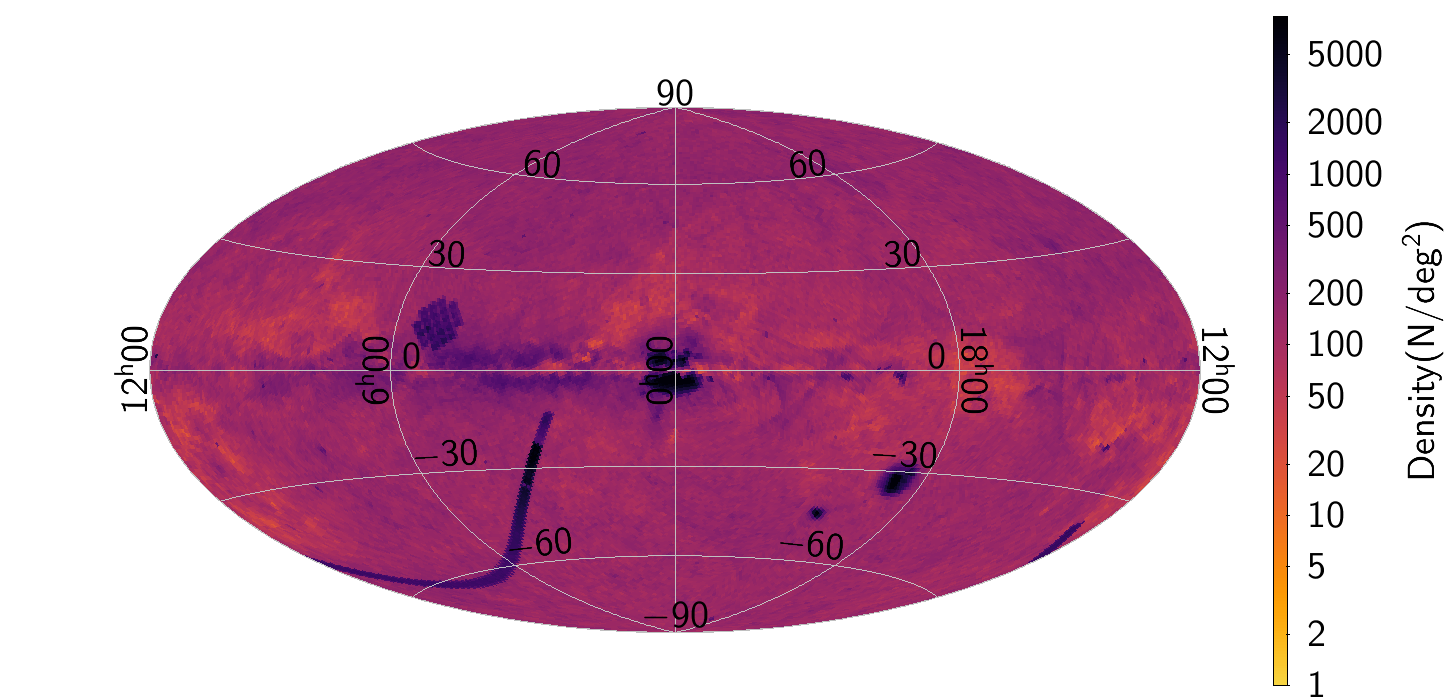}
  \caption{Sky density of all sources in the cross-match catalogue.}
  \label{Fig:catalogSkyDist}
\end{figure}

\begin{figure}
  \centering
  \includegraphics[width=\hsize]{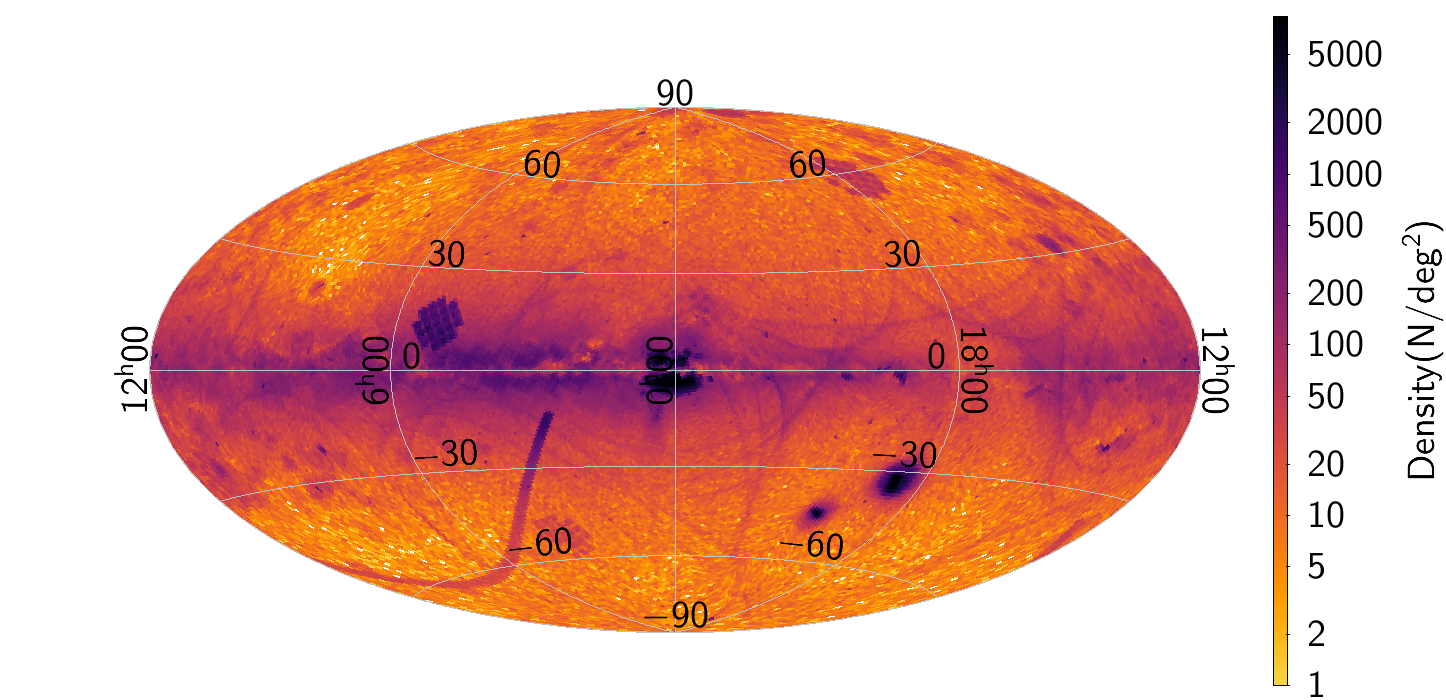}
  \caption{Sky density of 3\,157\,191 variable stars in the cross-match catalogue.}
  \label{Fig:catalogSkyDistVAR}
\end{figure}

\begin{figure}
  \centering
  \includegraphics[width=\hsize]{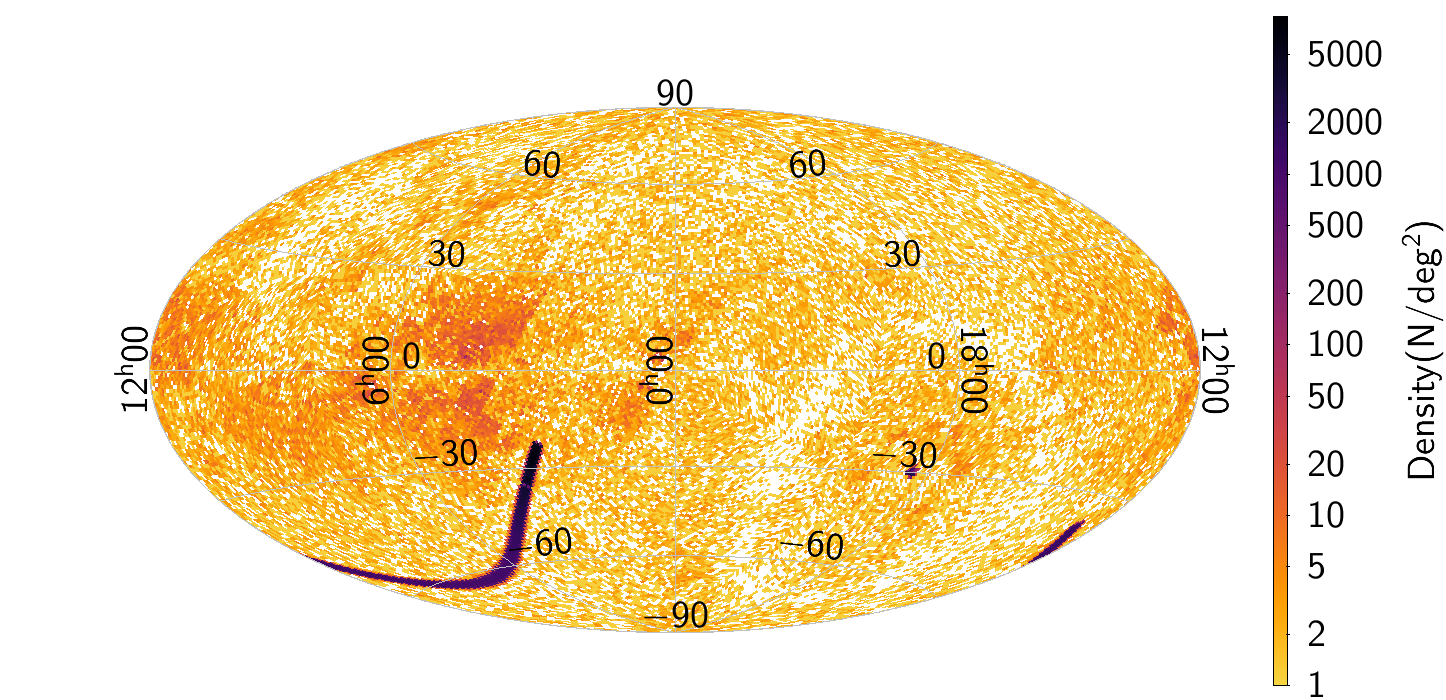}
  \caption{Sky density of 688\,960 constant sources in the cross-match catalogue.} 
  \label{Fig:catalogSkyDistCST}
\end{figure}

\begin{figure}
  \centering
  \includegraphics[width=\hsize]{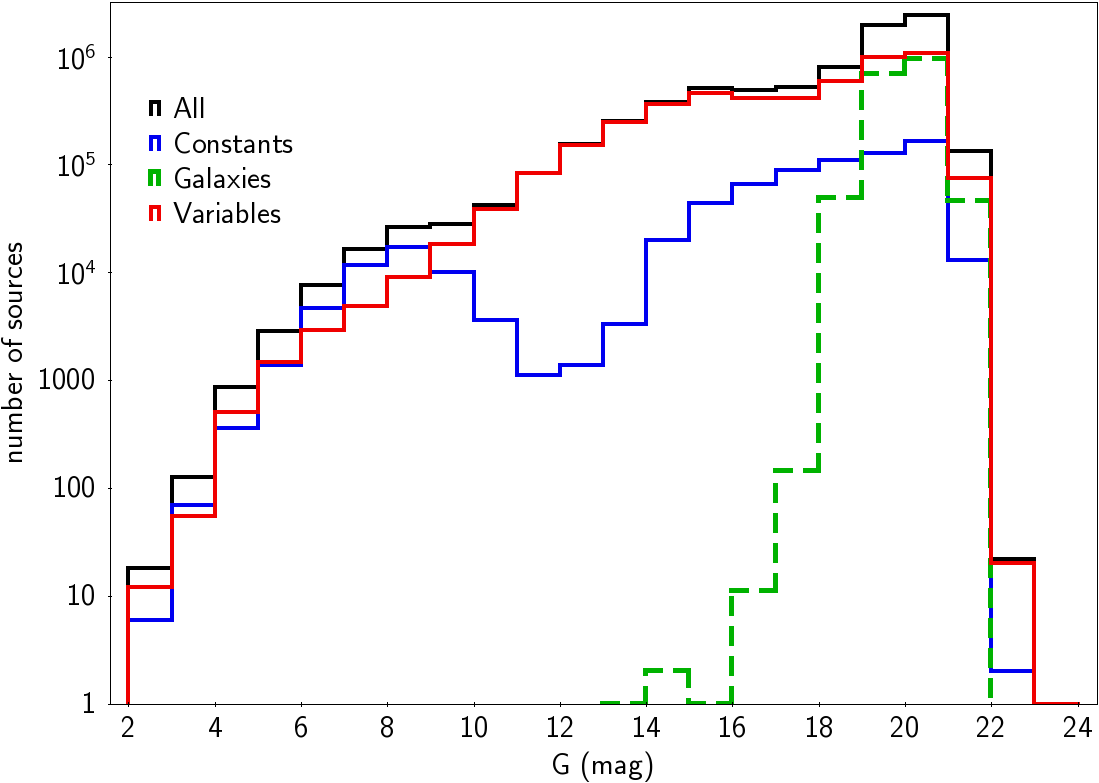}
  \caption{Magnitude distribution of the cross-match catalogue for galaxies, variables, and constants. }
  \label{Fig:catalogmags}
\end{figure}

\begin{figure}
  \centering
  \includegraphics[width=\hsize]{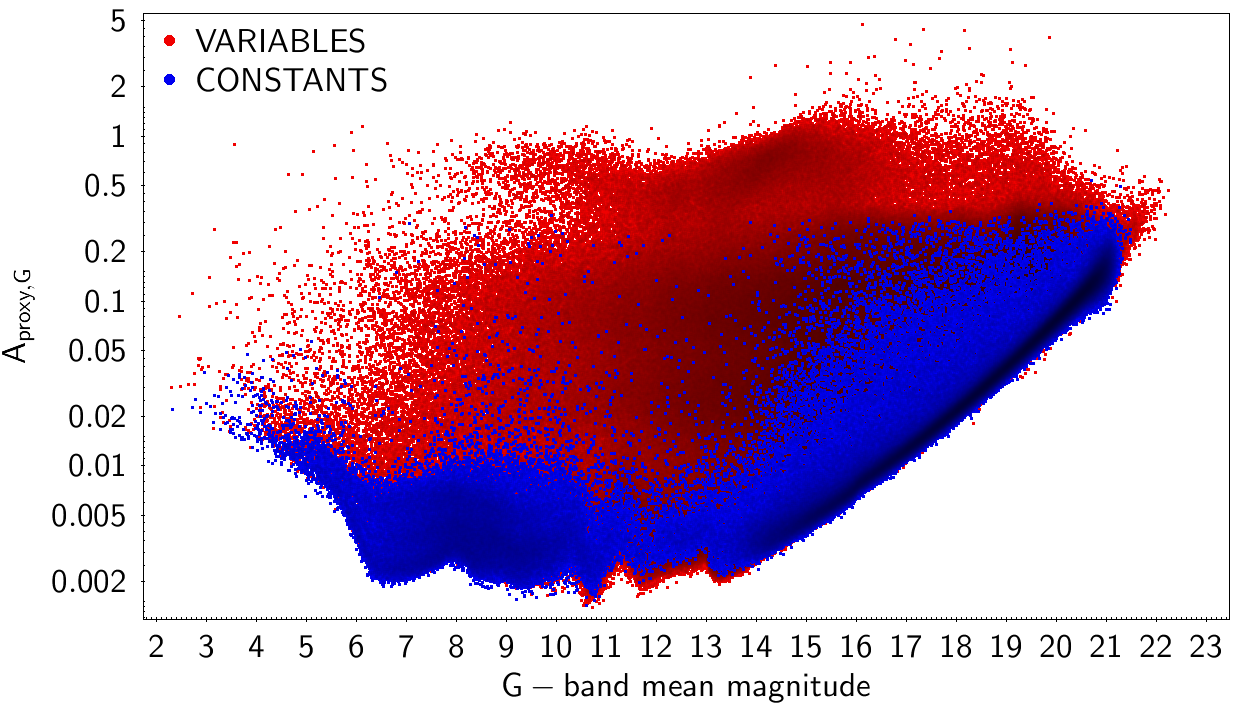}
  \caption{Amplitude proxy $G$ vs Mean $G$ magnitude for constant and variable stars. }
  \label{Fig:amplProxy}
\end{figure}

\begin{figure}
  \centering
  \includegraphics[width=\hsize]{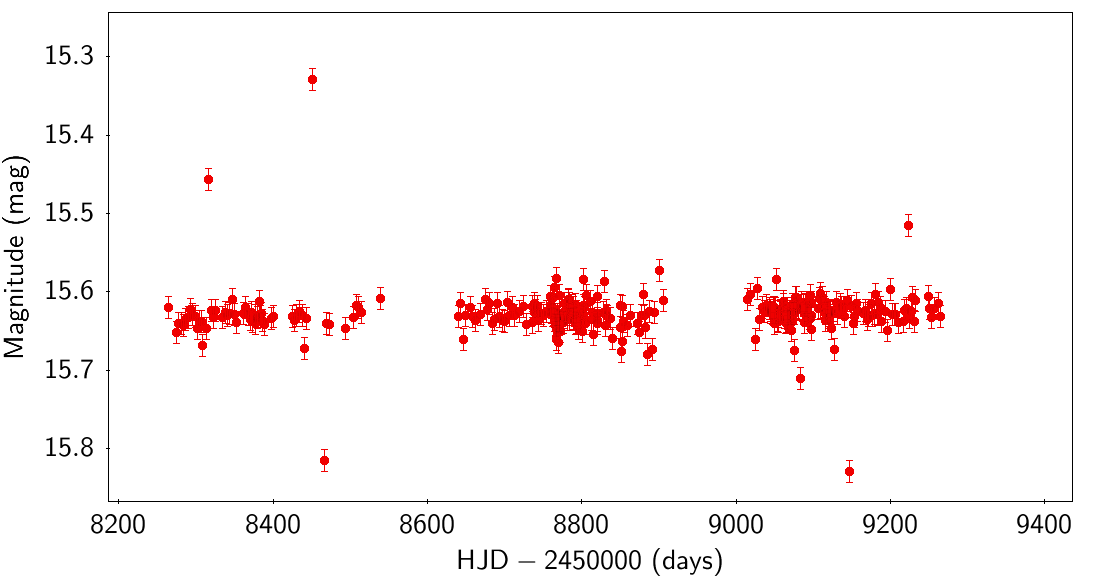}
  \caption{ZTF time series of \textit{Gaia}~DR3 source\_id~395015018457259904 selected as least variable but has large  $A_{proxy,G}$.}
  \label{Fig:cstExample}
\end{figure}

The sky distributions of cross-match sources are presented in Figs.~\ref{Fig:catalogSkyDist}--\ref{Fig:catalogSkyDistCST}, where all sources, only variable stars (without WD, CST, AGN, and GALAXY types), and only constant sources, respectively, are shown. The sky distributions for the extragalactic content are presented and discussed in Sect.~\ref{subsec:exragalactic}. The Galactic center, Magellanic Clouds, the Kepler fields, and the SDSS Stripe~82 are prominent as some literature catalogues are focused in those fields.
Figure~\ref{Fig:catalogmags} shows the distribution of magnitudes of all sources (black) and of constants (blue), variable objects (red), and galaxies (green dashed). 
The galaxies appear at the fainter end of the catalogue, while variable and constant sources 
are distributed along the full magnitude range. 

\begin{equation}
    \label{eq:aproxy}
    A_{proxy,G}\!=\!\!\sqrt{N_G} \frac{\varepsilon (I_G)}{I_G} ,
\end{equation}    
 Figure~\ref{Fig:amplProxy} presents the values of an amplitude proxy in $G$ 
 \citep[$A_{proxy,G}$;][]{2021A&A...648A..44M}
versus the mean $G$-band magnitude for the variable stars and constant sources. The amplitude proxy is a measure of the scatter in the light curve of each source.  $A_{proxy,G}$ is defined in eq.\ref{eq:aproxy} where $N_G$ is the number of observations contributing to G photometry, $I_G$ and $\varepsilon (I_G)$ are the G-band mean flux and its error.
A fraction of stars classified as variable in the literature have  low $A_{proxy,G}$, sometimes lower than constant sources. 
Such an example is source\_id 3328974248568180864, which has $A_{proxy,G}=0.0016$ and is classified as eclipsing binary with a period of 1.04 days by ASAS-SN (ASASSN~$-$~V~J061917.53+094328.8). 
The ASAS$-$SN time series\footnote{\url{https://asas-sn.osu.edu/variables/2d1b902b-d721-50e2-84a6-2fc2edb9f368}} shows a few points in eclipse, which could justify the low value of the amplitude proxy, especially if \textit{Gaia} missed measurements in eclipse.
On the other hand, there is a number of `constant' sources with large  $A_{proxy,G}$. The extreme case of source\_id 395015018457259904,  $A_{proxy,G}=0.29$ and mean magnitude $G=15.57$ is identified as constant in this work with data from ZTF.  
The time series from ZTF (Fig.~\ref{Fig:cstExample}) shows a constant source with a few bright and faint outlying observations and more than 300 points with small dispersion, resulting in a low MAD value. After removing the 4 outlying points, the standard deviation of the magnitude values is 0.016~mag. The large amplitude in the \textit{Gaia}~DR3 time series could be due to spurious measurements or transient events missed by ZTF. 
The discrepant sources may need to be filtered out depending on the requirements of each used case. 
As an example, the \textit{Gaia}~DR3 paper on classification of variables \citep{DR3-DPACP-165} describes the cleaning process applied to these cross-match sources before using them for training purposes.  
\section{Quality of the cross-matched sources} \label{Sec:Quality}
The following subsections assess the quality of the cross-matched sources per variability class. The assessment is based on a visual inspection of the variable sources loci in the CaMD with respect to a reference set defined with all the following criteria:
\begin{equation*}
\begin{aligned}
    & \texttt{phot\_g\_mean\_flux>0}\\
	& \texttt{phot\_bp\_mean\_flux/phot\_bp\_mean\_flux\_error>10} \\
	& \texttt{phot\_rp\_mean\_flux/phot\_rp\_mean\_flux\_error>10} \\
	&\texttt{phot\_bp\_n\_obs>10} \\
	& \texttt{phot\_rp\_n\_obs>10} \\
	& \texttt{parallax\_over\_error>10} \\
	& \texttt{visibility\_periods\_used>11} \\
	& \texttt{ruwe<1.2}
	\label{eqn:Constrains1 }
\end{aligned}
\end{equation*}
These criteria applied on all $\sim$1.8 billion  \textit{Gaia}~DR3 sources and the outcome was further reduced by sampling on their parallax.

The result of this process was a set of 4.2~million sources with high astrometric and photometric quality. This reference set serves as background in the CaMDs that follow in order to help the reader locate the areas the different variability types should exist.

Considering the significantly lower number of sources per class in the cross-match catalogue, less strict constraints were applied in order to select the sources of the various classes: 

\begin{equation*}
\begin{aligned}
	& \texttt{astrometric\_excess\_noise<0.5} \\
	& \texttt{parallax/parallax\_error>5} \\
	& \texttt{visibility\_periods\_used>5} \\
	& \texttt{ruwe<1.4}.
	\label{eqn:Constrains2 }
\end{aligned}
\end{equation*}
The sources within the Magellanic Clouds were excluded from the CaMDs. 
With these constraints, some rare types (like Black Hole X-ray Binaries (BHXB) and Small Amplitude Red Variables (SARV)) did not have sufficient representatives for the CaMD. 
In the following subsections, a short description of the properties of each class and discussion about the quality of the cross-match are given. More information about the various generic properties for each variability type 
can be found in the variable star type designations of the AAVSO VSX\footnote{\url{https://www.aavso.org/vsx/index.php?view=about.vartypessort}}.

\subsection{Pulsating Variables}
The cross-match catalogue contains many different classes of pulsating stars. 
Results are discussed separately for pulsating stars in dwarfs, sub-dwarfs, BLAPs, long period variables, semi-regulars, Cepheids, $\delta$\,Scuti, $\gamma$ Doradus, RR\,Lyrae stars, and other types. 
\subsubsection{White dwarfs, sub-dwarfs, and blue large amplitude pulsators}
\begin{figure}
  \centering
  \includegraphics[width=\hsize,clip]{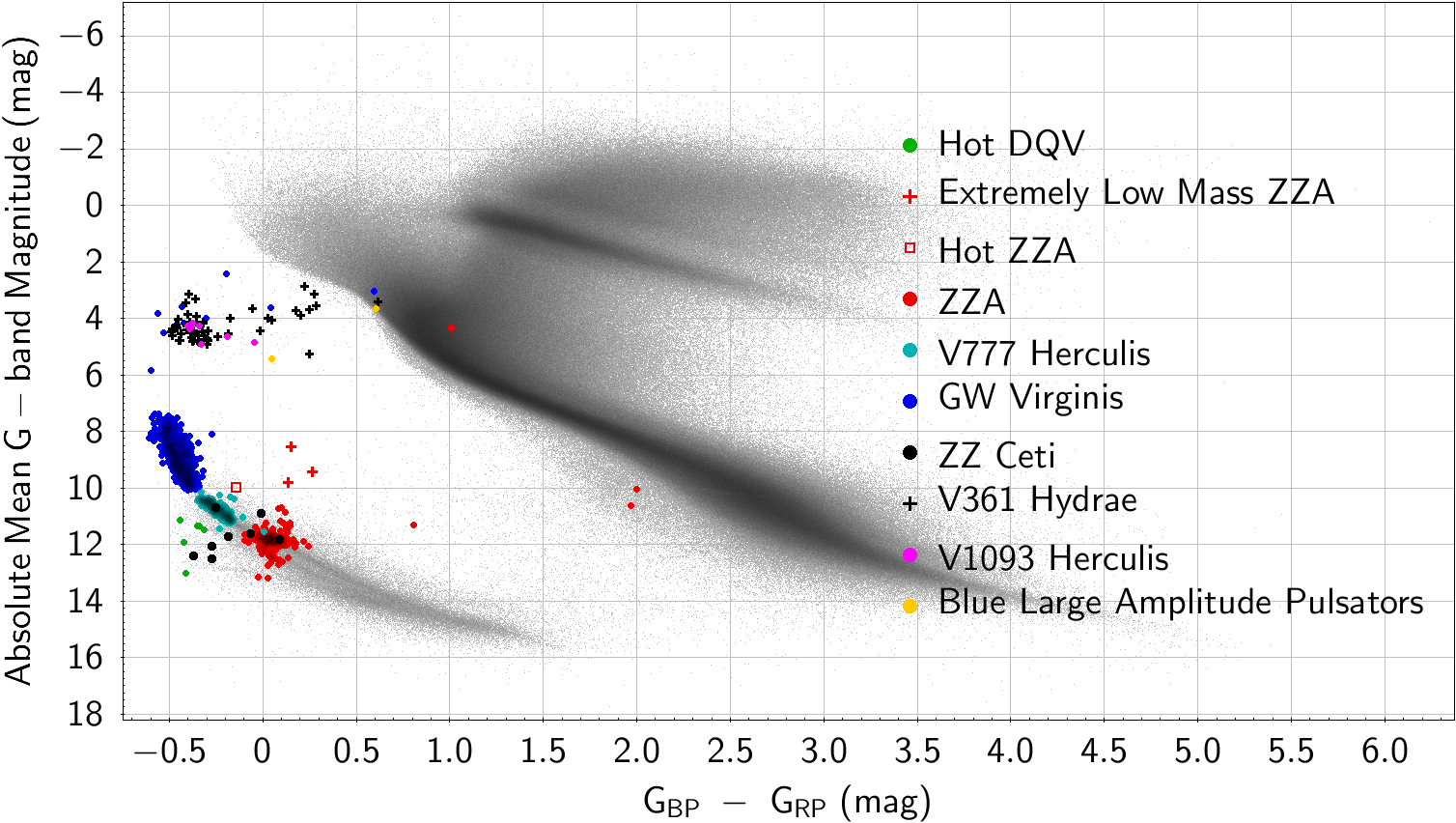}
  \caption{Colour-absolute magnitude diagram (CaMD) of white dwarfs, sub-dwarf variables, and blue large amplitude pulsators. The sources of the reference data set are plotted in grey-scale as background to facilitate locating the different areas.  }
  \label{Fig:cmdWD}
\end{figure}

There are ten different variability classes of variable white dwarfs (WD) and sub-dwarfs in the cross-match catalogue. Figure~\ref{Fig:cmdWD} shows the CaMD for these classes. 
As it is shown, several classes are overlapping or they are different sub-classes of a larger class, like the ZZ\,Ceti stars \citep[for a detailed review of pulsating white dwarfs, see][]{2019A&ARv..27....7C}.  Class labels are defined as follows. 

\begin{itemize}
    \item HOT\_DQV: These sources are DQ white dwarfs variables with C and H rich atmospheres.  In the CaMD plot it is clear that the majority of the stars are in the WD sequence below the area of V777\,Herculis stars. 
    \item ZZ\_Ceti: For ZZ\,Ceti types there are 6 subtypes in the catalogue, three of them concerning ZZA (DAV). But also there are few ZZ\,Ceti with no further subclassification. 
    \begin{itemize}
        \item ZZ: These are generic ZZ\,Ceti  without detailed class, they lay in the correct position in the CaMD.
        \item ZZA: ZZA (or DAV) are classical ZZ\,Ceti stars with DA spectral type with H atmospheres. They lay in the expected area in the CaMD of Fig.~\ref{Fig:cmdWD}, but there are 4 ZZA that seem to be well beyond the ZZ\,Ceti location. These sources originate from the VSX, which is very useful because of its diversity, but its cross-match is prone to mismatches.
        \item HOT\_ZZA: The only HOT-ZZA that survived the quality cuts for the CaMD is in the correct place with respect to ZZA and ZZB, as their effective temperature is in a similar range. 
        \item ELM\_ZZA: Extremely low mass (ELM) ZZA tend to have temperatures between 7\,800 and 10\,000K,  the difference between ELM-ZZA and ZZA is clear. The ELM-ZZA on the right is SDSS~J184037.78+642312.3, the first identified ELM-ZZA  \citep{2012ApJ...750L..28H}. 
        \item V777HER: The V777\,Herculis (or ZZB, DBV) are stars with He-rich atmospheres and their periods range between 100 and 1400~s \citep{2014A&A...570A.116B}. They are well defined in the CaMD, grouping in the WD sequence between the warmer GWVIR and the cooler ZZA.
        \item GWVIR: GW\,Virginis (or ZZO, DOV, PG1159) stars are a subtype of ZZ\,Ceti with  absorption lines of HeII and CIV, and it's the hottest known type of pulsating WD and pre-white dwarfs. The population in our catalogue is well defined. 
        There are some sources off the white dwarf sequence which lie closer to the horizontal branch.
    \end{itemize}
    \item Sub-dwarfs: 
    The cross-match catalogue contains two classes of sub-dwarf B stars: V361\,Hya and V1093\,Her.
    The two types are concentrated as expected in the extreme horizontal branch. Some of them (mostly V361\,Hya stars) can be redder than the main clump, but they follow the blue horizontal branch (BHB). Most of V361\,Hya stars are hotter (with effective temperatures in 28\,000--35\,000\,K) than V1093\,Her (23\,000--30\,000\,K), so the two populations are not distinct and overlap as predicted by their temperature range \citep{2016PASP..128h2001H}.
    \item BLAP: Blue large amplitude pulsators (BLAPs) have temperatures as hot as sub-dwarfs but with larger amplitudes \citep{2017NatAs...1E.166P}. In our catalogue, only two survived the astrometric cuts and they lay in the horizontal branch. 
\end{itemize}

\subsubsection{Long period and semi-regular variables}  
\begin{figure}
  \centering
  \includegraphics[width=\hsize]{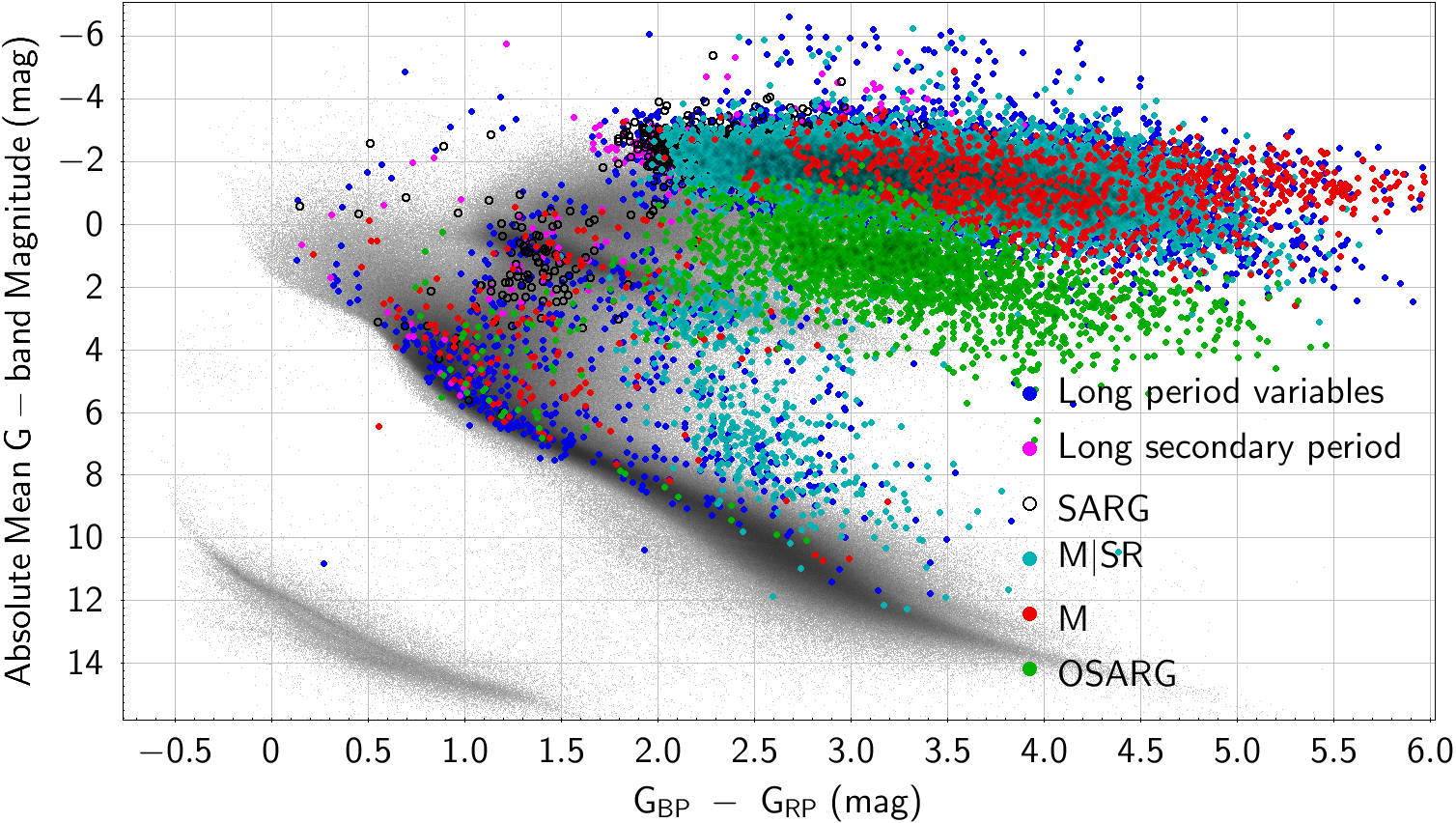}
  \caption{CaMD of long period variables.}
  \label{Fig:cmdLPV}
\end{figure}

\begin{figure}
  \centering
  \includegraphics[width=\hsize]{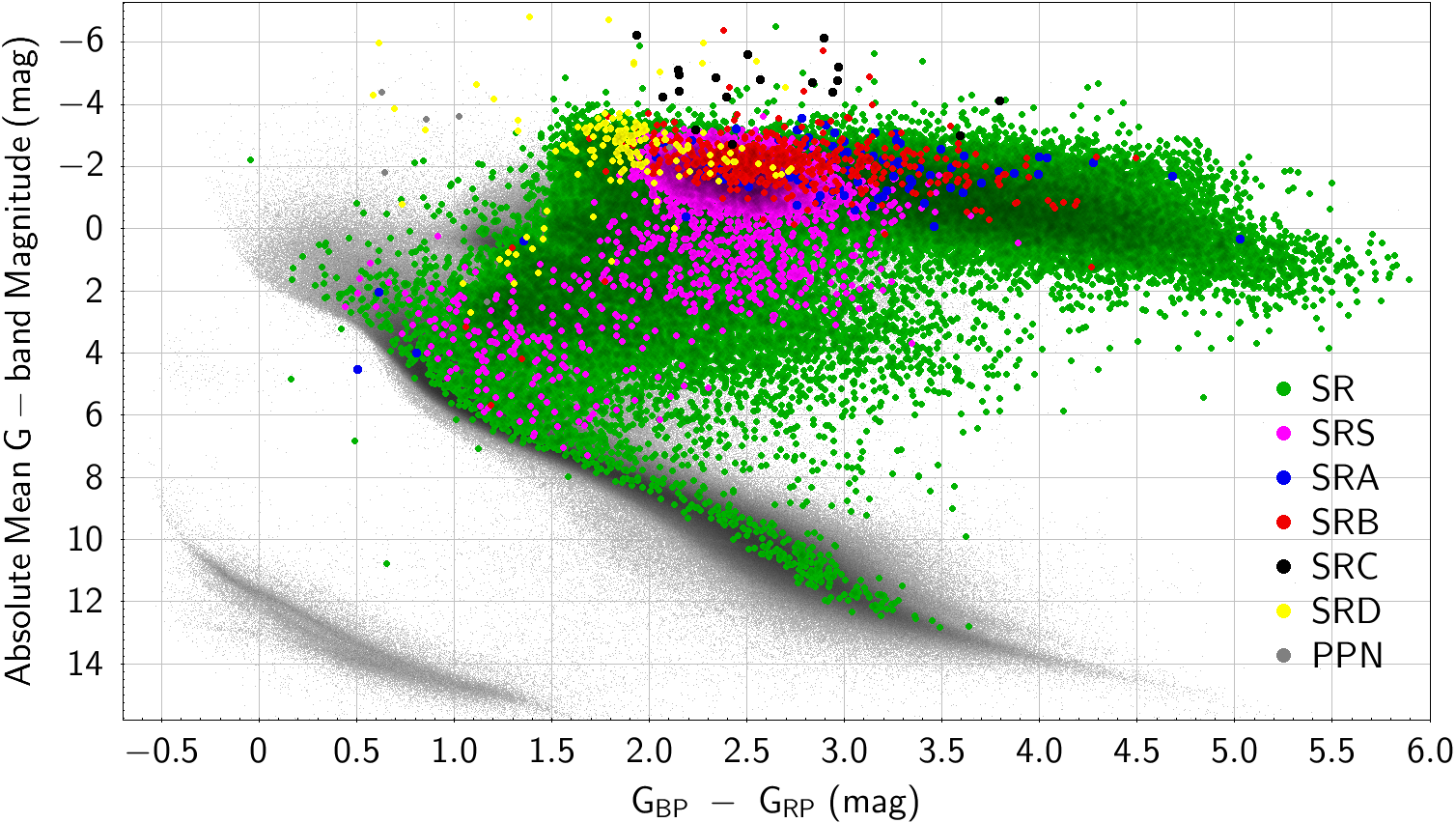}
  \caption{CaMD of semi-regular variables.}
  \label{Fig:cmdSR}
\end{figure}

The result of this work contains several classes of  long period variables. As for white dwarfs, several classes overlap or are subclasses of a generic one. Figures.~\ref{Fig:cmdLPV} and \ref{Fig:cmdSR}, present CaMD for long period and semi-regular varibles respectively. 
\begin{itemize}
    \item LPV: Long period variables include sources from surveys or catalogues that did not subclassify them.
    They are generally in the expected location in the CaMD, among the red giants, however $\sim$3\% of them are found in the main sequence. Some of them are due to literature misclassifications (with periods of less than a day), others might be mis-matched.
    \item M: M (or Mira, o\,Ceti) variables are late type stars with periods between 80 and 1000~days. They are very well defined in the red part of the CaMD  with little contamination.
    \item M|SR: These are long period variables that includes Mira and Semi-regular stars identified by classification in \textit{Gaia}~DR2 \citep{2019A&A...625A..97R}. The majority of this class occupies the expected area in the CaMD, overlapping the regions of Miras and LPVs, but there are some contaminants falling on the main sequence likely Young Stellar Objects \citep{2018A&A...618A..58M}.
    \item SARG: They are small amplitude red giants pulsating with periods from 10 to 100 days, a large fraction of them with long secondary periods and laying in the RGB or AGB branches. Most of them are well defined in Fig.~\ref{Fig:cmdLPV}, but there is a minority that is too blue or falls in the main sequence.
    \item OSARG: They are SARGs from OGLE; their location in the red giant branch has very little contamination. 
    \item LSP: Long secondary period variables are luminous red giants stars which have secondary period an order of magnitude longer than their primary \citep{1999IAUS..191..151W}. One third of LPVs exhibit this type of behaviour \citep{2007ApJ...660.1486S} and their periods range from 200 to 1500~days. In Fig.~\ref{Fig:cmdLPV} they lay into a well defined expected area overlapping with other LPVs, although some outliers extend to the main sequence.
    \end{itemize}
    
Semi-regular variables, generally, are giants or supergiants that exhibit irregular periods that vary, and some of them even show time lags of constancy. Figure~\ref{Fig:cmdSR} presents the CaMD for such classes.
\begin{itemize}
    \item SR: Semi-regular variables are  giants or supergiants of late type with no strict periodicity.
    Most of the ones that are in the main sequence are imported from ZTF\_Periodic\_Variables \citep{2020ApJS..249...18C}, likely due to misclassifications rather than mis-matches, as several of them were verified to have the same periods in  the \textit{Gaia} counterparts.
   \item SRA and SRB: Late type giants, semi-regular variables. SRA stars tend to have periods of 35--1200~days, while the SRB stars are more irregular, with cycles of 20--2300~days and also time intervals that show no variability. These classes are well defined in the CaMD and with only few outliers, although overlapping with the other SR types. This is justified as typically they all are of the same spectral type.
   \item SRC: This subclass consists of late type supergiants with periods that fall into the interval 30 to thousands of days. In the CaMD, they occupy a well defined area above the SRA and SRB stars.
   \item SRD: They are giants and supergiants of types earlier than SRA, SRB, and SRC, with variability periods from 40 to 1100~days. In the CaMD, they are close to but separate from the other subclasses, towards earlier spectral types.
    \item SRS: They are red giants with shortest periods than other SR, varying from a few days up to a month. This class is  defined in the same area as the other SR stars in the CaMD, but they appear to have also several contaminants. All of the SRS stars originate from the VSX.
   \item PPN: Protoplanetary nebulae with yellow supergiant post-AGB stars, exhibiting variability that resembles the SRD variables. The few that survived the quality cuts are found in reasonable places in the CaMD.

\end{itemize}

\subsubsection{RR\,Lyrae stars}
\begin{figure}
  \centering
  \includegraphics[width=\hsize]{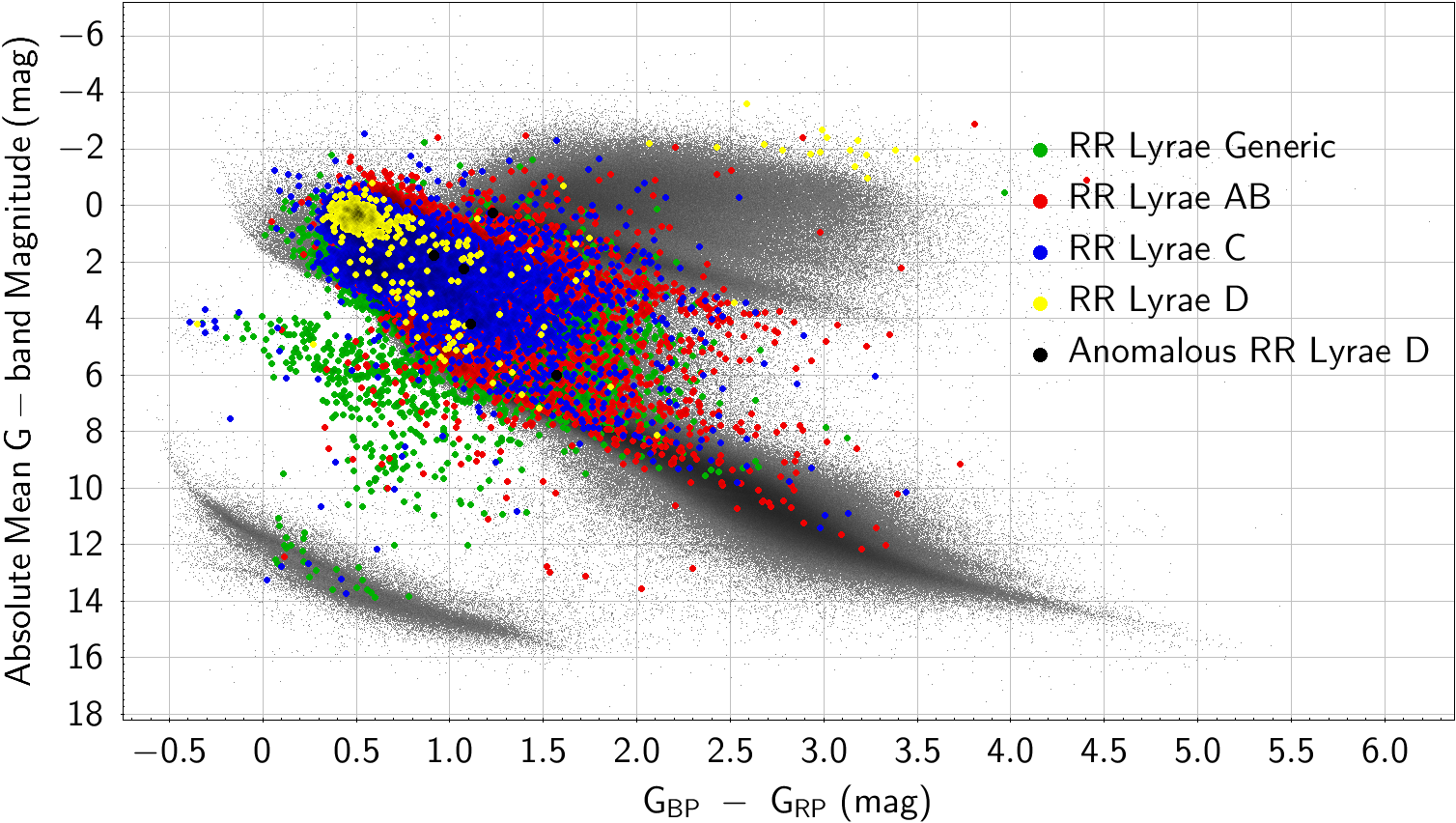}
  \caption{CaMD of RR Lyrae stars.}
  \label{Fig:cmdRR}
\end{figure}

Many input catalogues contain RR\,Lyrae stars, allowing us to construct a significant sample of this type of variable stars and of its subclasses. They are A to F type stars showing periodicity of less than a day and amplitudes that can reach 2 magnitudes in the optical. The RR\,Lyrae variables in the cross-match are divided into four subclasses and a generic one for the catalogues that do not provide detailed classification. Figure~\ref{Fig:cmdRR} shows that the majority of the sources falls into the expected place, but a significant fraction does not. Many sources are located in the lower part of the main sequence and some are between the main and white dwarf sequences. Visual inspection shows that some of them were mismatched sources. When in dense regions, two \textit{Gaia} sources may have a similar angular distance to the literature target and the most compatible magnitude associated with the incorrect counterpart. The user is encouraged to verify sources of these classes, especially if not filtering input catalogues.
\begin{itemize}
    \item RR: This is the generic class of RR\,Lyrae stars from catalogues that do not provide subclasses. The majority of those stars are from PS1\_RRL\_SESAR\_2017 \citep{2017AJ....153..204S}, which contains several problematic cases of sources laying in the white dwarf sequence or between the white dwarf and the main sequence, as expected because no selection based on score was applied to those candidates (unlike in PS1\_RRL\_SESAR\_SELECTION\_2017).
    \item RRAB: This is the most common RR\,Lyrae class, with asymmetric light curves and periods between 0.3 and 1~day. The majority of them lay in the expected place in the CaMD, with a few outliers towards the white dwarf sequence.
    \item RRC: They have symmetric and sinusoidal light curves and shorter periods than RRAB stars. In the CaMD, their majority has $G_{\rm BP} \ - \ G_{\rm RP} \sim 0.5$~mag, but extend also in the main sequence.
    \item RRD: They are double mode pulsators, which occupy a well defined region at  $G_{\rm BP} \ - \ G_{\rm RP} \sim 0.5$~mag,  but there are also some very red outliers.
    \item ARRD: They are anomalous RRD, double-mode pulsators that are similar to RRD but their ratio of periods is different. Very few ARRD survived the quality cuts for the CaMD and they are scattered towards the red part of the main sequence.
\end{itemize}

\begin{figure}
  \centering
  \includegraphics[width=\hsize]{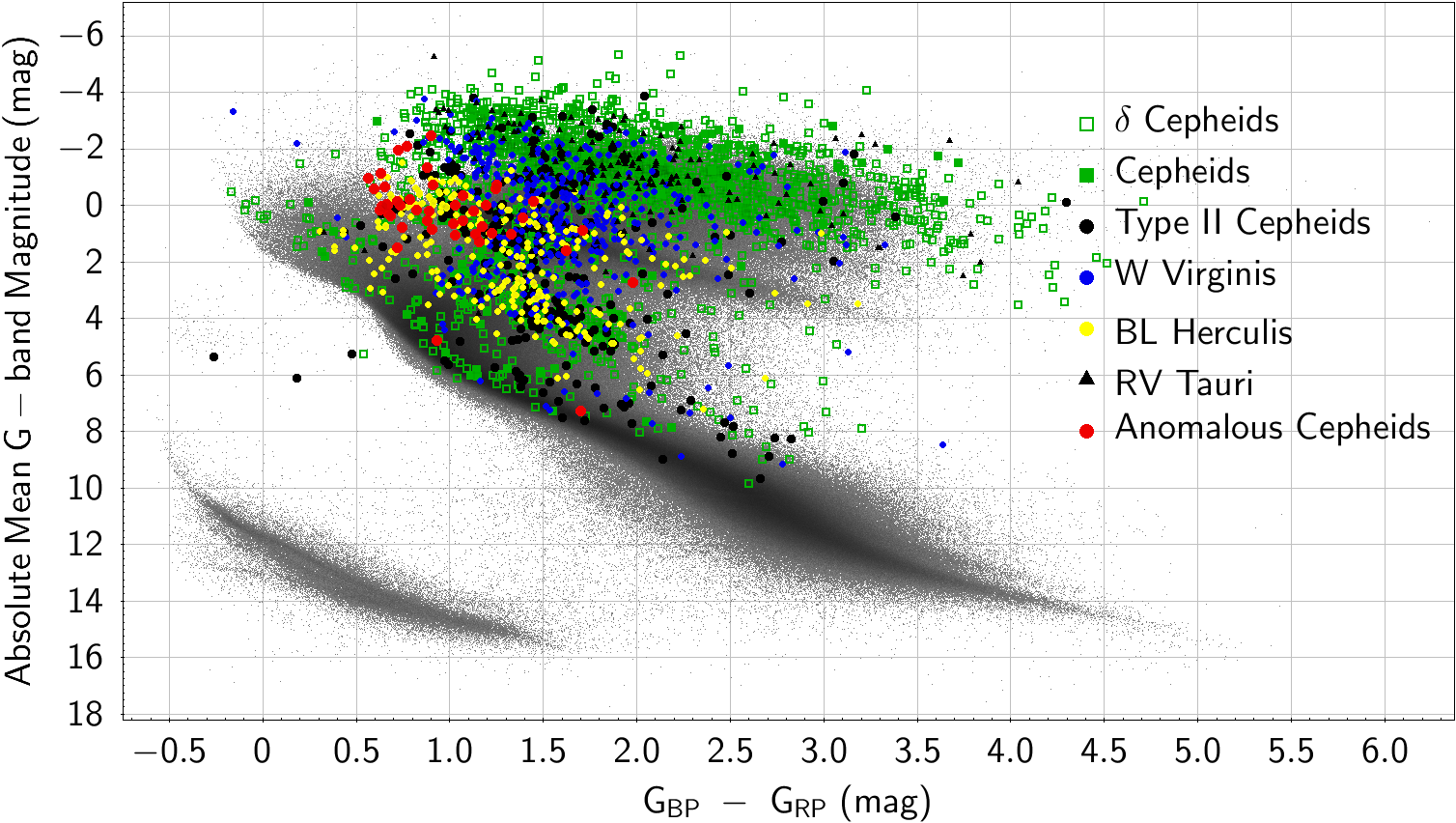}
  \caption{CaMD of the different types of Cepheids.}
  \label{Fig:camdCep}
\end{figure}

\subsubsection{Cepheids}
In our catalogue, we have included several types of Cepheids and the relevant CaMD is presented in Fig.~\ref{Fig:camdCep}. 

\begin{figure}
  \centering
  \includegraphics[width=\hsize]{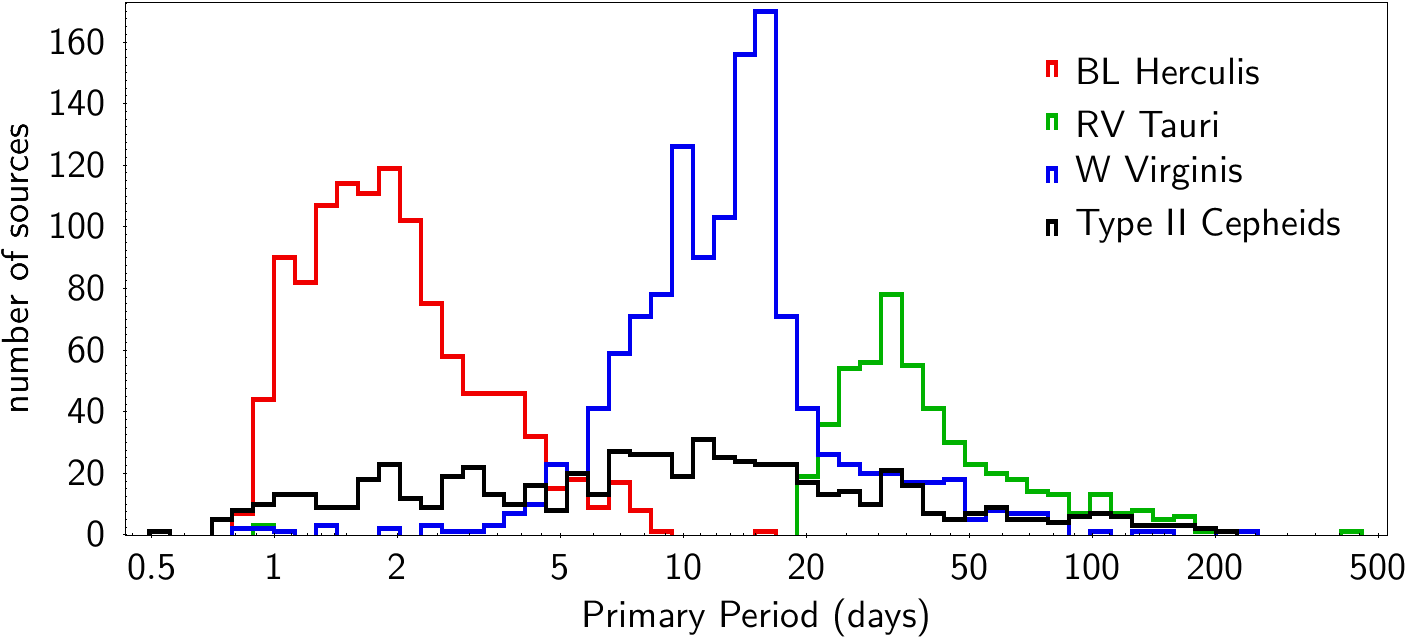}
  \caption{Period distribution of the Type~II Cepheids, as reported in the literature.The different colours show the three  subclasses and the generic class.}
  \label{Fig:pdistT2Cep}
\end{figure}

\begin{itemize}
    \item CEP: Cepheids are radial pulsating giants and supergiants with a large range of periodicities from $\sim$1 to more than 100~days. Their spectral type varies depending on their phase from F to K.  This class label includes all types of Cepheids from catalogues that do not provide detailed classification. Only a small number of sources of this class is included in the CaMD, half of them lay in a reasonable location, while the others fall on the main sequence.
    \item ACEP: Anomalous Cepheids, or BL\,Bootis, are pulsating variables that lay on the instability strip. Typically they have periods from a few hours to 2~days.
    The ACEP found in the CaMD occupy the expected position.
    \item DCEP: $\delta$ Cepheids, or classical Cepheids, tend to be brighter than the Type~II Cepheids, although there are sources in the cross-match that fall on the main sequence.
    \item T2CEP: This is a generic class of Type~II Cepheids from catalogues that don't provide further details about their subclass. They are pulsating variables with periods in the interval from 1 to more than 50~days. They are similar to classical Cepheids but with lower masses and luminosities, and tend to be older. They can be divided in 3 subclasses: BLHER, CW, and RV\,TAU. These subclasses have different period range, as shown in Fig.~\ref{Fig:pdistT2Cep}, while the generic class spreads in the full range of periods in the plot.
    \begin{itemize}
        \item BLHER: BL\,Herculis (or CWB) are the Type~II Cepheids variables with the shortest periods among the different subclasses. They have periods from 1 to 4~days and they lay in the region between the horizontal branch and the asymptotic giant branch. Only 90 BLHER survived after the quality cuts, some of them are found to be redder than expected.
        \item CW: W\,Virginis stars have periods between 10 and 20~days and are crossing the instability strip. They expand to the areas of BL\,Her and RV\,Tau in the CaMD. Their period distribution reported in the literature has tails that extend to the full range shown in Fig.~\ref{Fig:pdistT2Cep}.
        \item RV: RV\,Tauri variables are radially pulsating supergiants that change their spectral type along with their magnitude. Their spectral type span from F--G class to K--M, depending on their phase. Their periods are longer than 30~days, with typical values between 40--50~days. The RV\,Tau stars in the cross-match catalogue fall into expected CaMD location.
    \end{itemize}
\end{itemize}

\subsubsection{$\delta$\,Scuti and $\gamma$\,Doradus variables}
\begin{figure}
  \centering
  \includegraphics[width=\hsize]{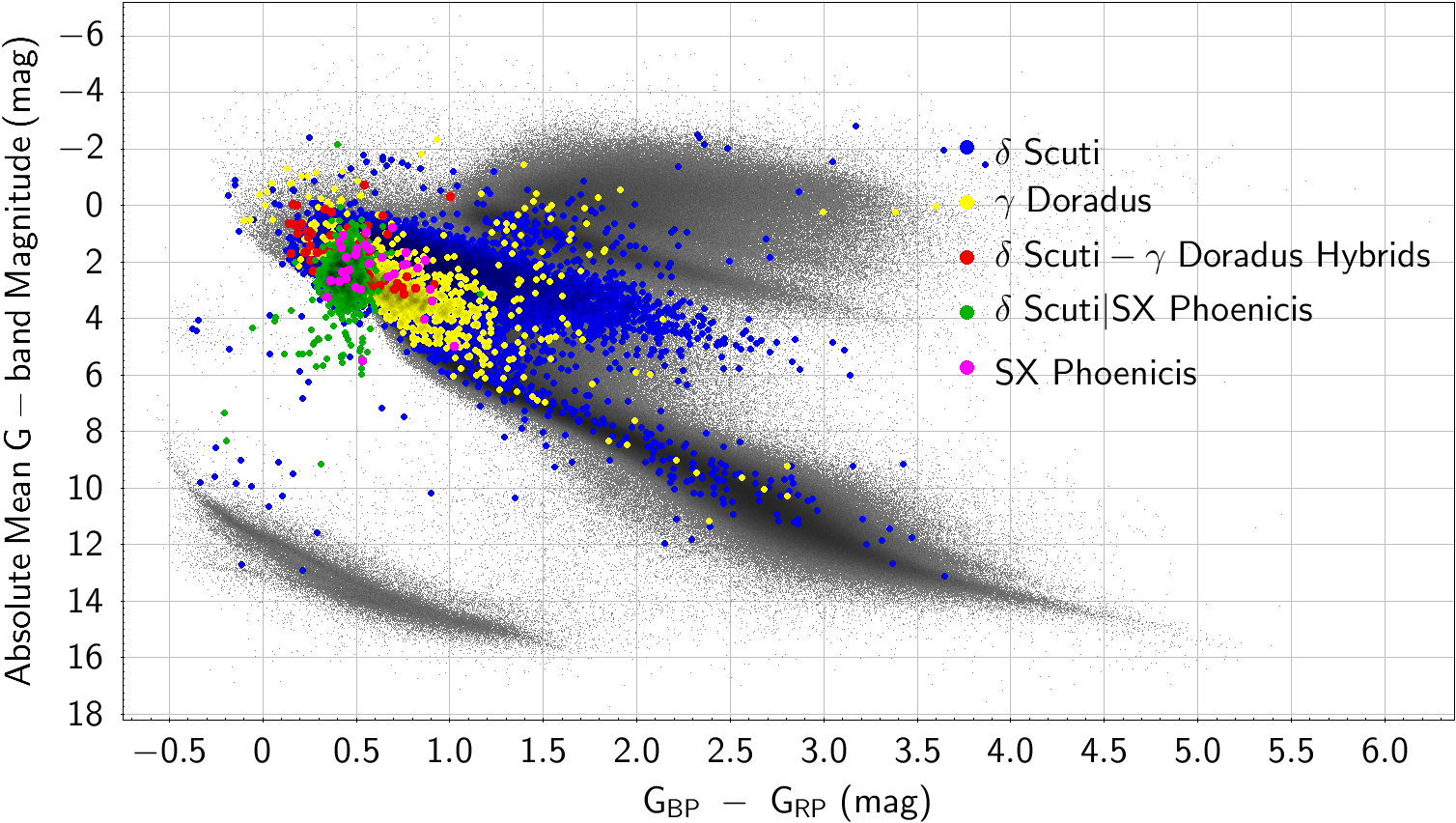}
  \caption{CaMD of the different types of $\delta$ Scuti and $\gamma$ Doradus  stars.}
  \label{Fig:camdDsct}
\end{figure}

Since $\delta$\,Scuti and $\gamma$\,Doradus variables are closely related and can also be hybrids, they are presented together in Fig.~\ref{Fig:camdDsct}.
\begin{itemize} 
    \item DSCT: $\delta$\,Scuti are pulsating variables similar to $\delta$\,Cepheids. They follow the same period-luminosity relation, but they have shorter periods (from 0.01 to 0.2~days). Their brightness varies with amplitudes between 0.003 to 0.9~magnitudes. Their spectral type is between A0 and F5. Usually,  $\delta$\,Scuti stars lie on the instability strip. Figure~\ref{Fig:camdDsct} shows several contaminants as the $\delta$\,Scuti representatives cover a large fraction of the main sequence, with some sources on the white dwarf sequence.
    \item SXPHE:  SX\,Phoenicis are considered similar to $\delta$\,Scuti stars that are sub-dwarfs with periods typically in the lower part of the DSCT range. They lay in the expected place of the CaMD. 
    \item DSCT|SXPHE: $\delta$\,Scuti or SX\,Phoenicis stars  identified by the classification of the variable sources of \textit{Gaia} data release 2. Generally they occupy a correct region, although there are some outliers.
    \item GDOR: $\gamma$\,Doradus are dwarfs with late A to late F spectral type that exhibit variability with non-radial pulsations. They usually have periods around 1~day. In Fig.~\ref{Fig:camdDsct}, they occupy the expected region, except for some outliers.
    \item DSCT+GDOR: The $\delta$\,Scuti and $\gamma$\,Doradus hybrids are variable stars that exhibit both g (GDOR) and p (DSCT) mode pulsations. They are found in the expected location of the CaMD.
\end{itemize}

\subsubsection{Other pulsating variables}
\begin{figure}
  \centering
  \includegraphics[width=\hsize]{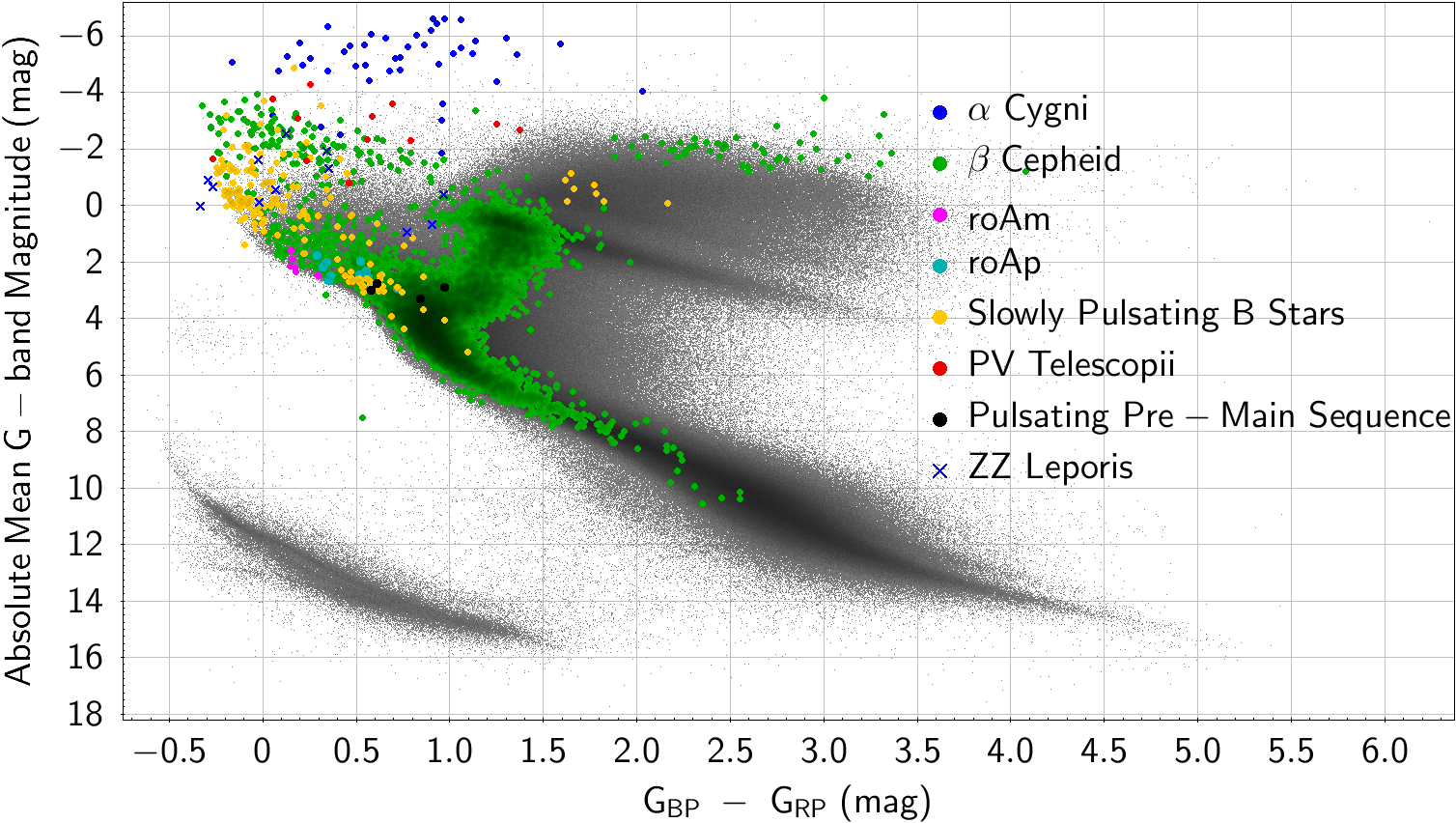}
  \caption{CaMD of the other types of pulsating variable stars.}
  \label{Fig:camdpulsRest}
\end{figure}

Additional pulsating types are presented in this subsection.
\begin{itemize}
    \item ACYG: $\alpha$\,Cygni stars are B--A supergiants exhibiting non-radial pulsations with a large range of periods. Their typical amplitude of photometric variability is about 0.1~magnitudes. In Fig.~\ref{Fig:camdpulsRest}, they may spread more than anticipated for A or B type stars.
    \item BCEP: $\beta$\,Cepheid stars are main sequence stars of O8--B6 spectral type exhibiting photometric and radial velocity variability  with short periods between 0.1 and 0.6~days. A large number of BCEP stars was cross-matched, with the majority originating from KEPLER\_VAR\_DEBOSSCHER\_2011, without applying probability thresholds, so the vast majority are misclassified sources and none of the ones in the CaMD lays in the expected region. If the {\tt{selection}} flag is not active, we encourage to reject unfiltered BCEP stars with {\tt{primary\_var\_type}} originating from  KEPLER\_VAR\_DEBOSSCHER\_2011 and also from ASAS\_VAR\_RICHARDS\_2012, which  fall on the RGB.
    \item SPB: Slowly pulsating B~stars that are pulsating in high radial mode with periods from 1 to 4~days \citep{2021MNRAS.501L..65S} and amplitudes up to 0.1~magnitudes.  Not all sources are compatible with the B type  colour (reddened or not) in Fig.~\ref{Fig:camdpulsRest}, with contaminants lower in the main sequence or among red giants. 
    \item ROAP: Rapidly Oscillating Ap stars are Ap/Fp stars that show photometric and radial velocity variability. Their period is shorter than 24~minutes \citep{2022MNRAS.510.5743B} and their amplitudes are lower than 0.01~magnitudes. The ROAP stars identified in the cross-match occupy the expected area in the CaMD, as seen in Fig.~\ref{Fig:camdpulsRest}.
    \item ROAM: Rapidly Oscillating Am stars, like ROAP, are chemical peculiar A~stars, but their spectral type is Am. They oscillate with periods between 8 and 22~minutes, and small amplitudes up to 0.01~magnitudes. They occupy a similar place in the CaMD next to ROAP.
    \item PVTEL: PV\,Telescopii are supergiants of several spectral types with hydrogen deficiency. They are divided into 3 sub-classes of different period ranges, from 0.5 to 100~days.
    In the CaMD, they lay in the expected region and their extended range of colours corresponds to the 3 sub-classes, where the hottest is of type~II with the shortest periods and the coolest stars are of type~III and exhibit the longest periods in the range. 
    \item ZZLEP: ZZ\,Leporis stars are central stars of planetary nebulae that exhibit photometric variations. They are O-type stars with periods that range from hours to days. The ZZ\,Lep stars  in Fig.~\ref{Fig:camdpulsRest} are compatible with O-type stars, some of which are reddened.
    \item PULS-PMS: Pulsating pre-Main Sequence stars are Herbig Ae/Be stars of B or A types that are in their PMS phase and have the right combination of physical parameters to become vibrationally unstable \citep{2006A&A...457..237Z}. In the CaMD of Fig.~\ref{Fig:camdpulsRest}, half of them seem to have the expected colour.
\end{itemize}

\subsection{Cataclysmic variables}
\begin{figure}
  \centering
  \includegraphics[width=\hsize]{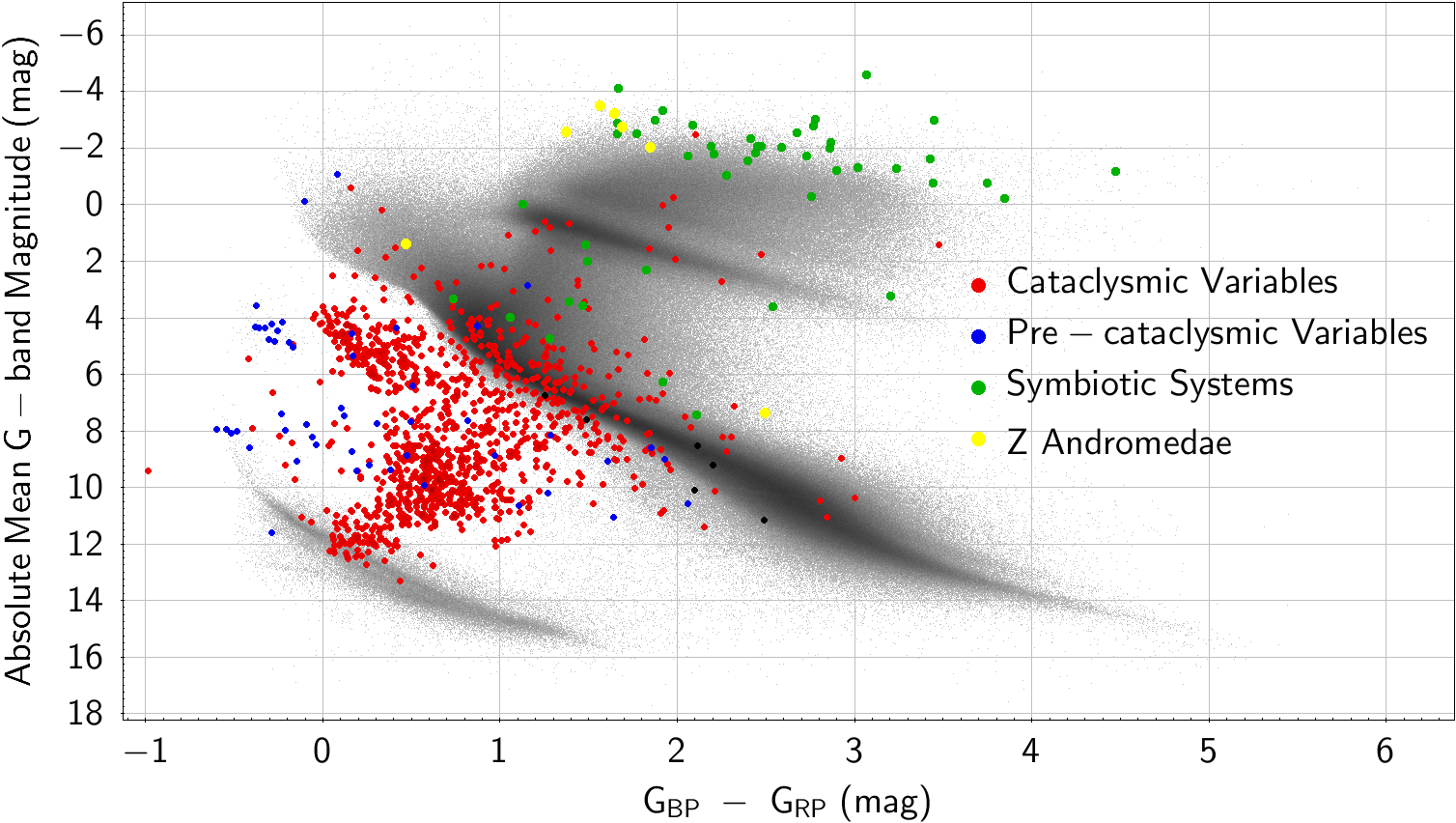}
  \caption{CaMD of Cataclysmic variable stars.}
  \label{Fig:cmdCataclysmic}
\end{figure}

A few types of cataclysmic variables are included in the cross-match. The most important ones are shown in Fig.~\ref{Fig:cmdCataclysmic} and discussed in this section.
\begin{itemize}
    \item PCEB: Pre-Cataclysmic variables or Post-Common Envelope binaries are binaries of a white dwarf and a main sequence star or a brown dwarf. In Fig.~\ref{Fig:cmdCataclysmic}, most of them lay in the extreme horizontal branch.
    \item CV: Generic type of cataclysmic variables including novae and dwarf novae, typically fall between the main sequence and the white dwarf sequence in the CaMD.
    \item ZAND:  Z\,Andromedae stars  include inhomogeneous types of symbiotic binary variables stars composed of a giant and a white dwarf. They display irregular variability with large amplitudes. Among the few cases that are present in Fig.~\ref{Fig:cmdCataclysmic}, the majority lay in the AGB branch.
    \item SYST: Symbiotic stars, which, like ZAND, form a heterogeneous group of objects, usually with a red giant or AGB star and a white dwarf. Most of them fall on the AGB branch in the CaMD.
\end{itemize}

\subsection{Eclipsing Binaries, Double Periodic Variables, and stars with exoplanet}\label{sec:eb}
\begin{figure}
  \centering
  \includegraphics[width=\hsize]{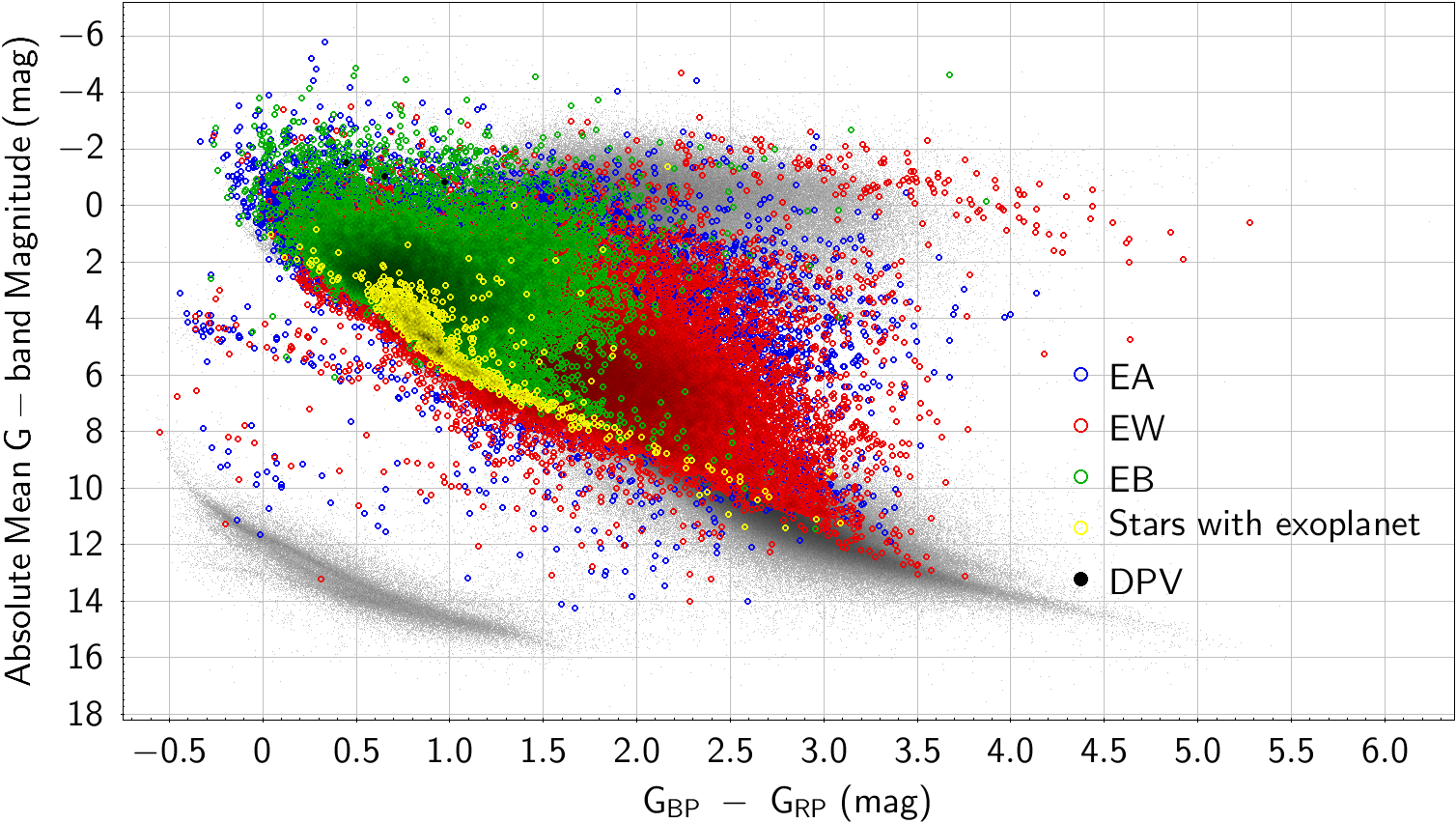}
  \caption{CaMD of eclipsing binaries and stars with transiting planets.}
  \label{Fig:cmdEB}
\end{figure}

\begin{figure}
  \centering
  \includegraphics[width=\hsize]{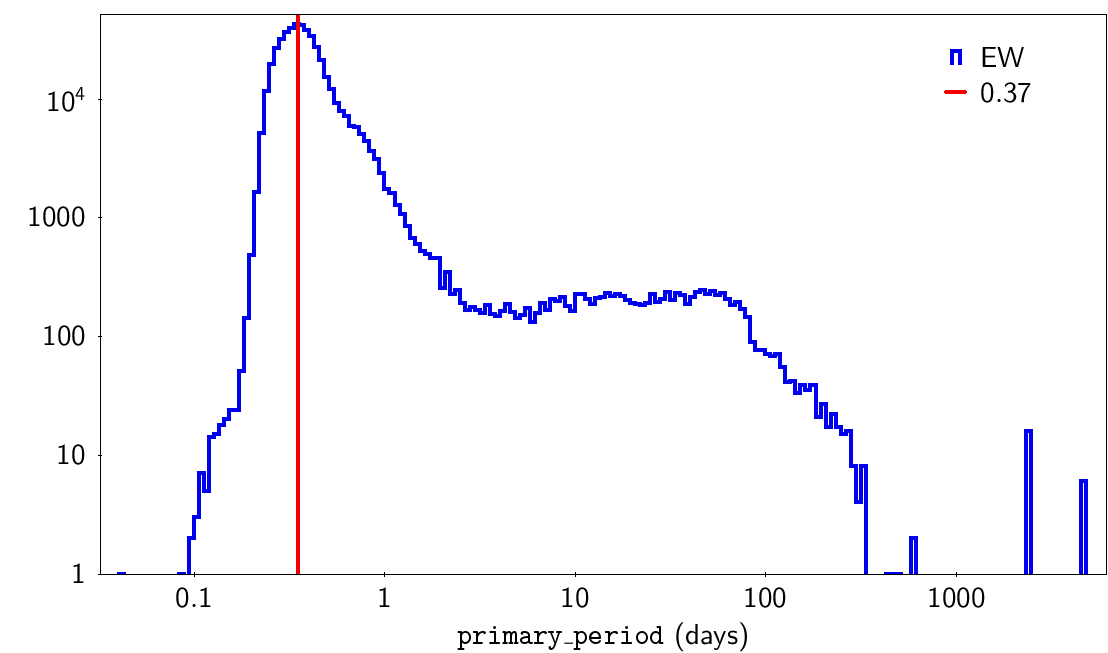}
  \caption{Distribution of periods from the literature for EW type eclipsing binaries.}
  \label{Fig:perEW}
\end{figure}

The CaMD for the eclipsing binary stars and stars with exoplanets in the data set is presented in Fig.~\ref{Fig:cmdEB}. The eclipsing binaries can be scattered throughout the HR diagram as shown in the figure.
\begin{itemize}
    \item EA: Algol ($\beta$\,Persei) type eclipsing binaries have stars with spherical or only slightly elliptical shape and the secondary eclipse is not always present in the time series. In Fig.~\ref{Fig:cmdEB}, it is clear that this type of objects can be anywhere in the CaMD.
    \item EB: $\beta$\,Lyrae eclipsing binaries have elliptical components and the secondary minimum is always visible in their light curve. The majority of such eclipsing binaries have periods larger than half a day. They usually cover the upper part of the main sequence and extend to the giants. 
    \item EW: W\,UMa-type eclipsing binaries are composed of two stars of similar spectral type between A and K with most of them being F or G. They have short periods, typically between 0.25 and 1~day. There are many red stars in Fig.~\ref{Fig:cmdEB} and $\sim$3\% have periods longer than 2~days in the literature, $\sim$30\% of which have different classifications (e.g., ROT, YSO) in other catalogues. The literature period distribution of this class is shown in Fig.~\ref{Fig:perEW} with a strong peak at around 0.37~days, as expected \citep{2012MNRAS.421.2769J}, but also with a tail extending to more than 200~days. 
     \item DPV: Double periodic variables are semi-detached interacting eclipsing binaries that exhibit photometric variability with two distinct periods. 
    Only 3 of them are shown in Fig.~\ref{Fig:cmdEB}.
\end{itemize}

Although not a system of binary stars, stars with transiting exoplanets (EP) are added to Fig.~\ref{Fig:cmdEB}, where they lie on the main sequence.

\subsection{Eruptive}\label{sec:eruptive}
\begin{figure}
  \centering
  (a)
  \includegraphics[width=\hsize]{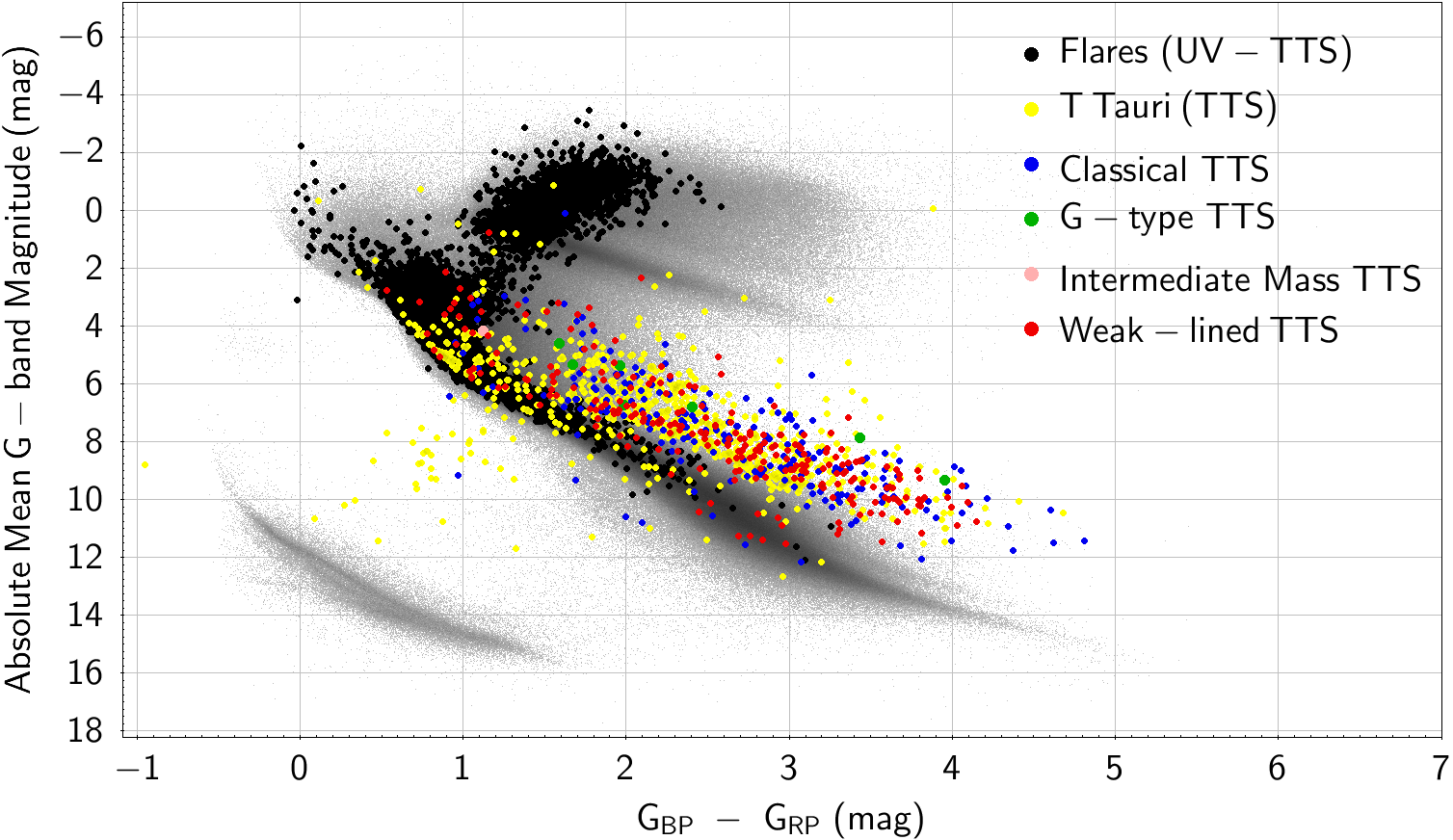}
  (b)
   \includegraphics[width=\hsize]{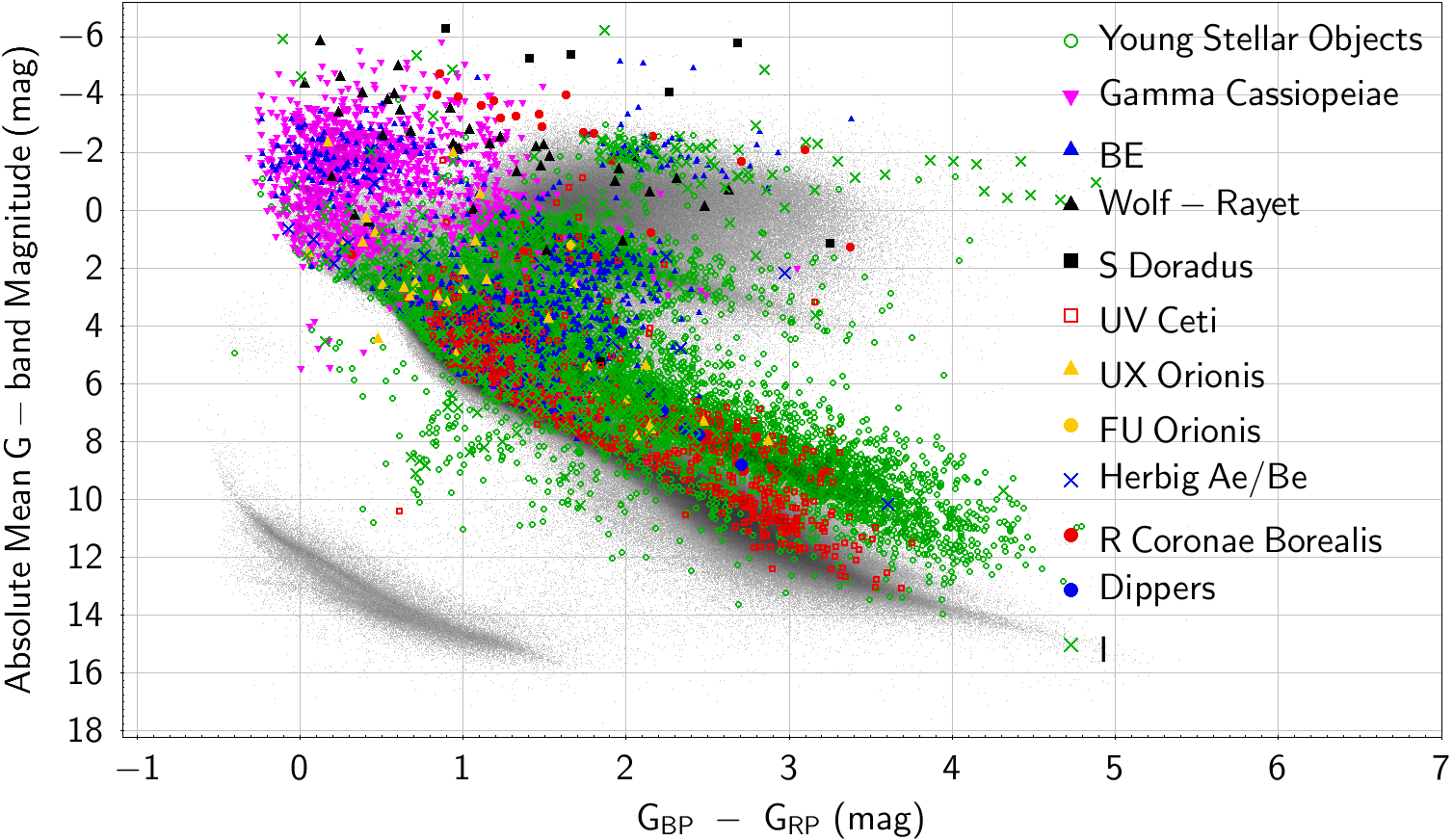}
  \caption{CaMD of eruptive type variable stars. The upper plot (a) shows the various TTS types and flare stars while the lower plot (b) the rest of eruptive type stars.}
  \label{Fig:cmdTTS}
\end{figure}

The compilation of variables from the literature contains 18  eruptive variability types. Many of them are different subtypes of T\,Tauri stars (TTS), which are plotted in Fig.~\ref{Fig:cmdTTS}(a) separately from other eruptive types in Fig.~\ref{Fig:cmdTTS}(b). In both plots there are stars fainter and bluer than the expected pre-main sequence locus. It is likely due to the circumstellar disks of these stars at high inclination. Thus the photosphere is strongly extincted,  and their optical colours are bluer due to the light scattered by the disk atmosphere. 
 A short discussion of the properties of all available eruptive stars follows.
\begin{itemize}
    \item TTS: T\,Tauri is the generic class of pre-main sequence objects. They are generally low to intermediate mass stars in a stage between protostars and low-mass main sequence stars. In Fig.~\ref{Fig:cmdTTS}(a), they occupy the expected region, however there is a small fraction bluer than the main sequence or falling on the main sequence. Most of these stars are in the Orion Molecular Cloud. 
    There are several TTS subclasses in the cross-match catalogue, depending on their spectra \citep{1994AJ....108.1906H,1999AJ....118.1043H}.
    \begin{itemize}
        \item CTTS: Classical TTS are well-studied stars. They are young accreting stars in their late stages of their evolution from protostars to the main sequence. They are well defined in Fig.~\ref{Fig:cmdTTS}, although there are a few misplaced sources, half of which are part of the Orion Molecular Cloud.
        \item GTTS: G-type TTS are G and K0 type TTS. Only a few GTTS are available and they fall in the expected place in the CaMD.
        \item WTTS: Weak-lined or `naked' TTS have little or even no accretion disk. They also follow the TTS trend in the CaMD with a few exceptions. 
        \item IMTTS: Intermediate mass TTS have masses between 1 and 4$M_\odot$ and are considered precursors to the PMS Herbig Ae/Be stars \citep{2017A&A...608A..77L}. 
    \end{itemize}
    \item Flares: This is a generic type, encapsulating several other types exhibiting flares due to magnetic activity (UV\,Ceti, TTS, etc.). In the CaMD, it is evident that they are spread all over the main sequence and in the RGB.
    \item HAEBE: Herbig Ae/Be variables are young stars of spectral types A or B. There are only a few but mostly in the expected location of the CaMD, including some very reddened ones.
    \item FUOR: FU Orionis variables are pre-main sequence stars closely related to the evolutionary stages of T\,Tauri stars. They are characterised by rapid and strong photometric and spectral variability. There are only 3 FUOR stars surviving the quality cuts, out of the 9 in the cross-match catalogue, and they are in reasonable places in the corresponding CaMD.
    \item UV: UV\,Ceti flare stars have spectral types K or M. Figure~\ref{Fig:cmdTTS}(b) shows a lot of stars spreading in the main sequence up to earlier spectral types.
    \item GCAS: $\gamma$\,Cassiopeiae stars are of O9--A0 type and thus occupy the expected place in the CaMD. Some are of later types but with no obvious problems. Some of them have been assigned  different types in other input catalogues (e.g., some of those in the extreme horizontal branch are also classified as CV).
    \item BE: B-type emission line variables which could be $\gamma$\,Cas stars or $\lambda$\,Eri. 
    In the cross-match catalogue, there are many misclassified objects in the lower part of the main sequence. 
    \item UXOR: UX\,Orionis stars are a subclass of Herbig Ae/Be stars, consistent with their location in the CaMD.
    \item RCB: R\,Coronae Borealis stars can have several diverse spectral types. In this work, a large fraction of them are post-AGB stars that exhibit RCB variations but they share the same origin.
    \item SDOR: S\,Doradus stars (or Luminous Blue Variables) are evolved stars characterised by large amplitude variations of Bpec to Fpec spectral type, as confirmed by their position in the CaMD. 
    \item YSO: Young Stellar Objects of generic type, many of them are TTS. The majority of them lay on the TTS region, but there are others scattered all around.
    \item I: Irregular stars that are mostly Young Stellar Objects, which is confirmed by their distribution in the CaMD. 
    \item Dippers: These pre-main-sequence stars of K and M spectral type exhibit dips in their light curves and don't have  a strict periodicity. The few Dippers presented in Fig.~\ref{Fig:cmdTTS}(b) are in the expected colour range this type of stars. 
    \item WR: Wolf-Rayet is a group of massive stars that present broad emission lines. They have high temperatures and luminosities and considered as descendants of O-type stars. Their variability is not periodic. In the CaMD, few WR stars lay at the expected area and others appear to be reddened.   
\end{itemize}

\subsection{Rotational}
\begin{figure}
  \centering
  \includegraphics[width=\hsize]{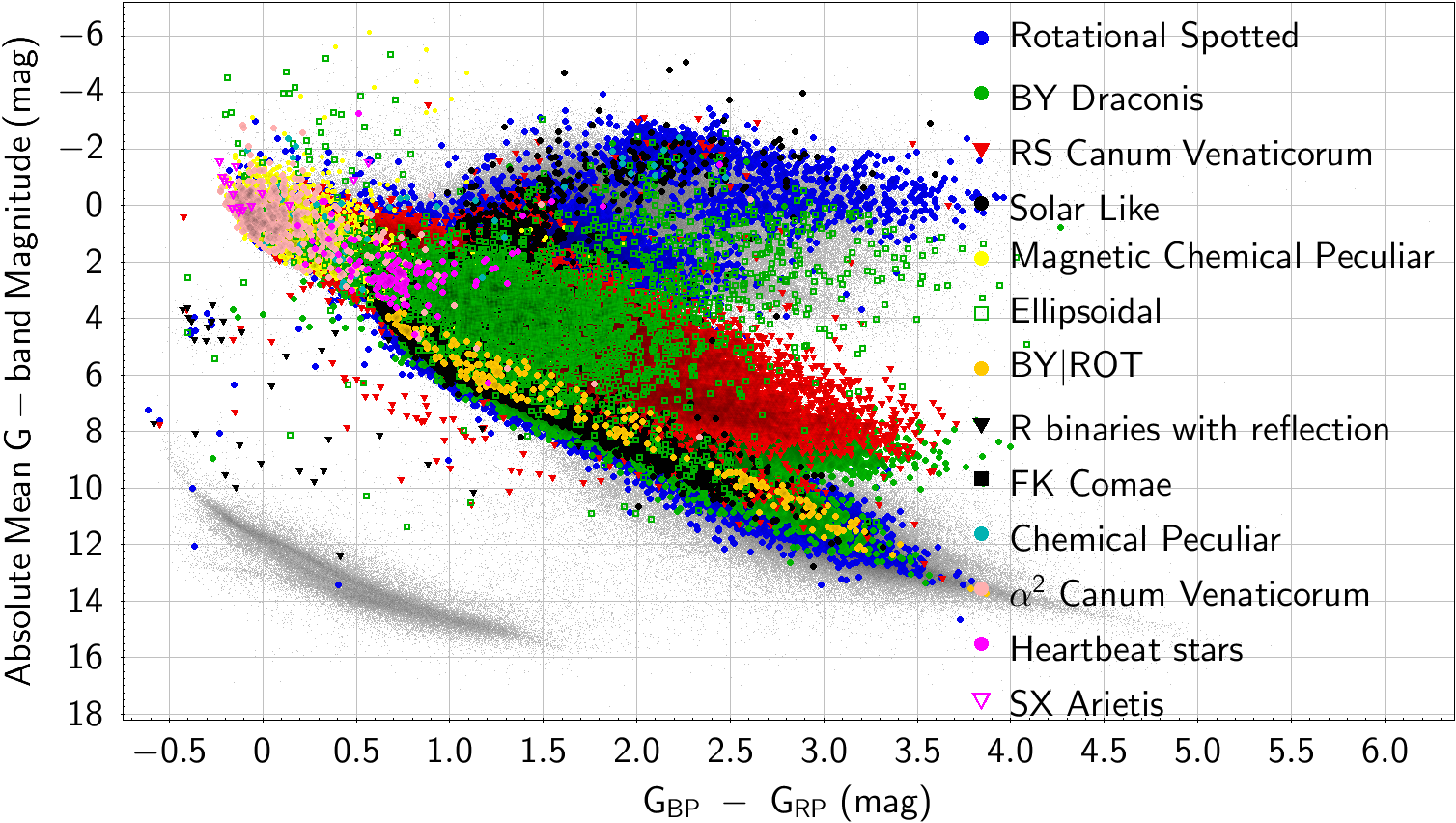}
  \caption{CaMD of rotational types variable stars.}
  \label{Fig:cmdRot}
  \end{figure}

Rotational variables are stars whose variability is caused by their rotation and asymmetries in shape or non-uniform surface brightness. The cross-match catalogue contains 13 types of rotational variables (some of them overlapping) and their CaMD is shown in Fig.~\ref{Fig:cmdRot}. 
  
  \begin{itemize}
      \item ROT: This is a generic class of spotted stars and are scattered everywhere in the CaMD.
      \item RS: RS\,Canum~Venaticorum variables are close binary systems of late spectral type that  exhibit chromospheric activity. 
      As shown in Fig.~\ref{Fig:cmdRot}, they extend to all of the main sequence and many of them are very red. The vast majority of these sources originates from the automatic classification of ZTF\_PERIODIC\_CHEN\ 2020.
      \item ACV: $\alpha^2$\,Canum~~Venaticorum variables are chemical peculiar main sequence stars of B8p--A7p type with strong magnetic fields. They have periods that vary from 0.5 to more than 100~days. Most of the ACV stars fall into the expected region of the CaMD with a few red outliers. A large fraction these outliers are listed in the VSX and originate from \cite{2009A&A...506..569K}.
      \item SXARI: SX\,Arietis are B-type chemical peculiar stars with strong magnetic fields and periods of about 1~day. They are similar to ACV stars but with higher temperatures, therefore there is some overlap with their distribution in the CaMD. Our list includes a few SXARI stars whose period is much longer than 1~day and thus their class is spurious.
      \item MCP: Magnetic Chemical Peculiar stars 
      that include Ap, HgMn, and Am types. They have a natural overlap with ACV and SXARI variables (and many of them are classified as ACV in other catalogues).
      \item CP: This is a generic class of chemically peculiar variables originating from \cite{2012ApJS..203...32R}, which were selected using a Random Forest classifier. It includes mostly hot stars but also few cooler stars of G and later spectral type, with \mbox{$G_{\rm BP}-G_{\rm RP}$}>1.5.
      \item FKCOM: FK\,Comae~Berenices variables are G to K giants that rotate rapidly and have strong magnetic fields. Only a few FKCOM variables exist in the cross-match catalogue but they have the expected colour and absolute magnitude for their type. 
      \item BY: BY\,Draconis stars are dwarfs that have inhomogeneous surface brightness and exhibit chromospheric activity. They have periodic variability with periods that can vary from less than a day to more than 120~days. The cross-match catalogue has a large number of BY variables that fall on the main sequence, however a significant fraction of them has been classified as other classes as well.
      \item ELL: Ellipsoidal variables are close binaries whose light curves do not contain an eclipse but their variability is due to the distortion of their shape from the mutual gravitational fields, thus the observed light varies because of varying projected surface towards the observer. The sample of ellipsoidal variables is scattered in all the CaMD with the majority laying on the main sequence.
      \item HB: Heartbeat variables are binary star systems with eccentric orbits that cause both variations of stellar shapes and vibrations induced by such changes.
      There are about 150 heartbeat stars in the cross-match catalogue,  $91\%$ of which is also classified as eclipsing binary in various catalogues. The majority of the ones that passed the quality cuts for the CaMD have colour 0.1<\mbox{$G_{\rm BP}-G_{\rm RP}$}<1.0~mag, only a few of them are redder than that.
      \item SOLAR\_LIKE: These stars exhibit chromospheric activity and include BY, ROT, and Flares types. In the CaMD, most representatives fall on the main sequence, although there are other sources in the red giant branch.
      \item R: close binaries that  exhibit strong reflection in their light curves (re-radiation of light of the hotter star from the surface of the cooler one). Most of these stars fall in the region between the main sequence and the white dwarfs; some are found in the extreme horizontal branch too.
      \item BY|ROT: similar to SOLAR\_LIKE, it includes stars of types BY or ROT as defined before.  
  \end{itemize}
  
  \subsection{Extragalactic content}
\begin{figure}
  \centering
  \includegraphics[width=\hsize]{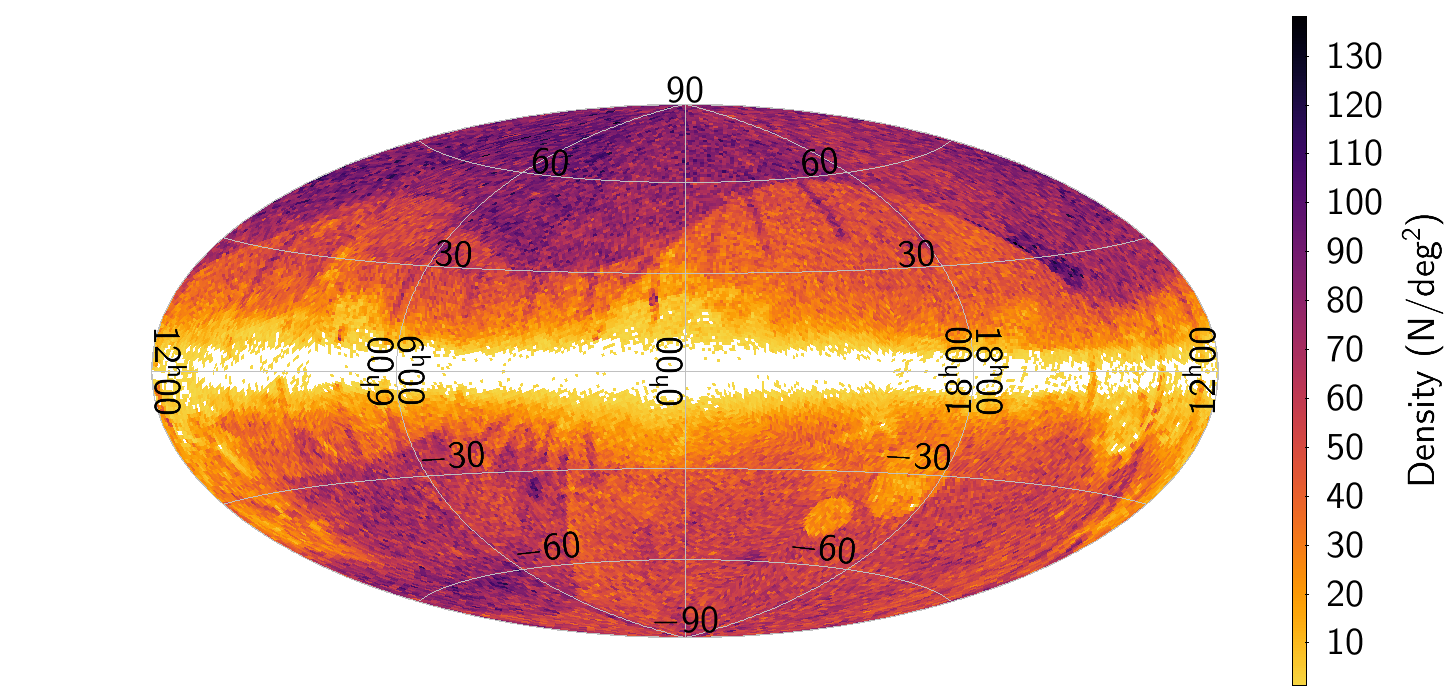}
  \caption{Sky map of 1\,801\,094 active galactic nuclei, blazars, and quasars in general.}
  \label{Fig:skyQso}
  \end{figure}

\begin{figure}
  \centering
  \includegraphics[width=\hsize]{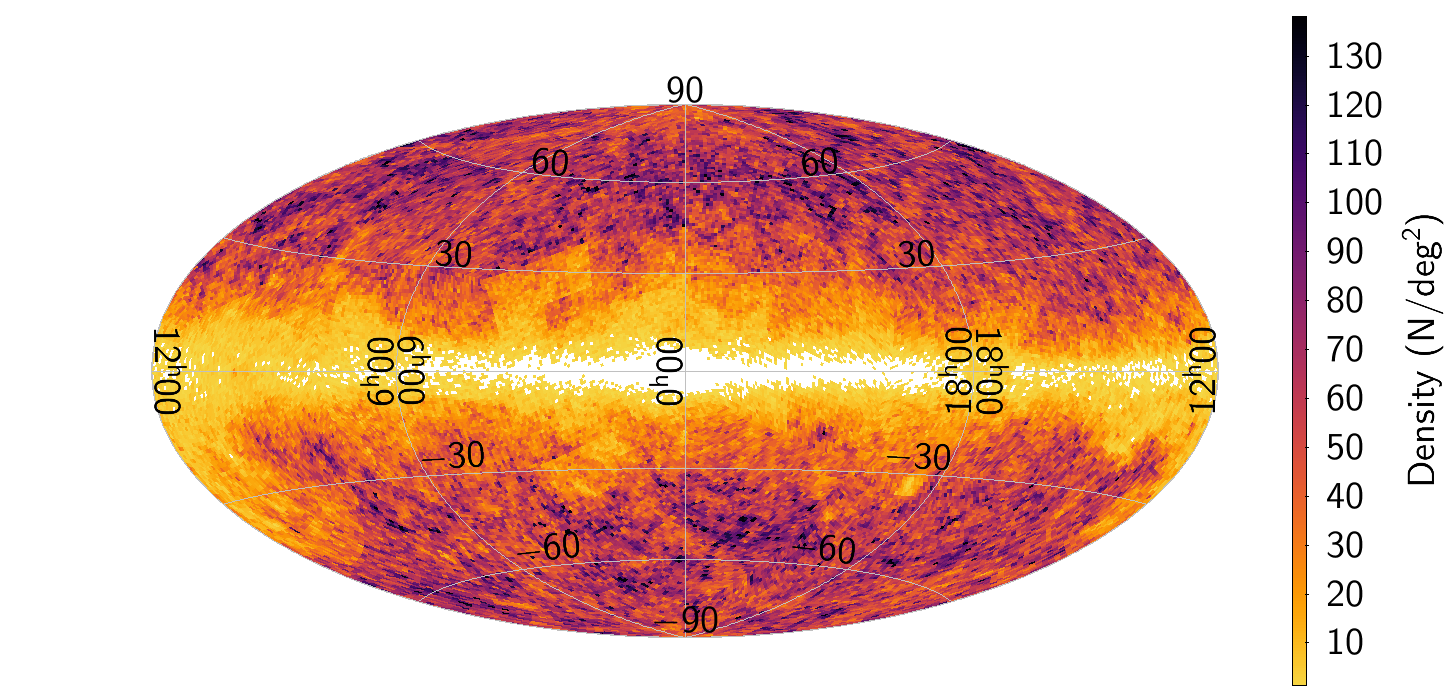}
  \caption{Sky map of the 1\,746\,224 galaxies in the cross-match catalogue.}
  \label{Fig:skyGal}
  \end{figure}

{\label{subsec:exragalactic}}
  A large number of sources in the cross-match catalogue concerns galaxies and active galactic nuclei. Most of input catalogues used to cross-match both galaxies and AGN were internal \textit{Gaia} catalogues and  their content could be identified by their source ids. The catalogues containing AGN had various levels of detail in their classification (AGN, BLAZAR, BLLAC, QSO), although the majority the sources were grouped as QSO as a generic class. For galaxies, no subclasses were reported. Figures~\ref{Fig:skyQso} and~\ref{Fig:skyGal} show the sky distribution all types of quasars and galaxies with darker colours indicating  higher density of objects. Both figures show that the galactic plane is avoided. 
  As the main contributions for both galaxies and quasars are from  \textit{Gaia} products, their properties are discussed in detail in their corresponding papers \citep{floisvos,DR3-DPACP-133}
\subsection{Class overlaps} 
\begin{figure}
  \centering
  \includegraphics[width=\hsize]{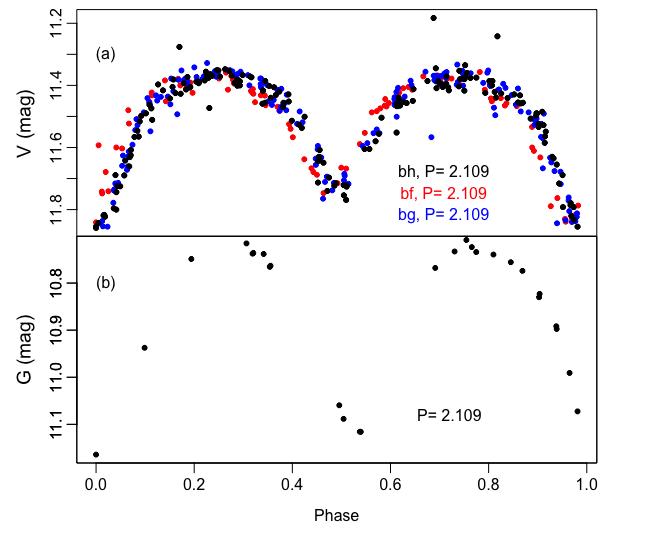}
  \caption{Folded light curve for \textit{Gaia}~DR3 4066039874096072576 using different colours for each ASAS-SN camera. (a) The recovered period is  almost always the same at 1.0545~days in the 3 different cameras provided by ASAS-SN, which is half of the period referred to in the literature. The folded light curve is plotted with twice the period recovered in each camera.
  (b) The same source using data and the period found in \textit{Gaia}~DR3. }
  \label{Fig:foldedLc}
  \end{figure}

{\label{subsec:overlaps}}
Due to the large number of catalogues that contributed to this cross-match, different classes might be associated with the same sources. Table~8 (available through the Centre de Donn\'ees astronomiques de Strasbourg website) shows the number of sources that overlap based on their superclasses, alphabetically ordered. The first column shows the {\tt{primary\_superclass}} and the 51 columns that follow, the overlapping superclasses taken from {\tt{var\_types}}. Not to confuse with the same classes, the numbers of sources classified as the same type in different catalogues (i.e., the diagonal of the table) have been set to zero. Some reasons that lead to class overlaps are listed below.
\begin{itemize}
    \item Mismatches: Due to the statistical approach used and the fact that each catalogue was treated separately, it is possible that \textit{Gaia} sources are erroneously assigned to input catalogue counterparts. This problem may occur more frequently in crowded regions and depends also on the astrometric accuracy of each catalogue.
    \item Misclassifications: Input catalogues might include misclassified sources, especially when generated by automatic methods. An example is presented in Table~\ref{table:missclass} for \textit{Gaia} DR3 source\_id 4066039874096072576, which is matched in 4 catalogues. This table lists the input catalogues, the identifiers of the source in each catalogue (with an additional online information, if present), the coordinates, variability types, and periods. ASAS-SN classified this source as a semi-regular (with a classification probability of 0.537) and identified a variability period of $\sim$18~days. However, the other catalogues classified it as an eclipsing binary (EW, EB, and ECL) with a period of $\sim$2~days.
    The ASAS-SN database\footnote{\url{https://asas-sn.osu.edu/variables/252221}} was used to download the photometric data of the source.
    Running a Lomb-Scargle \citep{1976Ap&SS..39..447L,1982ApJ...263..835S} period search \citep[using the R implementation of the \texttt{lomb} package;][]{lomb} for each camera separately, it was found that the periods were consistent with each other, and after doubling them \citep[as often needed for eclipsing binaries, as they have two minima per cycle instead of only one, as targeted by the sine function in this period search method; see fig.1 of][]{2014EAS....67..299H}, they corresponded to the 2.1091~day period identified in the other catalogues (see Fig.~\ref{Fig:foldedLc}a). This source is published also in \textit{Gaia}~DR3 as an eclipsing binary with the same period. The period provided from ASAS-SN was recovered as a secondary peak in the frequencygramme, but the folded light curve was worse, suggesting that the correct type is EW rather than SR.
    \item Multiple classes: Some classes are not excluding other ones and sources could be identified in the literature as a combination of two (or more) classes, such as Cepheids in eclipsing binaries, BY\,Draconis stars with flares of UV\,Ceti variables, etc. 
\end{itemize}

Table 8 shows that the most overlapped class is ECL as {\tt{primary\_superclass}} with  RR Lyrae stars with 58\,811 cases. However, the rate of overlap is a low as the cross-match catalogue contains more than 1.1~million sources which their {\tt{primary\_superclass}} is ECL. Looking at the {\tt{catalogue\_labels}} of these $\sim$59K sources reveals that 56\,361 of these are in PS1\_RRL\_SESAR\_2017. This catalogue contains sources without
filtering on the class probability. The high probability sources of this catalogue are provided in PS1\_RRL\_SESAR\_SELECTION\_2017 and only 1\,553 cases overlap with ECL. Moreover, the shapes of the light curves of EW and RRC stars are very similar and prone to confusion \citep{2009AJ....138..466H}. Indeed of the 1\,553 overlapped sources in PS1\_RRL\_SESAR\_SELECTION\_2017, 1\,162 are classified as RRC and EW.

Another significant overlap is between AGN and CST sources where 19\,434 cases exist. In this case the main contributor is GAIA\_WD\_GENTILEFUSILLO\_2019 with 18\,001 sources while the rest are from SDSS\_CST\_IVEZIC\_2007. Regarding the first catalogue no filtering was applied, selecting only the reliable sources \citep[see][]{2019MNRAS.482.4570G} reveals that 441 sources are overlapping. Also here it should noted that the overlap rate is very low as there are $\sim$1.8~million sources with AGN as {\tt{primary\_superclass}}.

\section{Selection of the least variable sources in ZTF and TESS}\label{sec:leastvariables}

\subsection{Least variable sources in ZTF}\label{subsec:ztf}
\begin{figure}
   \centering
   \includegraphics[width=\hsize]{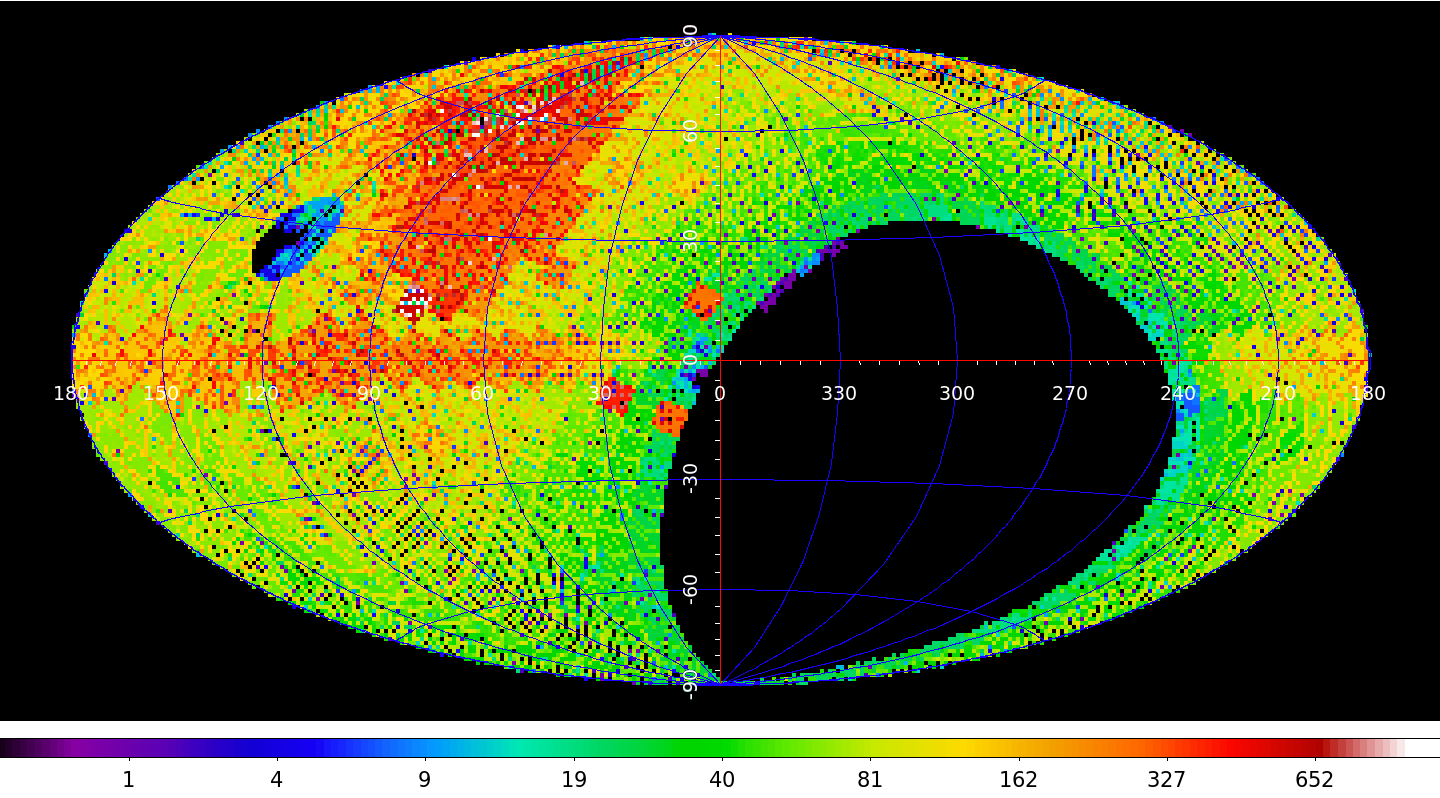}
      \caption{ZTF $r$ filter sky depth-of-coverage in Galactic coordinates. The colour scale corresponds to the number of observation epochs per approximate CCD-quadrant footprint. Image from
      \href{https://irsa.ipac.caltech.edu/data/ZTF/docs/releases/dr02/ztf_release_notes_dr02.pdf}{ZTF}.
    }
    \label{FigZTFcoverage}
   \end{figure}

In order to increase the number of constant stars and widen their sky distribution, it was decided to take advantage of the wealth of the 
\href{https://www.ztf.caltech.edu}{Zwicky Transient Facility} \citep{2019PASP..131a8003M} (hereafter ZTF). 
ZTF is a project started in 2017 in Palomar observatory.  Its goal is to provide a high cadence data stream, enhancing science in stellar astrophysics, supernovae, active galactic nuclei, etc. Each image is captured by a  47 square degree field camera mounted on the 48 inch Schmidt telescope. On average, ZTF observes the entire Northern sky more than 300 times per year  (see Fig.~\ref{FigZTFcoverage}) and makes a data release every two months.
 ZTF data release 2 has become available in December 2019 containing $\sim$2.3 billion light curves.
   
The idea was to obtain the ZTF photometric data and any statistic that is available, in order to detect the least variable stars.
However, there is no need to download all sources from the ZTF database as the aim is not a comprehensive detection of constant sources in ZTF. For this reason, a dense grid of points scattered all over the ZTF observable sky has been created and extracted ZTF sources by performing a cone search with a radius of 2\arcmin. The grid contained ~36\,000 points limited to $\delta>-30{^\circ}$ and it was created by getting healpix with depth 6 and nside 64. 

\begin{figure}
   \centering
   \includegraphics[width=\hsize]{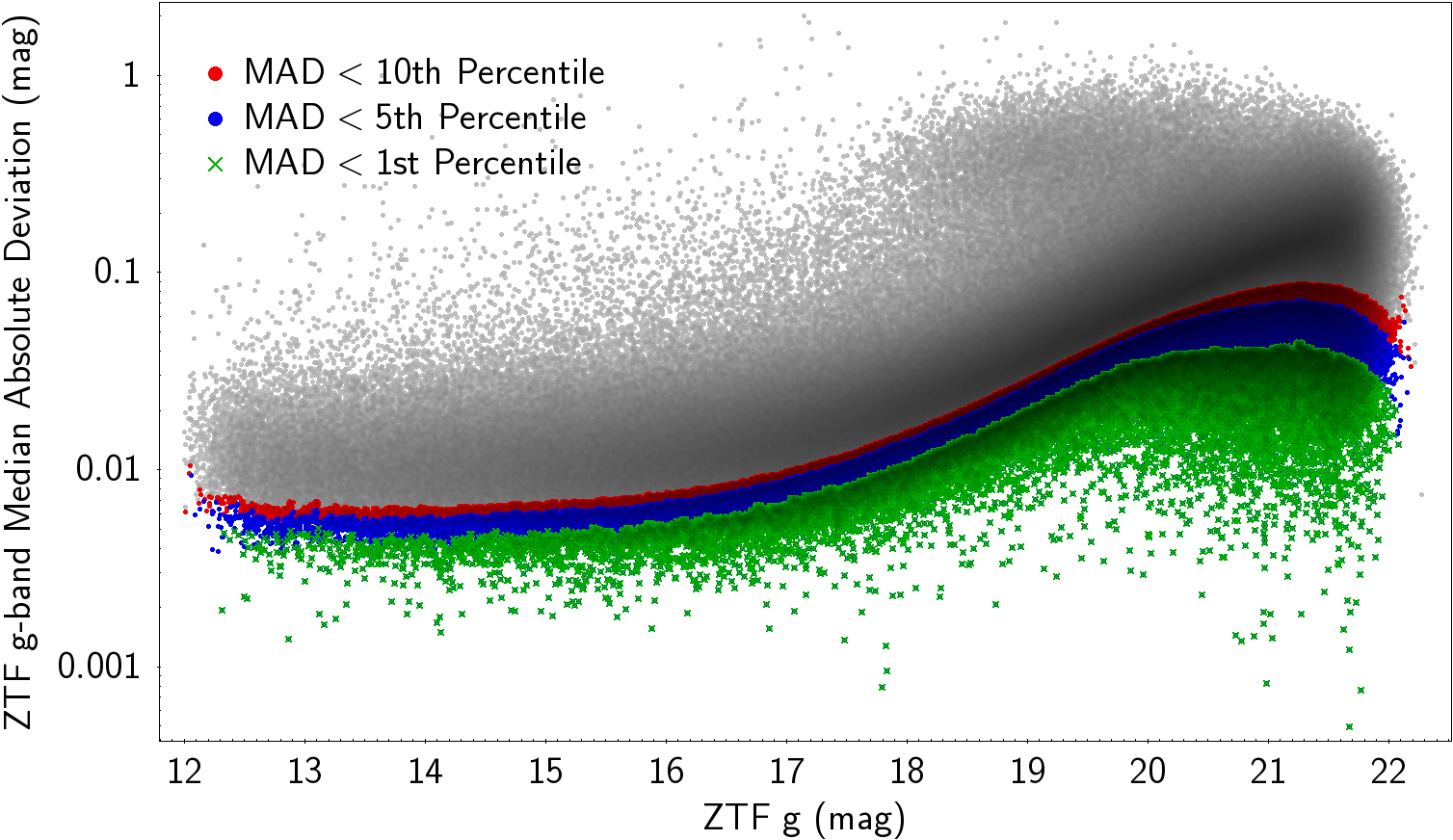}
      \caption{MAD versus magnitude per percentile cut for ZTF sources in this work. Sources between the 10th and 5th percentiles are in red, those with MAD between then 5th and 1st percentiles are in blue, and  the ones under the 1st percentile are in green crosses.}
   \label{FigMAD}
   \end{figure}

\begin{figure}
   \centering
   \includegraphics[width=\hsize]{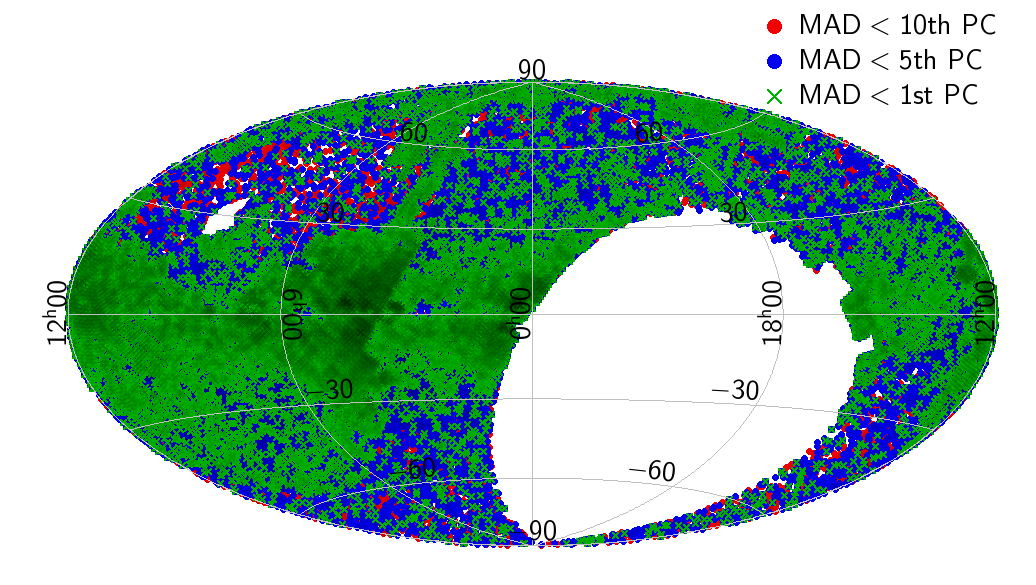}
      \caption{Sky map of the least variable sources. The same colour coding of Fig.~\ref{FigMAD} for percentile (pc) thresholds has been used.}
   \label{FigSkyConst}
   \end{figure}

In total, about 3~million ZTF sources were extracted.
In all those sources, the median absolute deviation of their photometric time series was already available and was used to select the least variable stars. The 3~million source sample was divided in 250 magnitude bins, and the sources with MAD under the 10th, 5th, and the 1st percentile of the MAD distribution of each bin were selected. Figure~\ref{FigMAD} shows the ZTF median $g$ magnitude versus the time series MAD. The sources with MAD over the 10th percentile per magnitude bin are shown in grey, the ones with MAD between the 10th--5th and 5th--1st percentiles are in red and blue, respectively, while sources with MAD less than 1st percentile are in green. Figure~\ref{FigSkyConst} shows the spatial distribution of the selected sources per percentile.
   
\begin{figure}
  \centering
  \includegraphics[width=\hsize]{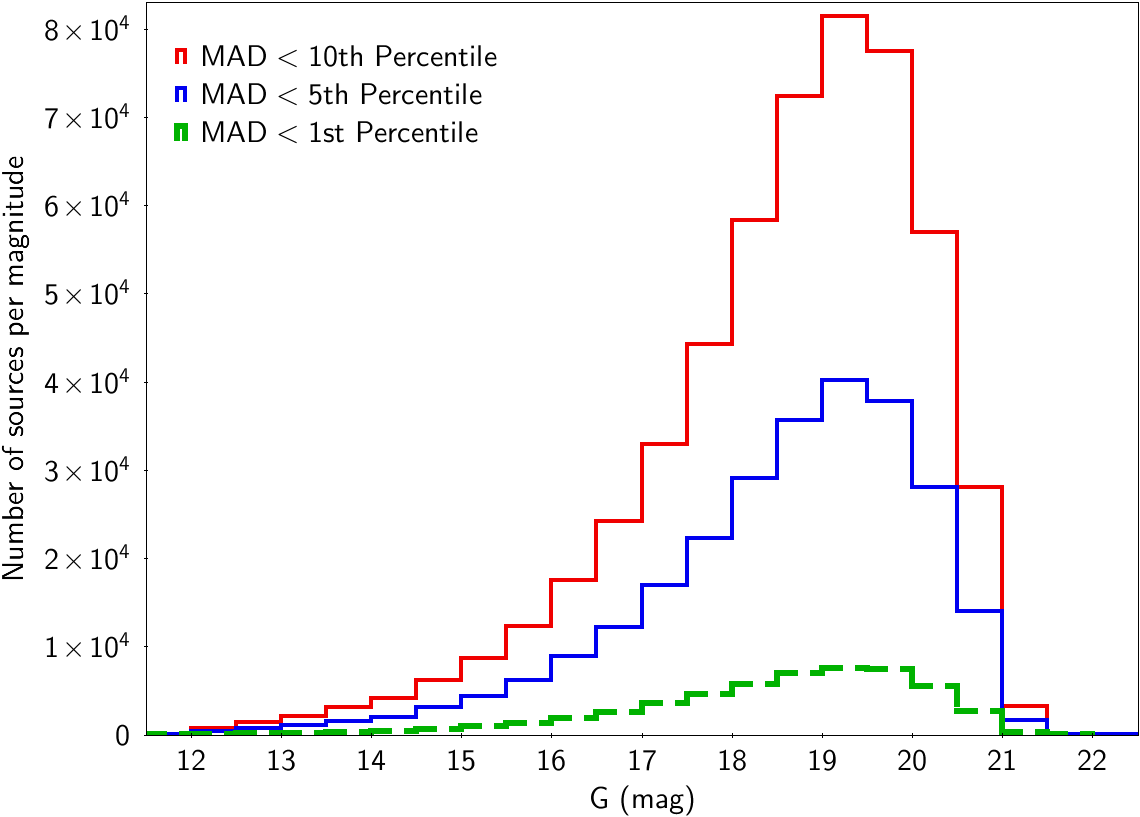}
  \caption{\textit{Gaia} $G$ magnitude distribution of the selected least variable sources per percentile threshold, after cross-match with \textit{Gaia}~DR3 data. The colour schema is the same as in fig.\ref{FigMAD}, the line for MAD below 1st percentile is dashed.}
  \label{FigMagDist}
\end{figure}

\begin{figure}
  \centering
  \includegraphics[width=\hsize]{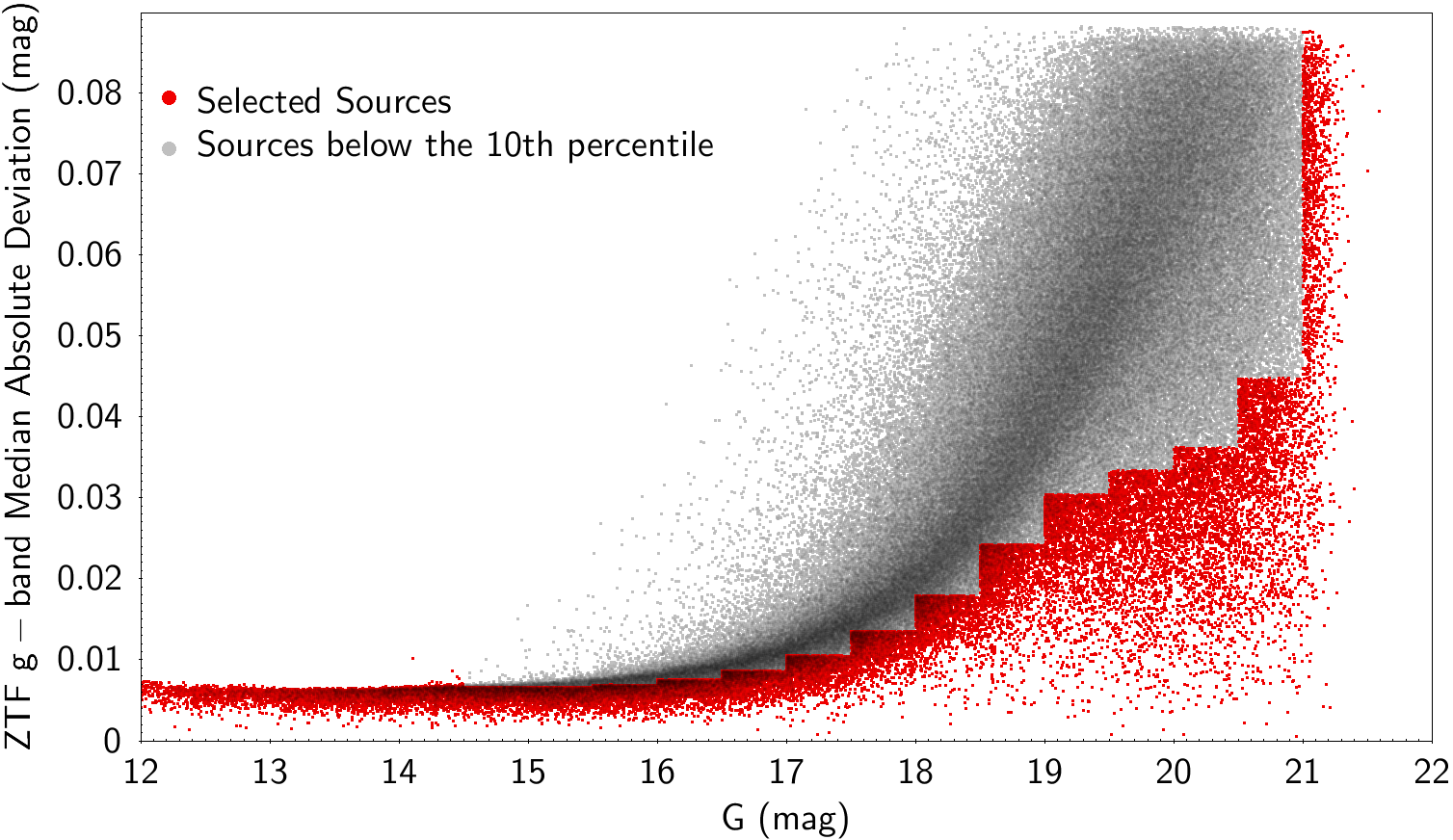}
  \caption{MAD vs $G$ magnitude of sources below the 10th percentile of the MAD distribution. We highlight with red colour the sources that we selected as least variables.
}
  \label{FigReduced}
\end{figure}

\begin{figure}
  \centering
  \includegraphics[width=\hsize]{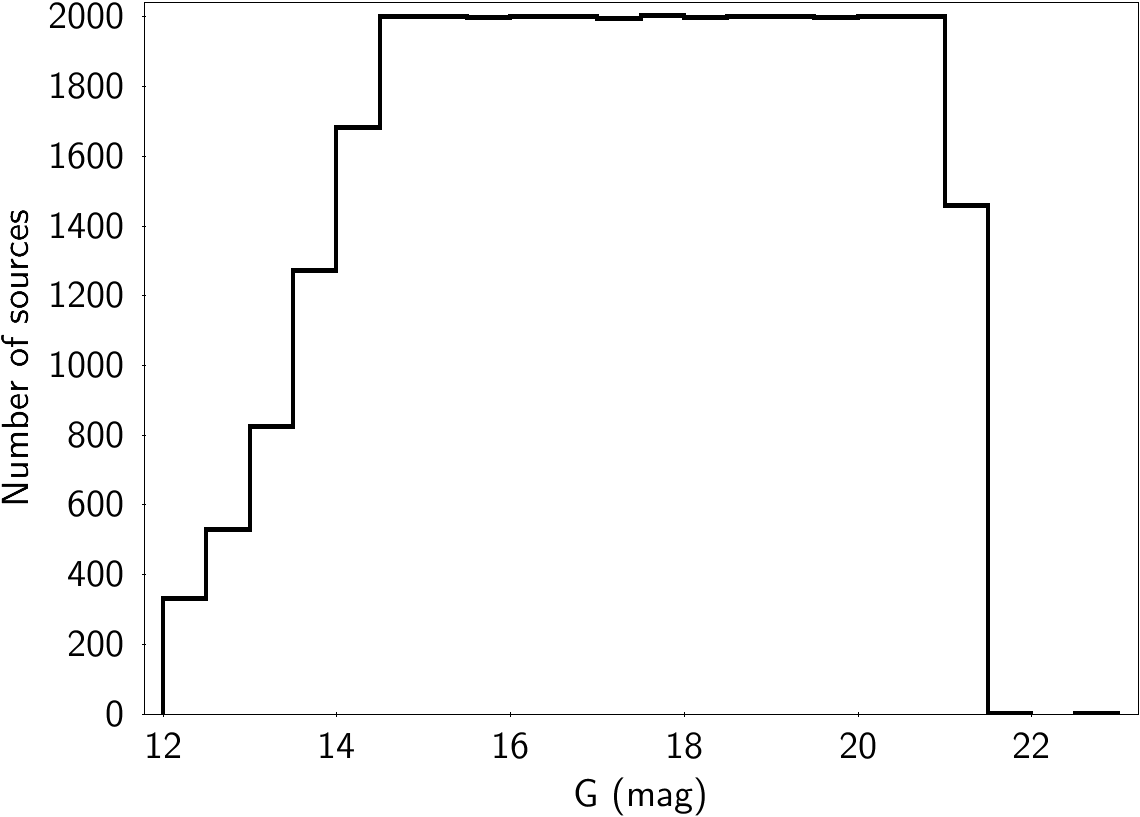}
  \caption{$G$ magnitude distribution of the final selection of sources with photometric MAD below the 10th percentile, which highlights the ZTF cross-match representation as a function of magnitude.}
  \label{FigmagReduced}
\end{figure}

 The next step was to cross-match the selected sources with the \textit{Gaia}~DR3 data set, which was performed with the same method as the rest of the catalogues in this document. At the end of this cross-match process, 267\,784, 133\,112 and 26\,217 sources for the 3 different cut-offs (10th, 5th , and 1st~percentiles) were left. Figure~\ref{FigMagDist} shows the $G$ magnitude distribution of these sources depending on their corresponding percentile range. Due to the very low number of sources at the bright end, it was decided to select an upper limit for the number of stars per bin (for a more fair representation of all magnitudes). 
 Figure~\ref{FigReduced} shows the MAD versus $G$ magnitude of the selected stars (depicted in red),
 with MAD less than the 10th percentile and including up to 2000 sources per 0.5 magnitude bin.
Figure~\ref{FigmagReduced} shows the $G$ magnitude distribution of the final selection of sources.
   
\subsection{Selection of least variable sources from TESS} \label{subsec:TESS}
\begin{figure}
  \centering
  \includegraphics[width=\hsize]{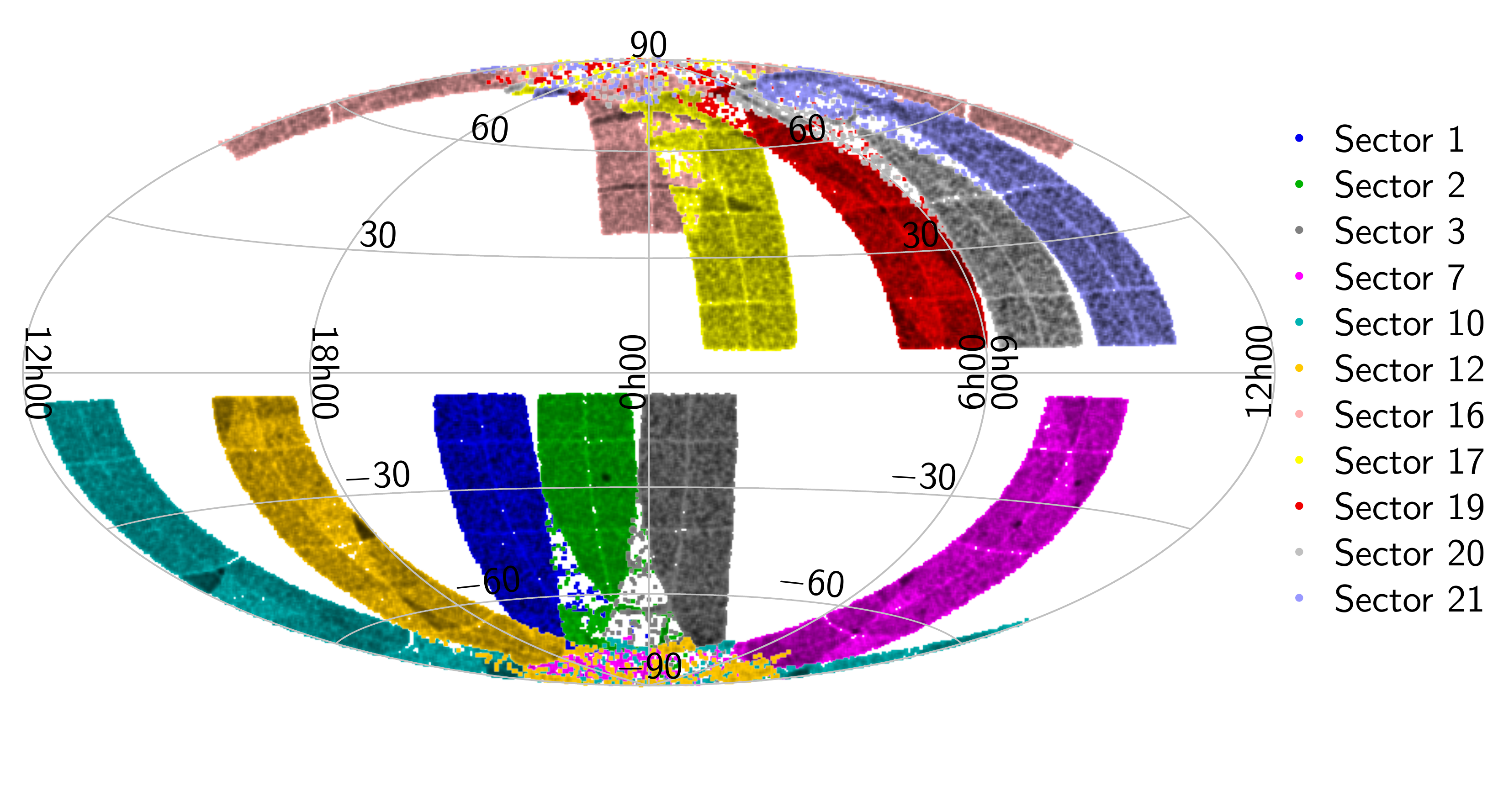}
  \caption{Sky coverage in Ecliptic coordinates of the TESS sectors that were used in this work.}
  \label{Fig:SectorsUsed}
\end{figure}

\begin{figure}
  \centering
  \includegraphics[width=\hsize]{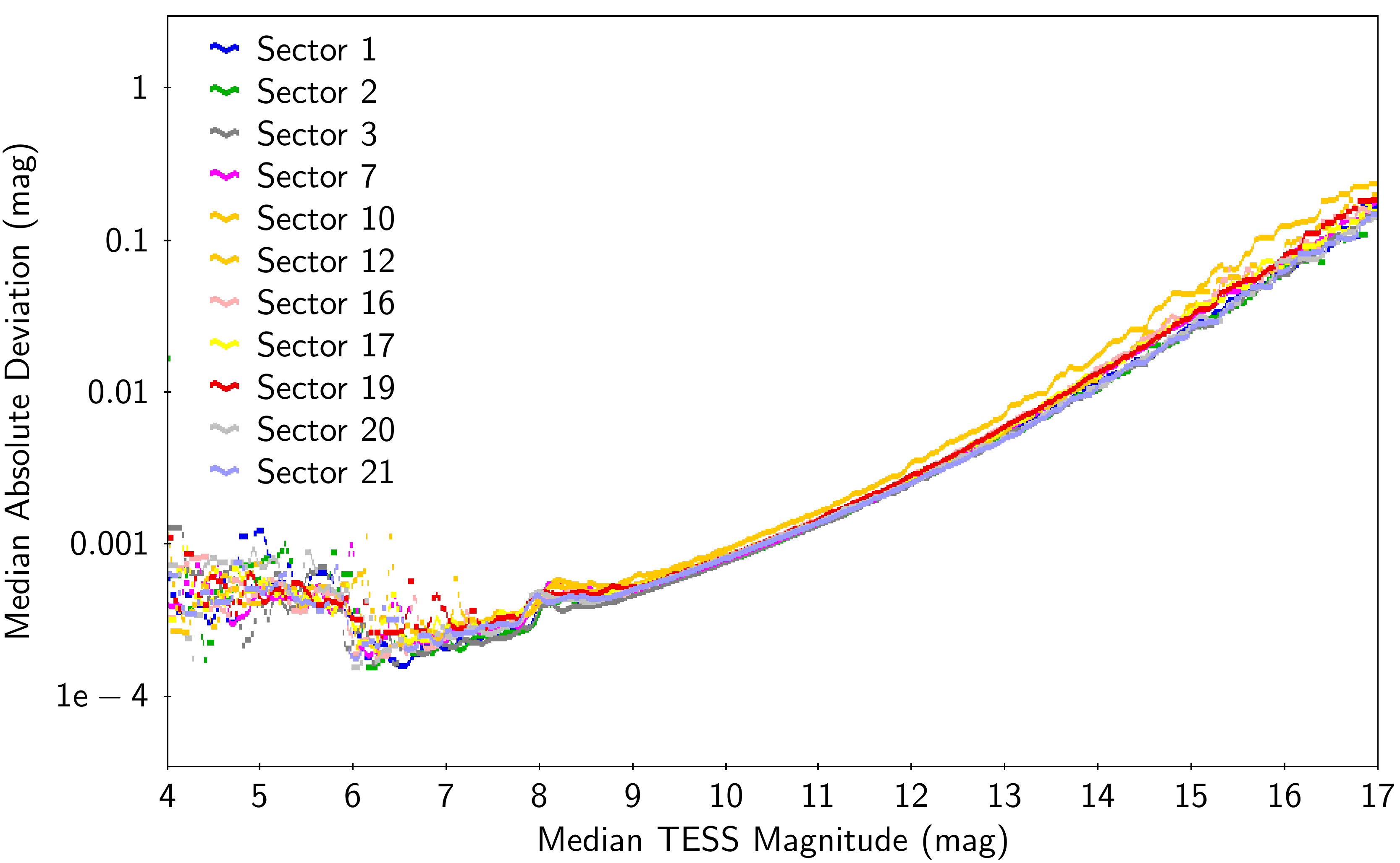}
  \caption{TESS magnitude versus MAD thresholds (of the 10th percentile) for the various sectors used.}
  \label{Fig:MadPerSector}
\end{figure}

\begin{figure}
  \centering
  \includegraphics[width=\hsize]{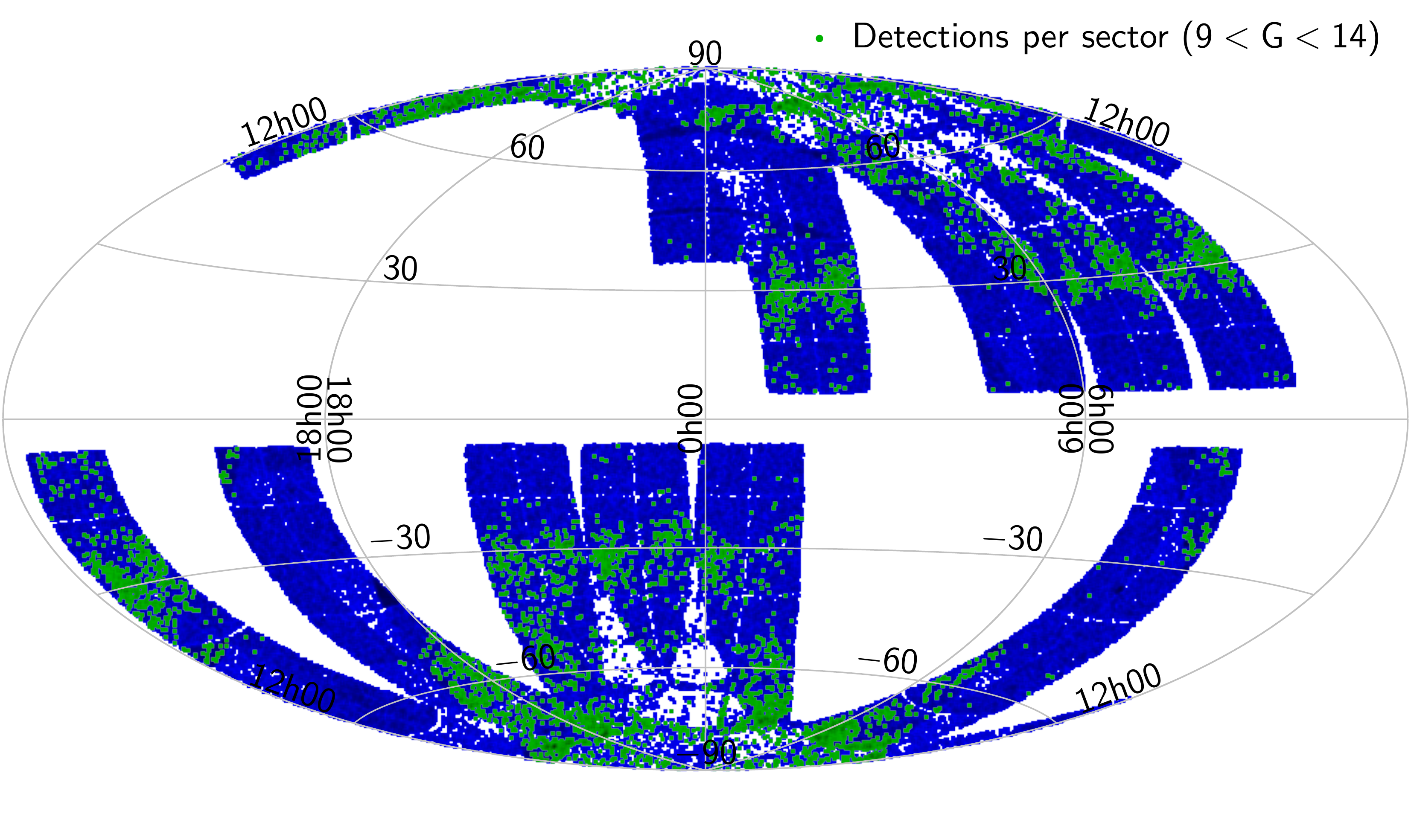}
  \caption{The 10\% least variable stars (green) selected, plotted over the whole sample of TESS targets (blue).}
  \label{Fig:persectorDetection}
\end{figure}

\begin{figure}
  \centering
  \includegraphics[width=\hsize]{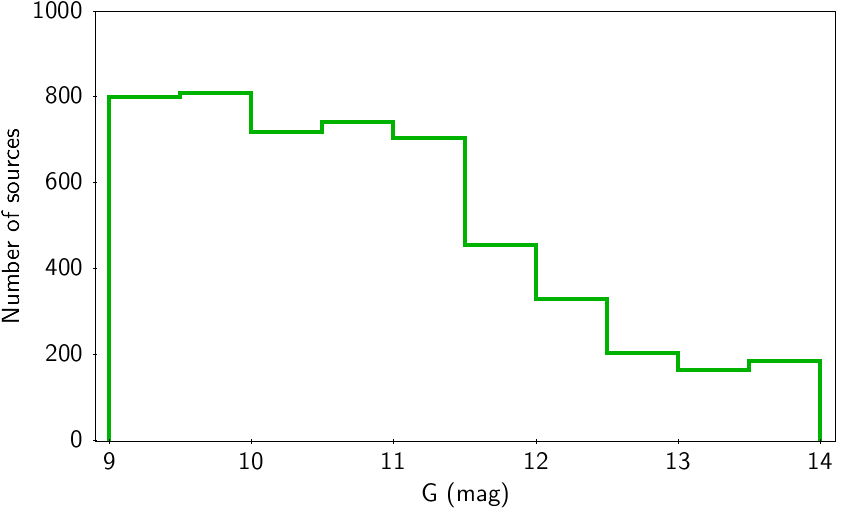}
  \caption{\textit{Gaia} $G$ magnitude distribution of the 10\% least variable stars in TESS, after cross-matching with \textit{Gaia} sources. }
  \label{Fig:tessmagdistribution}
\end{figure}

The Transiting Exoplanet Survey Satellite \citep[hereafter TESS;][]{2015JATIS...1a4003R} is a NASA space telescope with primary goal to search for exoplanets. It observes both hemispheres divided in 26 sectors and its targets are bright stars with the majority being brighter than $T\sim 12$ mag.
 In order to overcome the lack of constant sources with magnitude around G$\sim$12 and considering the targets TESS observes, it was decided to apply the same process described in Sect.~\ref{subsec:ztf} to TESS sources. The time series of
 $\sim$99~thousand unique stars
 covering 11 sectors were used (see Fig.~\ref{Fig:SectorsUsed}). We remind that our aim was not to cross-match the full TESS targets but to identify a sufficient number of least variable stars in a specific magnitude range. The light curves were downloaded from the TESS \href{https://archive.stsci.edu/tess/bulk_downloads/bulk_downloads_ffi-tp-lc-dv.html}{bulk download website}, where a script that extracts data per sector is provided.
 About half of the source were duplicated from sector overlaps at the Ecliptic poles and thus they were removed. 
 These photometric time series contained the Simple Aperture Photometry (SAP) and the 
 Pre-Search Data Conditioned Simple Aperture Photometry (PDCSAP) corrected flux of each object. SAP flux is the raw flux while in PDCSAP flux long term trends have been removed. This removal must be taken with caution as it can alter the true flux changes of variable sources. Fluxes have been converted to magnitudes using a preliminary zero point magnitude (L\'{a}szl\'{o} Moln\'{a}r, private communication) and the MAD is calculated for each source. The same procedure as in Sect.~\ref{subsec:ztf} has been followed in order to select those sources with lower MAD per magnitude bin. Figure~\ref{Fig:MadPerSector} shows the MAD per magnitude per sector, revealing that  sectors can have different MAD thresholds, so the 10\% least variable stars are selected per sector. 
 Figure~\ref{Fig:persectorDetection} shows the final spacial distribution of the least variable stars in TESS.  After the cross-matching with \textit{Gaia}, 5100 sources were selected. The magnitude distribution of the selected sources is shown in Fig.~\ref{Fig:tessmagdistribution} and it covers the magnitude gap of non-variable objects from \textit{Hipparcos} and SDSS Stripe~82. 

\section{Conclusions} \label{Sec:Conclusion}
We present the creation of a large and diverse cross-match catalogue with \textit{Gaia} from variable and constant sources in the literature. 
In total 152 different input catalogues from the literature were cross-matched with \textit{Gaia}~DR3 in order to find the counterpart sources, compiling a large data set of more than 7.8 million sources and 112 different types of variables, constants and galaxies. Each input was cross-matched individually performing an epoch propagation and using a synthetic distance that encapsulates the astrometric distance and the photometric difference between targets and counterparts. The catalogue is available online to the scientific community through the Centre de Donn\'ees astronomiques de Strasbourg website. Users of this catalogue 
might still need to verify or filter out some objects of interest depending on the purpose of the analysis. 
This catalogue is a valuable resource for the studies of variable sources as it provides a single data set for the \textit{Gaia} mission containing with uniform photometry and astrometry. 
 
\begin{acknowledgements}
This work has made use of data from the European Space Agency (ESA) mission {\it Gaia} (\url{https://www.cosmos.esa.int/gaia}), processed by the {\it Gaia}
Data Processing and Analysis Consortium (DPAC,
\url{https://www.cosmos.esa.int/web/gaia/dpac/consortium}). Funding for the DPAC has been provided by national institutions, in particular the institutions participating in the {\it Gaia} Multilateral Agreement which include, for Switzerland, the Swiss State Secretariat for Education, Research and Innovation through the Activités Nationales Complémentaires (ANC). This research has made use of the VizieR catalogue access tool, CDS, Strasbourg, France. We acknowledge with thanks the variable star observations from the AAVSO International Database contributed by observers worldwide and used in this research.
Based on observations obtained with the Samuel Oschin 48-inch Telescope at the Palomar Observatory as part of the Zwicky Transient Facility project. ZTF is supported by the National Science Foundation under Grant No. AST-1440341 and a collaboration including Caltech, IPAC, the Weizmann Institute for Science, the Oskar Klein Center at Stockholm University, the University of Maryland, the University of Washington, Deutsches Elektronen-Synchrotron and Humboldt University, Los Alamos National Laboratories, the TANGO Consortium of Taiwan, the University of Wisconsin at Milwaukee, and Lawrence Berkeley
National Laboratories. Operations are conducted by COO, IPAC, and UW.
This paper includes data collected by the TESS mission. Funding for the TESS mission is provided by the NASA's Science Mission Directorate.\\
This work made use of software from Postgres-XL (\url{https://www.postgres-xl.org}), Java (\url{https://www.oracle.com/java/}), Q3C \citep{2006ASPC..351..735K} and TOPCAT/STILTS \citep{2005ASPC..347...29T}.
      
\end{acknowledgements}

\bibliographystyle{aa}
\bibliography{GAIA_DR3_Crossmatch}
\onecolumn
\input{refTable_v10.0}
\clearpage
\input{ranking_v10.0}
\clearpage
\input{EB_naming_conv_table}
\input{fieldNames}
\input{superclassesTable}
\clearpage
\input{comparison10.0_class}
\input{overlapsExample}
\end{document}

%% file: refTable_v10.0.tex
\begin{longtable}{lll}
\caption{\label{table:catalogList} List of input catalouges used. First column is the {\tt{catalogue\_label}} of the input catalogues used, then the number of objects found in the cross-match follows, and finally references of the catalogue in literature.}\\
\hline 
\hline 
Catalogue\_label & \textnumero ~Objects & Reference \\
\hline 
\endfirsthead 
2MASS\_GAIA\_WISE\_SYST\_AKRAS\_2019&326&{\cite{2019ApJS..240...21A}}\\
2MASS\_GDOR\_ALICAVUS\_2016&51&{\cite{2016MNRAS.458.2307K}}\\
2MASS\_LPV\_DEMERS\_2007&103&{\cite{2007A&A...473..143D}}\\
2WHSP\_BLAZARS\_CHANG\_2017&815&{\cite{2017A&A...598A..17C}}\\
ALMA\_BLAZARS\_BONATO\_2018&1019&{\cite{2018MNRAS.478.1512B}}\\
ASAS3\_MCP\_BERNHARD\_2015&300&{\cite{2015A&A...581A.138B}}\\
ASASSN\_DIPPERS\_BREDALL\_2020&11&{\cite{2020MNRAS.496.3257B}}\\
ASASSN\_VAR\_JAYASINGHE\_2019&345133&{\cite{2019MNRAS.485..961J}}\\
& &{\cite{2014ApJ...788...48S}}\\
& &{\cite{2018MNRAS.477.3145J}}\\
& &{\cite{2019MNRAS.486.1907J}}\\
ASAS\_KEPLER\_VAR\_PIGULSKI\_2009&794&{\cite{2009AcA....59...33P}}\\
ASAS\_SN\_ASAS\_2019&387&{\cite{2014ApJ...788...48S}}\\
ASAS\_SOLAR\_LIKE\_MESSINA\_2010&221&{\cite{2010A&A...520A..15M}}\\
& &{\cite{2011A&A...532A..10M}}\\
ASAS\_VAR\_POJMANSKI\_2002&14994&{\cite{2002AcA....52..397P}}\\
ASAS\_VAR\_RICHARDS\_2012&19117&{\cite{2012ApJS..203...32R}}\\
ATLAS\_VAR\_HEINZE\_2018&356621&{\cite{2018AJ....156..241H}}\\
BULGE\_SOLAR\_LIKE\_DRAKE\_2006&2385&{\cite{2006AJ....131.1044D}}\\
BZCAT\_BLAZARS\_MASSARO\_2015&2686&{\cite{2015Ap&SS.357...75M}}\\
CAHA\_SOLARLIKE\_MARTINEZ\_2010&149&{\cite{2010A&A...520A..79M}}\\
CATALINA\_CSS\_RRAB\_DRAKE\_2013&2007&{\cite{2013ApJ...765..154D}}\\
CATALINA\_CV\_DRAKE\_2014&545&{\cite{2014MNRAS.441.1186D}}\\
CATALINA\_MLS\_RRAB\_DRAKE\_2013&1201&{\cite{2013ApJ...765..154D}}\\
CATALINA\_RRAB\_DRAKE\_2013&12098&{\cite{2013ApJ...763...32D}}\\
CATALINA\_RRAB\_TORREALBA\_2015&10452&{\cite{2015MNRAS.446.2251T}}\\
CATALINA\_SEKBO\_RR\_DRAKE\_2013&550&{\cite{2013ApJ...765..154D}}\\
CATALINA\_VAR\_DRAKE\_2014&46226&{\cite{2014ApJS..213....9D}}\\
CATALINA\_VAR\_DRAKE\_2017&36584&{\cite{2017MNRAS.469.3688D}}\\
COMP\_AP\_RENSON\_2009&3055&{\cite{2009A&A...498..961R}}\\
COMP\_BHXRB\_CORRALSANTANA\_2016&50&{\cite{2016A&A...587A..61C}}\\
COMP\_CABS\_EKER\_2008&260&{\cite{2008MNRAS.389.1722E}}\\
COMP\_CV\_RITTER\_2013&1096&{\cite{2003A&A...404..301R}}\\
COMP\_DSCT\_GDOR\_DEBOSSCHER\_2007&68&{\cite{2007A&A...475.1159D}}\\
COMP\_EP\_SOUTHWORTH\_2020&1048&\cite{2011MNRAS.417.2166S}, \href{https://www.astro.keele.ac.uk/jkt/tepcat/html-catalogue.html}{ URL link}\\
COMP\_GAL\_BCEP\_STANKOV\_2005&41&{\cite{2005ApJS..158..193S}}\\
COMP\_M15\_RRL\_CLEMENTINI&113&{\cite{2008AJ....135.1459C}}\\
COMP\_M3\_RRL\_CLEMENTINI&206&{\cite{2006MNRAS.372.1657B}}\\
COMP\_MICROLENSING\_GAIA&115&Internal\_Microlensing\_Kruszy{\'n}ska\\
COMP\_NGC6388\_RRL\_CLEMENTINI&33&{\cite{2002AJ....124..949P}}\\
& &{\cite{2006AJ....132.1014C}}\\
& &{\cite{2015A&A...573A.103S}}\\
COMP\_NGC6441\_RRL\_CLEMENTINI&43&{\cite{2003AJ....126.1381P}}\\
& &{\cite{2006AJ....132.1014C}}\\
COMP\_OMEGACEN\_RRL\_CLEMENTINI&183&{\cite{2016AJ....152..170B}}\\
COMP\_PCB\_RITTER\_2013&555&{\cite{2003A&A...404..301R}}\\
COMP\_SPB\_BCEP\_DECAT\_PR&69&Internal\_SPB\_BCEP\_DeCat\\
COMP\_SYST\_BELCZYNSKI\_2000&140&{\cite{2000A&AS..146..407B}}\\
COMP\_UFD\_RRL\_CLEMENTINI&64&{\cite{2009MNRAS.398.1757W}}\\
& &{\cite{2006ApJ...649L..83S}}\\
& &{\cite{2012ApJ...756..121M}}\\
& &{\cite{2013ApJ...767...62G}}\\
& &{\cite{2006ApJ...653L.109D}}\\
& &{\cite{2009ApJ...695L..83M}}\\
& &{\cite{2012ApJ...752...42D}}\\
& &{\cite{2013AJ....146...94B}}\\
& &{\cite{2014ApJ...793..135S}}\\
COMP\_VAR\_VSX\_2019&779348&{\cite{2006SASS...25...47W}}\\
\hline
\\
\caption{continued.}\\
\hline 
\hline 
Catalogue\_label & \textnumero ~Objects & Reference \\ 
\hline
COMP\_WD\_CORSICO\_2019&258&{\cite{2019A&ARv..27....7C}}\\
COMP\_XB\_RITTER\_2003&49&{\cite{2003A&A...404..301R}}\\
COMP\_YSO\_VARGA-VEREBELYI\_2020&11036&{\cite{2020IAUS..345..378V}}\\
COMP\_ZZ\_NGUYEN\_2020&31&{\cite{2011ApJ...733L..19D}}\\
& &{\cite{2007ApJS..171..219Q}}\\
& &{\cite{2013MNRAS.436.3573H}}\\
& &{\cite{2013MNRAS.432.1632K}}\\
& &{\cite{2013ApJ...765..102H}}\\
& &{\cite{2012ApJ...750L..28H}}\\
& &{\cite{2010ApJ...720L.159D}}\\
& &{\cite{2014MNRAS.442.2278K}}\\
& &{\cite{2016ApJ...817...27W}}\\
& &{\cite{2009ApJ...690..560N}}\\
& &{\cite{1999ApJ...516..887B}}\\
& &{\cite{2005ApJ...631.1100G}}\\
& &{\cite{2020A&A...638A..82B}}\\
COROT\_DSCT\_GDOR\_SARRO\_2013&712&{\cite{2013A&A...550A.120S}}\\
COROT\_ROT\_DEMEDEIROS\_2013&3862&{\cite{2013A&A...555A..63D}}\\
COROT\_SOLAR\_LIKE\_MEDHI\_2007&174&{\cite{2007A&A...469..713M}}\\
DECAM\_VAR\_VIVAS\_2020&119&{\cite{2020MNRAS.492.1061V}}\\
EROS2\_BEATCEP\_MARQUETTE\_2009&399&{\cite{2009A&A...495..249M}}\\
EROS2\_LPV\_SPANO\_2011&41762&{\cite{2011A&A...536A..60S}}\\
EROS2\_VAR\_KIM\_2014&129954&{\cite{2014A&A...566A..43K}}\\
EROSITA\_AGN\_LIU\_2021&8550&\cite{2021arXiv210614522L}\\
& & \cite{2021arXiv210614523B}\\
& & \cite{2021arXiv210614520S}\\
GAIA\_BY\_DISTEFANO\_2019&24968&Internal\_BY\_Distefano\\
GAIA\_CEP\_RIPEPI\_2019&1339&{\cite{2019A&A...625A..14R}}\\
GAIA\_CEP\_ZAK\_2018&18&Internal\_CEP\_Zak\\
GAIA\_DR2\_CLASSIFIER\_VARIABLES\_2018&360580&{\cite{2018A&A...618A..30H}}\\
GAIA\_DR2\_CLASS\_DSCT\_SXPHE\_SELECTION&1905&Internal\_DSCT\_DeRidder\\
GAIA\_DR2\_SOS\_VARIABLES\_2018&387510&{\cite{2018A&A...618A..30H}}\\
GAIA\_ECL\_RYBIZKI\_2018&2189&{Internal\_ECL\_Rybizki}\\
GAIA\_GALAXY\_CLEMENTINI\_2020&769&{\cite{2019A&A...622A..60C}}\\
GAIA\_GAL\_GAIA\_2018&1748322&{\cite{floisvos}}\\
GAIA\_ICRF2\_QSO\_GAIA\_2017&2079&{\cite{2013yCat.1323....0M}}\\
GAIA\_M31\_CEP\_GAIA\_2018&747&Internal\_CEP\_M31\\
GAIA\_QSO\_GAIA\_CRF3&1614173&\cite{DR3-DPACP-133}\\
GAIA\_ROT\_GAIA\_2017&3725&Internal\_ROT\_Distefano\\
GAIA\_RRL\_GAROFALO\_SELECTION&1569&Internal\_RRL\_Garofalo\\
GAIA\_TRANSIENTS\_ALERTS\_2019&6843&Internal\_Gaia\_Science\_Alerts\\
GAIA\_WD\_GENTILEFUSILLO\_2019&468880&{\cite{2019MNRAS.482.4570G}}\\
GAIA\_ZZ\_EYER\_2019&1461&{\cite{2019arXiv191207659E}}\\
GALEX\_ZZ\_ROWAN\_2019&62&{\cite{2019MNRAS.486.4574R}}\\
HALO\_MSR\_MAURON\_2020&419&{\cite{2019A&A...626A.112M}}\\
HATNET\_COMP\_PLEIADES\_SOLAR\_LIKE\_HARTMAN\_2010&18&{\cite{2010MNRAS.408..475H}}\\
HATNET\_MEMBERS\_PLEIADES\_SOLAR\_LIKE\_HARTMAN\_2010&372&{\cite{2010MNRAS.408..475H}}\\
HATNET\_NONMEMBERS\_PLEIADES\_SOLAR\_LIKE\_HARTMAN\_2010&1724&{\cite{2010MNRAS.408..475H}}\\
HIPPARCOS\_LPV\_BERGEAT\_2001&361&{\cite{2001A&A...369..178B}}\\
HIP\_VAR\_ESA\_1997&48932&{\cite{1997ESASP1200.....E}}\\
HMQ\_QSO\_FLESCH\_2015&294253&{\cite{2015PASA...32...10F}}\\
INTEGRAL\_VAR\_ALFONSOGARZON\_2012&3516&{\cite{2012A&A...548A..79A}}\\
IRAS\_LPV\_C\_SUH\_2017&490&{\cite{2017JKAS...50..131S}}\\
IRAS\_LPV\_O\_SUH\_2017&1498&{\cite{2017JKAS...50..131S}}\\
IRAS\_LPV\_SIC\_SUH\_2017&15&{\cite{2017JKAS...50..131S}}\\
IRAS\_LPV\_SUH\_2017&231&{\cite{2017JKAS...50..131S}}\\
IUE\_FKCOM\_BOPP\_1981&3&{\cite{1981ApJ...247L.131B}}\\
IUE\_SPB\_NIEMCZURA\_2003&40&{\cite{2003A&A...404..689N}}\\
KEPLER\_GAIA\_BY\_ROT\_DISTEFANO\_2020&173&Intenal\_BY\_ROT\_Distefano\\
KEPLER\_DSCT\_GDOR\_BRADLEY\_2015&314&{\cite{2015AJ....149...68B}}\\
\hline
\\
\caption{continued.}\\
\hline 
\hline 
Catalogue\_label & \textnumero ~Objects & Reference \\ 
\hline
KEPLER\_DSCT\_GDOR\_DEBOSSCHER\_2011&1023&{\cite{2011A&A...529A..89D}}\\
KEPLER\_DSCT\_GDOR\_UYTTERHOEVEN\_2011&436&{\cite{2011A&A...534A.125U}}\\
KEPLER\_ECL\_KIRK\_2016&2797&{\cite{2016AJ....151...68K}}\\
KEPLER\_FLARES\_COMPILATION&654&{\cite{2011AJ....141...50W}}\\
& &{\cite{2013ApJS..209....5S}}\\
& &{\cite{2015ApJ...798...92W}}\\
KEPLER\_GDOR\_VANREETH\_2015&68&{\cite{2015ApJS..218...27V}}\\
KEPLER\_MCP\_HUMMERICH\_2018&53&{\cite{2018A&A...619A..98H}}\\
KEPLER\_ROAP\_HEY\_2019&6&{\cite{2019MNRAS.488...18H}}\\
KEPLER\_ROT\_HOWELL\_2016&19&{\cite{2016ApJ...831...27H}}\\
KEPLER\_ROT\_REINHOLD\_2015&19891&{\cite{2015A&A...583A..65R}}\\
KEPLER\_SHORTP\_SLAWSON\_2011&42&{\cite{2011AJ....142..160S}}\\
KEPLER\_VAR\_DEBOSSCHER\_2011&145181&{\cite{2011A&A...529A..89D}}\\
LAMOST\_RAD\_VEL\_VAR\_TIAN\_2020&78629&{\cite{2020ApJS..249...22T}}\\
LINEAR\_VAR\_PALAVERSA\_2013&6851&{\cite{2013AJ....146..101P}}\\
M31\_SEL\_LPV\_MOULD\_2004&442&{\cite{2004ApJS..154..623M}}\\
M33\_CEP\_PELLERIN\_2011&311&{\cite{2011ApJS..193...26P}}\\
MILLIQUAS\_QSO\_FLESCH\_2019&1099699&{\cite{2019arXiv191205614F}}\\
MMT\_M37\_FLARES\_CHANG\_2015&214&{\cite{2015ApJ...814...35C}}\\
NSVS\_RRAB\_KINEMUCHI\_2006&1078&{\cite{2006AJ....132.1202K}}\\
NSVS\_VAR\_WILLIAMS\_2004&5239&{\cite{2004AJ....128.2965W}}\\
OGLE3\_VAR\_OGLE3\_2012&404247&{\cite{2008AcA....58..163S}}\\
& &{\cite{2008AcA....58..293S}}\\
& &{\cite{2009AcA....59....1S}}\\
& &{\cite{2009AcA....59..239S}}\\
& &{\cite{2009AcA....59..335S}}\\
& &{\cite{2010AcA....60....1P}}\\
& &{\cite{2010AcA....60...17S}}\\
& &{\cite{2010AcA....60..179P}}\\
& &{\cite{2010AcA....60...91S}}\\
& &{\cite{2010AcA....60..165S}}\\
& &{\cite{2011AcA....61....1S}}\\
& &{\cite{2011AcA....61..285S}}\\
& &{\cite{2011AcA....61..217S}}\\
& &{\cite{2013AcA....63...21S}}\\
& &{\cite{2013AcA....63..323P}}\\
OGLE4\_BLG\_CEP\_RR\_OGLE4\_2016&36342&{\cite{2017AcA....67..297S}}\\
OGLE4\_BLG\_RRL\_SOSZYNSKI\_2019&65304&{\cite{2019AcA....69..321S}}\\
OGLE4\_CEP\_OGLE\_2020&3692&{\cite{2014AcA....64..177S}}\\
& &{\cite{2015AcA....65..297S}}\\
& &{\cite{2016AcA....66..421P}}\\
& &{\cite{2016AcA....66..131S}}\\
& &{\cite{2017AcA....67..297S}}\\
& &{\cite{2018AcA....68..315U}}\\
& &{\cite{2020AcA....70..101S}}\\
OGLE4\_CEP\_RR\_OGLE\_2016&89511&{\cite{2014AcA....64..177S}}\\
& &{\cite{2015AcA....65..297S}}\\
& &{\cite{2016AcA....66..131S}}\\
& &{\cite{2017AcA....67..297S}}\\
OGLE4\_CV\_2016\_OGLE4&195&{\cite{2015AcA....65..313M}}\\
OGLE4\_GD\_RRL\_SOSZYNSKI\_2019&10008&{\cite{2019AcA....69..321S}}\\
OGLE4\_GSEP\_CST\_SOSZYNSKI\_2012&10495&{\cite{2012AcA....62..219S}}\\
OGLE4\_GSEP\_VAR\_SOSZYNSKI\_2012&6455&{\cite{2012AcA....62..219S}}\\
OGLE4\_LMC\_CEP\_RR\_OGLE4\_2016&43291&{\cite{2015AcA....65..297S}}\\
& &{\cite{2016AcA....66..131S}}\\
OGLE4\_LMC\_ECL\_OGLE4\_2017&38810&{\cite{2016AcA....66..421P}}\\
OGLE4\_M54\_VAR\_HAMANOWICZ\_2016&168&{\cite{2016AcA....66..197H}}\\
OGLE4\_MICROLENSING\_OGLE4\_2016&2707&{\cite{2019ApJS..244...29M}}\\
OGLE4\_SHORTP\_ECL\_SOSZYNSKI\_2014&161&{\cite{2015AcA....65...39S}}\\
\hline
\\
\caption{continued.}\\
\hline 
\hline 
Catalogue\_label & \textnumero ~Objects & Reference \\ 
\hline
OGLE4\_SMC\_CEP\_RR\_OGLE4\_2016&11342&{\cite{2016AcA....66..131S}}\\
& &{\cite{2015AcA....65..297S}}\\
OGLE4\_SMC\_ECL\_OGLE4\_2017&8282&{\cite{2016AcA....66..421P}}\\
OGLE4\_VAR\_OGLE\_2019&496569&{\cite{2014AcA....64..177S}}\\
& &{\cite{2015AcA....65..297S}}\\
& &{\cite{2016AcA....66..421P}}\\
& &{\cite{2016AcA....66..131S}}\\
& &{\cite{2017AcA....67..297S}}\\
& &{\cite{2018AcA....68..315U}}\\
OGLE\_BE\_MENNICKENT\_2002&1002&{\cite{2002A&A...393..887M}}\\
OGLE\_BE\_SABOGAL\_2005&2370&{\cite{2005MNRAS.361.1055S}}\\
OGLE\_BE\_SABOGAL\_2008&1408&{\cite{2008A&A...478..659S}}\\
OGLE\_BLAP\_PIETRUKOWICZ\_2017&14&{\cite{2017NatAs...1E.166P}}\\
OGLE\_BLG\_EB\_SOSZYNSKI\_2016&347267&{\cite{2016AcA....66..405S}}\\
PS1\_RRL\_SESAR\_2017&228876&{\cite{2017AJ....153..204S}}\\
PS1\_RRL\_SESAR\_SELECTION\_2017&60872&{\cite{2017AJ....153..204S}}\\
RAVE\_SOLAR\_LIKE\_ZERJAL\_2017&30830&{\cite{2017ApJ...835...61Z}}\\
SB9\_SB\_POURBAIX\_2004&3256&{\cite{2004A&A...424..727P}}\\
SDSS\_CST\_IVEZIC\_2007&597335&{\cite{2007AJ....134..973I}}\\
SDSS\_CV\_SZKODY\_2011&263&{\cite{2011AJ....142..181S}}\\
SDSS\_DQWD\_KOESTER\_2019&279&{\cite{2019A&A...628A.102K}}\\
SDSS\_DSCT\_RR\_SUVEGES\_2012&259&{\cite{2012MNRAS.424.2528S}}\\
SDSS\_PS1\_CATALINA\_RRL\_ABBAS\_2014&5558&{\cite{2014MNRAS.441.1230A}}\\
SDSS\_VAR\_IVEZIC\_2007&66120&{\cite{2007AJ....134..973I}}\\
SOAR\_ZZ\_ROMERO\_2019&14&{\cite{2019MNRAS.490.1803R}}\\
STEREO\_MAP\_WRAITH\_2012&75&{\cite{2012MNRAS.420..757W}}\\
TESSGAIA\_BY\_ROT\_DISTEFANO\_2020&110&Intenal\_BY\_ROT\_Distefano\\
TESS\_CST\_GAIA\_2020&5072&{\cite{DR3-DPACP-177}}\\
TESS\_ROT\_SIKORA\_2019&131&{\cite{2019MNRAS.487.4695S}}\\
TESS\_VAR\_CUNHA\_2019&51&{\cite{2019MNRAS.487.3523C}}\\
TRES\_ECL\_DEVOR\_2008&714&{\cite{2008AJ....135..850D}}\\
UKST\_LPV\_KUNKEL\_1997&445&{\cite{1997A&AS..122..463K}}\\
VVV\_VAR\_BRAGA\_2019&329&{\cite{2019A&A...625A.151B}}\\
ZTF\_CST\_GAIA\_2020&32053&{\cite{DR3-DPACP-177}}\\
ZTF\_CV\_SZKODY\_2020&251&{\cite{2020AJ....159..198S}}\\
ZTF\_PERIODIC\_CHEN\_2020&746570&{\cite{2020ApJS..249...18C}}\\
\hline
\end{longtable}

%% file: ranking_v10.0.tex
\begin{longtable}{l}
\caption{\label{table:rank} Rank-ordered list of literature catalogues.}\\
\hline
\hline
Catalogue  \\
\hline 
\endfirsthead 
GAIA\_CEP\_RIPEPI\_2019\\
GAIA\_DR2\_CLASS\_DSCT\_SXPHE\_SELECTION\\
GAIA\_CEP\_ZAK\_2018\\
GAIA\_GALAXY\_CLEMENTINI\_2020\\
GAIA\_RRL\_GAROFALO\_SELECTION\\
KEPLERGAIA\_BY\_ROT\_DISTEFANO\_2020\\
TESSGAIA\_BY\_ROT\_DISTEFANO\_2020\\
GAIA\_BY\_DISTEFANO\_2019\\
OGLE\_BLAP\_PIETRUKOWICZ\_2017\\
OGLE4\_CEP\_OGLE\_2020\\
OGLE4\_VAR\_OGLE\_2019\\
OGLE4\_BLG\_RRL\_SOSZYNSKI\_2019\\
OGLE4\_GD\_RRL\_SOSZYNSKI\_2019\\
CATALINA\_RRAB\_TORREALBA\_2015\\
OGLE4\_GSEP\_VAR\_SOSZYNSKI\_2012\\
NSVS\_RRAB\_KINEMUCHI\_2006\\
CATALINA\_CSS\_RRAB\_DRAKE\_2013\\
CATALINA\_MLS\_RRAB\_DRAKE\_2013\\
CATALINA\_RRAB\_DRAKE\_2013\\
CATALINA\_SEKBO\_RR\_DRAKE\_2013\\
SDSS\_DSCT\_RR\_SUVEGES\_2012\\
COMP\_OMEGACEN\_RRL\_CLEMENTINI\\
COMP\_M3\_RRL\_CLEMENTINI\\
COMP\_M15\_RRL\_CLEMENTINI\\
COMP\_NGC6388\_RRL\_CLEMENTINI\\
COMP\_NGC6441\_RRL\_CLEMENTINI\\
COMP\_UFD\_RRL\_CLEMENTINI\\
CATALINA\_VAR\_DRAKE\_2014\\
CATALINA\_VAR\_DRAKE\_2017\\
LINEAR\_VAR\_PALAVERSA\_2013\\
SDSS\_PS1\_CATALINA\_RRL\_ABBAS\_2014\\
OGLE4\_SHORTP\_ECL\_SOSZYNSKI\_2014\\
OGLE4\_M54\_VAR\_HAMANOWICZ\_2016\\
GAIA\_ECL\_RYBIZKI\_2018\\
OGLE3\_VAR\_OGLE3\_2012\\
OGLE\_BE\_SABOGAL\_2008\\
OGLE\_BE\_SABOGAL\_2005\\
OGLE\_BE\_MENNICKENT\_2002\\
ASASSN\_DIPPERS\_BREDALL\_2020\\
ASAS3\_MCP\_BERNHARD\_2015\\
ASAS\_KEPLER\_VAR\_PIGULSKI\_2009\\
COMP\_WD\_CORSICO\_2019\\
ASASSN\_VAR\_JAYASINGHE\_2019\\
ASAS\_VAR\_POJMANSKI\_2002\\
ATLAS\_VAR\_HEINZE\_2018\\
COMP\_SPB\_BCEP\_DECAT\_PR\\
IUE\_SPB\_NIEMCZURA\_2003\\
COMP\_GAL\_BCEP\_STANKOV\_2005\\
KEPLER\_DSCT\_GDOR\_BRADLEY\_2015\\
KEPLER\_GDOR\_VANREETH\_2015\\
KEPLER\_DSCT\_GDOR\_DEBOSSCHER\_2011\\
COROT\_DSCT\_GDOR\_SARRO\_2013\\
KEPLER\_DSCT\_GDOR\_UYTTERHOEVEN\_2011\\
2MASS\_GDOR\_ALICAVUS\_2016\\
EROS2\_BEATCEP\_MARQUETTE\_2009\\
ASAS\_SOLAR\_LIKE\_MESSINA\_2010\\
HATNET\_MEMBERS\_PLEIADES\_SOLAR\_LIKE\_HARTMAN\_2010\\
HATNET\_NONMEMBERS\_PLEIADES\_SOLAR\_LIKE\_HARTMAN\_2010\\
KEPLER\_ROAP\_HEY\_2019\\
\hline
\\
\caption{continued.}\\
\hline
\hline 
Catalogue  \\
\hline 
KEPLER\_MCP\_HUMMERICH\_2018\\
TESS\_VAR\_CUNHA\_2019\\
M31\_SEL\_LPV\_MOULD\_2004\\
M33\_CEP\_PELLERIN\_2011\\
GAIA\_M31\_CEP\_GAIA\_2018\\
KEPLER\_FLARES\_COMPILATION\\
KEPLER\_ROT\_REINHOLD\_2015\\
MMT\_M37\_FLARES\_CHANG\_2015\\
GAIA\_ROT\_GAIA\_2017\\
COROT\_ROT\_DEMEDEIROS\_2013\\
NSVS\_VAR\_WILLIAMS\_2004\\
HIP\_VAR\_ESA\_1997\\
SDSS\_CV\_SZKODY\_2011\\
ASAS\_SN\_ASAS\_2019\\
COMP\_CABS\_EKER\_2008\\
COMP\_AP\_RENSON\_2009\\
STEREO\_MAP\_WRAITH\_2012\\
TESS\_ROT\_SIKORA\_2019\\
CAHA\_SOLARLIKE\_MARTINEZ\_2010\\
KEPLER\_ROT\_HOWELL\_2016\\
VVV\_VAR\_BRAGA\_2019\\
IRAS\_LPV\_O\_SUH\_2017\\
IRAS\_LPV\_C\_SUH\_2017\\
IRAS\_LPV\_SUH\_2017\\
IRAS\_LPV\_SIC\_SUH\_2017\\
HALO\_MSR\_MAURON\_2020\\
2MASS\_LPV\_DEMERS\_2007\\
HIPPARCOS\_LPV\_BERGEAT\_2001\\
COMP\_SYST\_BELCZYNSKI\_2000\\
GAIA\_DR2\_SOS\_VARIABLES\_2018\\
GAIA\_DR2\_CLASSIFIER\_VARIABLES\_2018\\
2MASS\_GAIA\_WISE\_SYST\_AKRAS\_2019\\
GAIA\_ICRF2\_QSO\_GAIA\_2017\\
BZCAT\_BLAZARS\_MASSARO\_2015\\
ALMA\_BLAZARS\_BONATO\_2018\\
2WHSP\_BLAZARS\_CHANG\_2017\\
GAIA\_QSO\_GAIA\_CRF3\\
EROSITA\_AGN\_LIU\_2021\\
MILLIQUAS\_QSO\_FLESCH\_2019\\
HMQ\_QSO\_FLESCH\_2015\\
COMP\_ZZ\_NGUYEN\_2020\\
GAIA\_ZZ\_EYER\_2019\\
SOAR\_ZZ\_ROMERO\_2019\\
GALEX\_ZZ\_ROWAN\_2019\\
GAIA\_TRANSIENTS\_ALERTS\_2019\\
COMP\_VAR\_VSX\_2019\\
ZTF\_PERIODIC\_CHEN\_2020\\
IUE\_FKCOM\_BOPP\_1981\\
RAVE\_SOLAR\_LIKE\_ZERJAL\_2017\\
COROT\_SOLAR\_LIKE\_MEDHI\_2007\\
SDSS\_DQWD\_KOESTER\_2019\\
ZTF\_CV\_SZKODY\_2020\\
CATALINA\_CV\_DRAKE\_2014\\
OGLE4\_CV\_2016\_OGLE4\\
COMP\_MICROLENSING\_GAIA\\
GAIA\_GAL\_GAIA\_2018\\
COMP\_BHXRB\_CORRALSANTANA\_2016\\
OGLE4\_GSEP\_CST\_SOSZYNSKI\_2012\\
COMP\_YSO\_VARGA-VEREBELYI\_2020\\
\hline
\\
\caption{continued.}\\
\hline
\hline 
Catalogue  \\
\hline 
COMP\_DSCT\_GDOR\_DEBOSSCHER\_2007\\
DECAM\_VAR\_VIVAS\_2020\\
BULGE\_SOLAR\_LIKE\_DRAKE\_2006\\
ZTF\_CST\_GAIA\_2020\\
TESS\_CST\_GAIA\_2020\\
COMP\_EP\_SOUTHWORTH\_2020\\
PS1\_RRL\_SESAR\_SELECTION\_2017\\
KEPLER\_SHORTP\_SLAWSON\_2011\\
COMP\_XB\_RITTER\_2003\\
COMP\_CV\_RITTER\_2013\\
COMP\_PCB\_RITTER\_2013\\
KEPLER\_ECL\_KIRK\_2016\\
UKST\_LPV\_KUNKEL\_1997\\
SB9\_SB\_POURBAIX\_2004\\
SDSS\_CST\_IVEZIC\_2007\\
LAMOST\_RAD\_VEL\_VAR\_TIAN\_2020\\
GAIA\_WD\_GENTILEFUSILLO\_2019\\
PS1\_RRL\_SESAR\_2017\\
EROS2\_LPV\_SPANO\_2011\\
TRES\_ECL\_DEVOR\_2008\\
KEPLER\_VAR\_DEBOSSCHER\_2011\\
EROS2\_VAR\_KIM\_2014\\
ASAS\_VAR\_RICHARDS\_2012\\
SDSS\_VAR\_IVEZIC\_2007\\
INTEGRAL\_VAR\_ALFONSOGARZON\_2012\\
OGLE4\_MICROLENSING\_OGLE4\_2016\\
HATNET\_COMP\_PLEIADES\_SOLAR\_LIKE\_HARTMAN\_2010\\
OGLE4\_LMC\_CEP\_RR\_OGLE4\_2016\\
OGLE4\_BLG\_CEP\_RR\_OGLE4\_2016\\
OGLE4\_SMC\_CEP\_RR\_OGLE4\_2016\\
OGLE4\_LMC\_ECL\_OGLE4\_2017\\
OGLE4\_SMC\_ECL\_OGLE4\_2017\\
OGLE4\_CEP\_RR\_OGLE\_2016\\
OGLE\_BLG\_EB\_SOSZYNSKI\_2016\\
\hline
\end{longtable}

%% file: EB_naming_conv_table.tex
\begin{table}
\caption{\label{table:eb_naming_conv} Convention between different classes and sub-classes of eclipsing binaries and how they merged to 4 generic classes, ECL, EA, EB, and EW.}
\centering
\begin{tabular}{lllll}
\hline
\hline
\multicolumn{5}{c}{EA}\\
\hline
BETA\_PERSEI&detached&detachedEB&E/D &E/D/WR \\
E/DM&E/DS&E/DW & EA &EA/AR\\
EA/AR/RS&EA/AR: &EA/D &EA/D/R&EA/D/RS\\
EA/D/WR &EA/D:&EA/DM&EA/DM/RS&EA/DM: \\ 
EA/DS &EA/DS/RS&EA/DS:&EA/DW & EA/DW/RS \\
EA/GS&EA/GS+ &EA/GS/D &EA/GS/RS & EA/HW\\
EA/K&EA/KE &EA/KE: &EA/KW & EA/PN\\
EA/RS &EA/RS: &EA/WD&EA/WD/RS& EA/WR \\
EB\_ED&ECL/NC &ECLED&ECLNC & ED \\ 
NC&NONEC& \\
\multicolumn{5}{c}{}\\
\hline
\hline
\multicolumn{5}{c}{EB}\\
\hline
BETA\_LYRAE&E/SD&EB&EB/A & EB/AR/RS\\
EB/D&EB/D/G&EB/D/GS & EB/D:&EB/DM\\
EB/DM/WR&EB/DM: &EB/DS&EB/DW&EB/DW/RS\\
EB/GS &EB/GS/D&EB/RS&EB/SD&EB/SD: \\ 
EB/WR&EB\_ESD&ESD&near\_contac & semi-detached \\
\multicolumn{5}{c}{}\\
\hline
\hline 
\multicolumn{5}{c}{EW}\\
\hline
C&contac&contact&contactEB &E/KE\\
E/KW&EB/K&EB/KE &EB/KE:&EB/KW\\
EB\_EC&EC &ECL&ECL/C&ECLEC\\
EW & EW/K&EW/K&EW/KE:&EW/KW \\ 
EW/KW:&EW/RS&EW/WTTS&overcontact &W\_URSAE\_MAJ \\
\multicolumn{5}{c}{}\\
\hline
\hline
\multicolumn{5}{c}{ECL}\\
\hline 
CBF&CBH&DBF&DBH &E\\
E/GS &E/PSR&E/RS &E/WD&E/WR\\
EA/EB&EA/EL &EA/EW&EA/SD&EA/SD/RS\\
EA/SD: &EB/EA&EB/EW&EB\_ED\_ESD&EB\_Other \\ 
EC/ESD&EC=ESD&ED/ESD&ESD/EC &ESD/EC/ED\\
ESD=ED&EW/D&EW/DW &EW/DW/RS&EW/DW:\\
EW/EA& \\
\hline

\end{tabular}
\end{table} 

%% file: fieldNames.tex
\begin{table}
\centering
\caption{\label{table:catalog_labels} Definition of the fields in the cross-match catalogue. The fields in plural may contain multiple values separeted by ";", thus are regarded as Strings. }
\begin{tabular}{llll}
\hline
\hline
Field name & type & units & Definition\\
\hline
gaia\_dr3\_source\_id& Long&&\textit{Gaia}~DR3 source id\\
non\_gaia\_ids &String&& Identifications from literature catalogues.\\
non\_gaia\_coordinates &String&& Coordinates from literature catalogues.\\
primary\_ra&Double&degrees& Right Ascension from the higher ranked catalogue.\\
primary\_dec&Double&degrees& Declination from the higher ranked catalogue.\\
non\_gaia\_mags&String&& Magnitude(s) from literature\\
synthetic\_distances&Double& & Synthetic astrometric-photometric distance (see Eq.~\ref{eq:rsynth}). \\
primary\_superclass&String& & Generic superclass label corresponding to primary\_var\_type.\\
primary\_var\_type&String& & Variability type from the highest ranked catalogue for a given source.\\ 
var\_types&String& & Variability type(s) from literature, sorted by catalogue rank.\\
original\_var\_types&String&& Original variability (sub)type(s) from literature.\\
original\_alt\_var\_types&String&& Alternative variable type(s) from literature.\\
primary\_period&Double&days&Variability period from the highest ranked catalogue for a given source.\\
periods&String& & Variability period(s) from the literature.\\
other\_periods&String&& Other period(s) from the literature.\\
ref\_epochs&String&& Approximate reference epochs (J2000) of the literature.\\
primary\_catalogue\_label&String& & Label of the highest ranked catalogue for a given source.\\
catalogue\_labels&String&& Label(s) of the literature catalogue(s).\\
references&String&& Reference(s) of the literature.\\
selection&Boolean& & Selection of the most reliable catalogues or classes. \\ 
\hline
\end{tabular}
\end{table}

%% file: superclassesTable.tex
\begin{table}
\centering
\caption{\label{table:sourcesPerSuperclass} List of available types of {\tt{primary\_superclass}}. The second column shows the {\tt{primary\_var\_types}} that contribute to this {\tt{primary\_superclass}}. The last two columns give the number of sources for each of  {\tt{primary\_superclass}} in the full catalogue ({\tt{All}}) and when the {\tt{selection}} flag is true.}
\begin{tabular}{llrr}
\hline
\hline
{\tt{primary\_superclass}}&{\tt{primary\_var\_types}}&All&{\tt{selection}}\\ 
\hline 
ACYG&ACYG&64&64\\ 
AGN&AGN, BLAZAR, BLLAC, QSO&1801094&1646705\\ 
AHB1&AHB1&2&2\\ 
BCEP&BCEP&17677&184\\ 
BE&BE&4978&4869\\ 
BHXB&BHXB&45&45\\ 
BLAP&BLAP&14&14\\ 
CEP&ACEP, BLHER, CEP, CW, DCEP, RV, T2CEP&19482&19131\\ 
CP&ACV, CP, MCP, ROAM, ROAP&2949&2879\\ 
CST&CST, DQ, HOT\_DQ, WARM\_DQ, WD&1137487&688952\\ 
CV&CV, SN&3937&3936\\ 
DPV&DPV&5&5\\ 
DSCT&DSCT&37656&27605\\ 
DSCT+GDOR&DSCT+GDOR&128&128\\ 
DSCT|SXPHE&DSCT|SXPHE&8196&8196\\ 
ECL&EA, EB, ECL, EW&1128021&1091917\\ 
ELL&ELL&28405&27730\\ 
EP&EP&1053&1053\\ 
FKCOM&FKCOM&5&5\\ 
GALAXY&GALAXY&1746224&1746200\\ 
GCAS&GCAS&2000&2000\\ 
GDOR&GDOR&1757&1590\\ 
HB&HB&147&147\\ 
HMXB&HMXB&39&38\\ 
I&I&599&553\\ 
L&L&15754&15682\\ 
LPV& LPV, LSP, M, M|SR, OSARG, SARG, SR, SRA, SRB, SRC,&618988&595656\\ 
& SRD, SRS&&\\ 
MICROLENSING&MICROLENSING&2712&113\\ 
OMIT&OMIT&341218&null\\ 
PCEB&PCEB&81&81\\ 
PPN&PPN&13&13\\ 
PVTEL&PVTEL&14&14\\ 
R&R&49&49\\ 
RAD\_VEL\_VAR&RAD\_VEL\_VAR&74511&74511\\ 
RCB&RCB&75&70\\ 
RR&ARRD,RR,RRAB,RRC,RRD&393030&294544\\ 
S&S&2830&2828\\ 
SARV&SARV&8&8\\ 
SB&SB&1571&1571\\ 
SDOR&SDOR&21&21\\ 
SOLAR\_LIKE&BY, BY|ROT, FLARES, ROT, RS, SOLAR\_LIKE, UV&430575&420173\\ 
SPB&SPB&261&202\\ 
SXARI&SXARI&23&23\\ 
SXPHE&SXPHE&92&92\\ 
SYST&SYST, ZAND&268&268\\ 
V1093HER&V1093HER&10&10\\ 
V361HYA&V361HYA&57&57\\ 
WR&WR&45&44\\ 
X&X&42&41\\ 
YSO&CTTS, DIP, FUOR, GTTS, HAEBE, IMTTS, PULS\_PMS, TTS,&15786&15786\\ 
&UXOR, WTTS, YSO&&\\ 
ZZ&ELM\_ZZA, GWVIR, HOT\_DQV, HOT\_ZZA, PRE\_ELM\_ZZA,&1712&1712\\ 
&V777HER, ZZ, ZZA&&\\ 
ZZLEP&ZZLEP&13&13\\ 
\hline \\ 
\end{tabular}
\end{table} 

%% file: comparison10.0_class.tex
\begin{longtable}{lrrrr}
\caption{\label{table:sourcesPerclass} List of different classes of sources existing in the cross-match catalogue and the number of sources found in the {\tt{primary\_var\_type}} and {\tt{var\_type}} fields with such class. For the field {\tt{var\_type}} if a type is found multiple times for a given source id then it is counted once. The columns {\tt{All}} refer to all sources in our catalogue while {\tt{selection}} only to the cases where {\tt{selection}} flag is true.}\\
\hline \hline 
  & \multicolumn{2}{c}{primary\_var\_type} & \multicolumn{2}{c}{var\_type}\\
 \hline 
Class&All&{\tt{selection}}&All&{\tt{selection}}\\ 
 \hline
 \endfirsthead 
 \hline 
ACEP&600&599&872&871\\ 
ACV&607&607&1414&1414\\ 
ACYG&64&64&75&74\\ 
AGN&5616&4624&24313&23321\\ 
AHB1&2&2&188&188\\ 
ARRD&114&114&170&170\\ 
BCEP&17677&184&19118&1623\\ 
BE&4978&4869&5327&5218\\ 
BHXB&45&45&50&50\\ 
BLAP&14&14&14&14\\ 
BLAZAR&2990&2938&4875&4823\\ 
BLHER&1153&1153&1615&1615\\ 
BLLAC&11&11&650&650\\ 
BY&104699&104695&109833&109829\\ 
BY|ROT&283&283&283&283\\ 
CEP&510&485&961&936\\ 
CP&70&&234&164\\ 
CST&688960&688948&691015&691003\\ 
CTTS&443&443&647&647\\ 
CV&3063&3062&3263&3262\\ 
CW&1265&1265&1596&1596\\ 
DCEP&14601&14430&16115&15944\\ 
DIP&13&13&14&14\\ 
DPV&5&5&26&26\\ 
DQ&244&4&246&5\\ 
DSCT&37656&27605&40766&30715\\ 
DSCT+GDOR&128&128&202&202\\ 
DSCT|SXPHE&8196&8196&8956&8956\\ 
EA&415864&407326&424621&416057\\ 
EB&31796&21819&47003&36898\\ 
ECL&198020&185384&360501&347833\\ 
ELL&28405&27730&29613&28357\\ 
ELM\_ZZA&9&9&10&10\\ 
EP&1053&1053&1155&1155\\ 
EW&482341&477388&537269&531932\\ 
FKCOM&5&5&8&8\\ 
FLARES&9890&854&20567&11528\\ 
FUOR&7&7&13&13\\ 
GALAXY&1746224&1746200&1748871&1748847\\ 
GCAS&2000&2000&2130&2130\\ 
GDOR&1757&1590&2039&1872\\ 
GTTS&13&13&20&20\\ 
GWVIR&782&782&787&787\\ 
HAEBE&30&30&79&79\\ 
HB&147&147&147&147\\ 
HMXB&39&38&41&40\\ 
HOT\_DQ&3&&7&4\\ 
HOT\_DQV&6&6&6&6\\ 
HOT\_ZZA&3&3&3&3\\ 
I&599&553&941&894\\ 
IMTTS&2&2&4&4\\ 
L&15754&15682&21747&21651\\ 
LPV&91816&79980&130165&118323\\ 
LSP&1073&39&1846&812\\ 
M&13597&13304&22028&21596\\ 
\hline
\\
\caption{continued.}\\
\hline 
\hline 
  & \multicolumn{2}{c}{primary\_var\_type} & \multicolumn{2}{c}{var\_type}\\
 \hline 
Class&All&{\tt{selection}}&All&{\tt{selection}}\\ 
 \hline
MCP&2243&2243&3110&3110\\ 
MICROLENSING&2712&113&2740&141\\ 
M|SR&55621&55621&150252&150252\\ 
OMIT&341218&&604313&253449\\ 
OSARG&272570&268951&277507&270351\\ 
PCEB&81&81&81&81\\ 
PPN&13&13&16&16\\ 
PRE\_ELM\_ZZA&10&10&10&10\\ 
PULS\_PMS&5&5&5&5\\ 
PVTEL&14&14&20&20\\ 
QSO&1792477&1639132&1797208&1643863\\ 
R&49&49&71&71\\ 
RAD\_VEL\_VAR&74511&74511&78629&78629\\ 
RCB&75&70&107&102\\ 
ROAM&11&11&12&12\\ 
ROAP&18&18&50&50\\ 
ROT&202126&200830&231056&229760\\ 
RR&94128&122&229514&135000\\ 
RRAB&218723&217115&224482&222874\\ 
RRC&74761&71967&83624&80829\\ 
RRD&5304&5226&7568&7490\\ 
RS&80259&80193&84866&84800\\ 
RV&516&469&696&649\\ 
S&2830&2828&3150&3148\\ 
SARG&4757&&5512&755\\ 
SARV&8&8&8&8\\ 
SB&1571&1571&3256&3256\\ 
SDOR&21&21&23&23\\ 
SN&874&874&875&875\\ 
SOLAR\_LIKE&32341&32341&35836&35836\\ 
SPB&261&202&663&604\\ 
SR&170957&169208&198481&195201\\ 
SRA&190&179&560&548\\ 
SRB&765&735&1230&1193\\ 
SRC&23&22&46&45\\ 
SRD&249&247&393&391\\ 
SRS&7370&7370&8054&8054\\ 
SXARI&23&23&30&30\\ 
SXPHE&92&92&127&127\\ 
SYST&241&241&361&361\\ 
T2CEP&837&730&3173&3064\\ 
TTS&1376&1376&1699&1699\\ 
UV&977&977&1084&1084\\ 
UXOR&57&57&97&97\\ 
V1093HER&10&10&10&10\\ 
V361HYA&57&57&57&57\\ 
V777HER&347&347&348&348\\ 
WARM\_DQ&26&&26&\\ 
WD&448254&&468880&19701\\ 
WR&45&44&46&45\\ 
WTTS&414&414&537&537\\ 
X&42&41&57&56\\ 
YSO&13426&13426&16410&16410\\ 
ZAND&27&27&116&116\\ 
ZZ&10&10&20&20\\ 
ZZA&545&545&554&554\\ 
ZZLEP&13&13&13&13\\ 

 \hline 
 
\end{longtable}

%% file: overlapsExample.tex
\begin{table}
\caption{ Counterparts of \textit{Gaia}~DR3 source\_id 4066039874096072576 in different catalogues, discussed in \ref{subsec:overlaps}. The column {\tt{var\_type}} shows the how it was identified in the different literature catalogues it exists.}
\label{table:missclass}
\centering
\begin{tabular}{llllr}
\hline
\hline
            {\tt{catalogue\_label}}&Id&Coordinates&{\tt{var\_type}}&Period\\
            \hline
            ASASSN\_VAR\_JAYASINGHE\_2019 &  \href{https://asas-sn.osu.edu/variables/252221}{J180253.67-240945.3}&270.72363,-24.16259&SR & 18.31322\\
            COMP\_VAR\_VSX\_2019& \href{https://www.aavso.org/vsx/index.php?view=detail.top&oid=108780}{108780}&270.72383,-24.16267&ECL&2.10904\\
            ASAS\_VAR\_RICHARDS\_2012&180254-2409.7&270.72342,-24.16262&EW&2.10907\\
            INTEGRAL\_VAR\_ALFONSOGARZON\_2012& \href{http://cdsarc.u-strasbg.fr/ftp/cats/J/A+A/548/A79/img/IOMC6842000081.png}{6842000081}& 270.72354,-24.16225&EB&2.10914\\
            \hline
\end{tabular}
\end{table}

%% file: GAIA_DR3_Crossmatch.bbl
\begin{thebibliography}{205}
\expandafter\ifx\csname natexlab\endcsname\relax\def\natexlab#1{#1}\fi

\bibitem[{{Abbas} {et~al.}(2014){Abbas}, {Grebel}, {Martin}, {Burgett},
  {Flewelling}, \& {Wainscoat}}]{2014MNRAS.441.1230A}
{Abbas}, M.~A., {Grebel}, E.~K., {Martin}, N.~F., {et~al.} 2014, \mnras, 441,
  1230

\bibitem[{{Akras} {et~al.}(2019){Akras}, {Guzman-Ramirez}, {Leal-Ferreira}, \&
  {Ramos-Larios}}]{2019ApJS..240...21A}
{Akras}, S., {Guzman-Ramirez}, L., {Leal-Ferreira}, M.~L., \& {Ramos-Larios},
  G. 2019, \apjs, 240, 21

\bibitem[{{Alfonso-Garz{\'o}n} {et~al.}(2012){Alfonso-Garz{\'o}n}, {Domingo},
  {Mas-Hesse}, \& {Gim{\'e}nez}}]{2012A&A...548A..79A}
{Alfonso-Garz{\'o}n}, J., {Domingo}, A., {Mas-Hesse}, J.~M., \& {Gim{\'e}nez},
  A. 2012, \aap, 548, A79

\bibitem[{{Baade}(1926)}]{1926AN....228..359B}
{Baade}, W. 1926, Astronomische Nachrichten, 228, 359

\bibitem[{{Balona}(2022)}]{2022MNRAS.510.5743B}
{Balona}, L.~A. 2022, \mnras, 510, 5743

\bibitem[{{Beauchamp} {et~al.}(1999){Beauchamp}, {Wesemael}, {Bergeron},
  {Fontaine}, {Saffer}, {Liebert}, \& {Brassard}}]{1999ApJ...516..887B}
{Beauchamp}, A., {Wesemael}, F., {Bergeron}, P., {et~al.} 1999, \apj, 516, 887

\bibitem[{{Belczy{\'n}ski} {et~al.}(2000){Belczy{\'n}ski}, {Miko{\l}ajewska},
  {Munari}, {Ivison}, \& {Friedjung}}]{2000A&AS..146..407B}
{Belczy{\'n}ski}, K., {Miko{\l}ajewska}, J., {Munari}, U., {Ivison}, R.~J., \&
  {Friedjung}, M. 2000, \aaps, 146, 407

\bibitem[{{Benk{\H{o}}} {et~al.}(2006){Benk{\H{o}}}, {Bakos}, \&
  {Nuspl}}]{2006MNRAS.372.1657B}
{Benk{\H{o}}}, J.~M., {Bakos}, G.~{\'A}., \& {Nuspl}, J. 2006, \mnras, 372,
  1657

\bibitem[{{Bergeat} {et~al.}(2001){Bergeat}, {Knapik}, \&
  {Rutily}}]{2001A&A...369..178B}
{Bergeat}, J., {Knapik}, A., \& {Rutily}, B. 2001, \aap, 369, 178

\bibitem[{{Bernhard} {et~al.}(2015){Bernhard}, {H{\"u}mmerich}, {Otero}, \&
  {Paunzen}}]{2015A&A...581A.138B}
{Bernhard}, K., {H{\"u}mmerich}, S., {Otero}, S., \& {Paunzen}, E. 2015, \aap,
  581, A138

\bibitem[{{Boettcher} {et~al.}(2013){Boettcher}, {Willman}, {Fadely},
  {Strader}, {Baker}, {Hopkins}, {Tasnim Ananna}, {Cunningham}, {Douglas},
  {Gilbert}, {Preston}, \& {Sturner}}]{2013AJ....146...94B}
{Boettcher}, E., {Willman}, B., {Fadely}, R., {et~al.} 2013, \aj, 146, 94

\bibitem[{{Bogn{\'a}r} {et~al.}(2020){Bogn{\'a}r}, {Kawaler}, {Bell},
  {Schrandt}, {Baran}, {Bradley}, {Hermes}, {Charpinet}, {Handler}, {Mullally},
  {Murphy}, {Raddi}, {S{\'o}dor}, {Tremblay}, {Uzundag}, \&
  {Zong}}]{2020A&A...638A..82B}
{Bogn{\'a}r}, Z., {Kawaler}, S.~D., {Bell}, K.~J., {et~al.} 2020, \aap, 638,
  A82

\bibitem[{{Bogn{\'a}r} {et~al.}(2014){Bogn{\'a}r}, {Papar{\'o}}, {C{\'o}rsico},
  {Kepler}, \& {Gy{\H{o}}rffy}}]{2014A&A...570A.116B}
{Bogn{\'a}r}, Z., {Papar{\'o}}, M., {C{\'o}rsico}, A.~H., {Kepler}, S.~O., \&
  {Gy{\H{o}}rffy}, {\'A}. 2014, \aap, 570, A116

\bibitem[{{Boller} {et~al.}(2021){Boller}, {Schmitt}, {Buchner}, {Freyberg},
  {Georgakakis}, {Liu}, {Robrade}, {Merloni}, {Nandra}, {Malyali}, {Krumpe},
  {Salvato}, \& {Dwelly}}]{2021arXiv210614523B}
{Boller}, T., {Schmitt}, J.~H.~M.~M., {Buchner}, J., {et~al.} 2021, arXiv
  e-prints, arXiv:2106.14523

\bibitem[{{Bonato} {et~al.}(2018){Bonato}, {Liuzzo}, {Giannetti}, {Massardi},
  {De Zotti}, {Burkutean}, {Galluzzi}, {Negrello}, {Baronchelli}, {Brand},
  {Zwaan}, {Rygl}, {Marchili}, {Klitsch}, \& {Oteo}}]{2018MNRAS.478.1512B}
{Bonato}, M., {Liuzzo}, E., {Giannetti}, A., {et~al.} 2018, \mnras, 478, 1512

\bibitem[{{Bopp} \& {Stencel}(1981)}]{1981ApJ...247L.131B}
{Bopp}, B.~W. \& {Stencel}, R.~E. 1981, \apjl, 247, L131

\bibitem[{{Bradley} {et~al.}(2015){Bradley}, {Guzik}, {Miles}, {Uytterhoeven},
  {Jackiewicz}, \& {Kinemuchi}}]{2015AJ....149...68B}
{Bradley}, P.~A., {Guzik}, J.~A., {Miles}, L.~F., {et~al.} 2015, \aj, 149, 68

\bibitem[{{Braga} {et~al.}(2019){Braga}, {Contreras Ramos}, {Minniti},
  {Ferreira Lopes}, {Catelan}, {Minniti}, {Nikzat}, \&
  {Zoccali}}]{2019A&A...625A.151B}
{Braga}, V.~F., {Contreras Ramos}, R., {Minniti}, D., {et~al.} 2019, \aap, 625,
  A151

\bibitem[{{Braga} {et~al.}(2016){Braga}, {Stetson}, {Bono}, {Dall'Ora},
  {Ferraro}, {Fiorentino}, {Freyhammer}, {Iannicola}, {Marengo}, {Neeley},
  {Valenti}, {Buonanno}, {Calamida}, {Castellani}, {da Silva},
  {Degl'Innocenti}, {Di Cecco}, {Fabrizio}, {Freedman}, {Giuffrida}, {Lub},
  {Madore}, {Marconi}, {Marinoni}, {Matsunaga}, {Monelli}, {Persson},
  {Piersimoni}, {Pietrinferni}, {Prada-Moroni}, {Pulone}, {Stellingwerf},
  {Tognelli}, \& {Walker}}]{2016AJ....152..170B}
{Braga}, V.~F., {Stetson}, P.~B., {Bono}, G., {et~al.} 2016, \aj, 152, 170

\bibitem[{{Bredall} {et~al.}(2020){Bredall}, {Shappee}, {Gaidos}, {Jayasinghe},
  {Vallely}, {Stanek}, {Kochanek}, {Gagn{\'e}}, {Hart}, {Holoien}, {Prieto}, \&
  {Van Saders}}]{2020MNRAS.496.3257B}
{Bredall}, J.~W., {Shappee}, B.~J., {Gaidos}, E., {et~al.} 2020, \mnras, 496,
  3257

\bibitem[{{Chang} {et~al.}(2015){Chang}, {Byun}, \&
  {Hartman}}]{2015ApJ...814...35C}
{Chang}, S.~W., {Byun}, Y.~I., \& {Hartman}, J.~D. 2015, \apj, 814, 35

\bibitem[{{Chang} {et~al.}(2017){Chang}, {Arsioli}, {Giommi}, \&
  {Padovani}}]{2017A&A...598A..17C}
{Chang}, Y.~L., {Arsioli}, B., {Giommi}, P., \& {Padovani}, P. 2017, \aap, 598,
  A17

\bibitem[{{Chen} {et~al.}(2020){Chen}, {Wang}, {Deng}, {de Grijs}, {Yang}, \&
  {Tian}}]{2020ApJS..249...18C}
{Chen}, X., {Wang}, S., {Deng}, L., {et~al.} 2020, \apjs, 249, 18

\bibitem[{{Clementini} {et~al.}(2016){Clementini}, {Ripepi}, {Leccia},
  {Mowlavi}, {Lecoeur-Taibi}, {Marconi}, {Szabados}, {Eyer}, {Guy},
  {Rimoldini}, {Jevardat de Fombelle}, {Holl}, {Busso}, {Charnas}, {Cuypers},
  {De Angeli}, {De Ridder}, {Debosscher}, {Evans}, {Klagyivik}, {Musella},
  {Nienartowicz}, {Ord{\'o}{\~n}ez}, {Regibo}, {Riello}, {Sarro}, \&
  {S{\"u}veges}}]{2016A&A...595A.133C}
{Clementini}, G., {Ripepi}, V., {Leccia}, S., {et~al.} 2016, \aap, 595, A133

\bibitem[{{Clementini} {et~al.}(2019){Clementini}, {Ripepi}, {Molinaro},
  {Garofalo}, {Muraveva}, {Rimoldini}, {Guy}, {Jevardat de Fombelle},
  {Nienartowicz}, {Marchal}, {Audard}, {Holl}, {Leccia}, {Marconi}, {Musella},
  {Mowlavi}, {Lecoeur-Taibi}, {Eyer}, {De Ridder}, {Regibo}, {Sarro},
  {Szabados}, {Evans}, \& {Riello}}]{2019A&A...622A..60C}
{Clementini}, G., {Ripepi}, V., {Molinaro}, R., {et~al.} 2019, \aap, 622, A60

\bibitem[{{Corral-Santana} {et~al.}(2016){Corral-Santana}, {Casares},
  {Mu{\~n}oz-Darias}, {Bauer}, {Mart{\'\i}nez-Pais}, \&
  {Russell}}]{2016A&A...587A..61C}
{Corral-Santana}, J.~M., {Casares}, J., {Mu{\~n}oz-Darias}, T., {et~al.} 2016,
  \aap, 587, A61

\bibitem[{{C{\'o}rsico} {et~al.}(2019){C{\'o}rsico}, {Althaus}, {Miller
  Bertolami}, \& {Kepler}}]{2019A&ARv..27....7C}
{C{\'o}rsico}, A.~H., {Althaus}, L.~G., {Miller Bertolami}, M.~M., \& {Kepler},
  S.~O. 2019, \aapr, 27, 7

\bibitem[{{Corwin} {et~al.}(2008){Corwin}, {Borissova}, {Stetson}, {Catelan},
  {Smith}, {Kurtev}, \& {Stephens}}]{2008AJ....135.1459C}
{Corwin}, T.~M., {Borissova}, J., {Stetson}, P.~B., {et~al.} 2008, \aj, 135,
  1459

\bibitem[{{Corwin} {et~al.}(2006){Corwin}, {Sumerel}, {Pritzl}, {Smith},
  {Catelan}, {Sweigart}, \& {Stetson}}]{2006AJ....132.1014C}
{Corwin}, T.~M., {Sumerel}, A.~N., {Pritzl}, B.~J., {et~al.} 2006, \aj, 132,
  1014

\bibitem[{{Cunha} {et~al.}(2019){Cunha}, {Antoci}, {Holdsworth}, {Kurtz},
  {Balona}, {Bogn{\'a}r}, {Bowman}, {Guo}, {Ko{\l}aczek-Szyma{\'n}ski},
  {Lares-Martiz}, {Paunzen}, {Skarka}, {Smalley}, {S{\'o}dor}, {Kochukhov},
  {Pepper}, {Richey-Yowell}, {Ricker}, {Seager}, {Buzasi}, {Fox-Machado},
  {Hasanzadeh}, {Niemczura}, {Quitral-Manosalva}, {Monteiro}, {Stateva}, {De
  Cat}, {Garc{\'\i}a Hern{\'a}ndez}, {Ghasemi}, {Handler}, {Hey}, {Matthews},
  {Nemec}, {Pascual-Granado}, {Safari}, {Su{\'a}rez}, {Szab{\'o}}, {Tkachenko},
  \& {Weiss}}]{2019MNRAS.487.3523C}
{Cunha}, M.~S., {Antoci}, V., {Holdsworth}, D.~L., {et~al.} 2019, \mnras, 487,
  3523

\bibitem[{{Dall'Ora} {et~al.}(2006){Dall'Ora}, {Clementini}, {Kinemuchi},
  {Ripepi}, {Marconi}, {Di Fabrizio}, {Greco}, {Rodgers}, {Kuehn}, \&
  {Smith}}]{2006ApJ...653L.109D}
{Dall'Ora}, M., {Clementini}, G., {Kinemuchi}, K., {et~al.} 2006, \apjl, 653,
  L109

\bibitem[{{Dall'Ora} {et~al.}(2012){Dall'Ora}, {Kinemuchi}, {Ripepi},
  {Rodgers}, {Clementini}, {Di Fabrizio}, {Smith}, {Marconi}, {Musella},
  {Greco}, {Kuehn}, {Catelan}, {Pritzl}, \& {Beers}}]{2012ApJ...752...42D}
{Dall'Ora}, M., {Kinemuchi}, K., {Ripepi}, V., {et~al.} 2012, \apj, 752, 42

\bibitem[{{De Medeiros} {et~al.}(2013){De Medeiros}, {Ferreira Lopes},
  {Le{\~a}o}, {Canto Martins}, {Catelan}, {Baglin}, {Vieira}, {Bravo},
  {Cort{\'e}s}, {de Freitas}, {Janot-Pacheco}, {Maciel}, {Melo}, {Osorio},
  {Porto de Mello}, \& {Valio}}]{2013A&A...555A..63D}
{De Medeiros}, J.~R., {Ferreira Lopes}, C.~E., {Le{\~a}o}, I.~C., {et~al.}
  2013, \aap, 555, A63

\bibitem[{{de Vaucouleurs}(1978)}]{1978ApJ...223..351D}
{de Vaucouleurs}, G. 1978, \apj, 223, 351

\bibitem[{{Debosscher} {et~al.}(2011){Debosscher}, {Blomme}, {Aerts}, \& {De
  Ridder}}]{2011A&A...529A..89D}
{Debosscher}, J., {Blomme}, J., {Aerts}, C., \& {De Ridder}, J. 2011, \aap,
  529, A89

\bibitem[{{Debosscher} {et~al.}(2007){Debosscher}, {Sarro}, {Aerts}, {Cuypers},
  {Vandenbussche}, {Garrido}, \& {Solano}}]{2007A&A...475.1159D}
{Debosscher}, J., {Sarro}, L.~M., {Aerts}, C., {et~al.} 2007, \aap, 475, 1159

\bibitem[{{Demers} \& {Battinelli}(2007)}]{2007A&A...473..143D}
{Demers}, S. \& {Battinelli}, P. 2007, \aap, 473, 143

\bibitem[{{Devor} {et~al.}(2008){Devor}, {Charbonneau}, {O'Donovan},
  {Mandushev}, \& {Torres}}]{2008AJ....135..850D}
{Devor}, J., {Charbonneau}, D., {O'Donovan}, F.~T., {Mandushev}, G., \&
  {Torres}, G. 2008, \aj, 135, 850

\bibitem[{{Drake}(2006)}]{2006AJ....131.1044D}
{Drake}, A.~J. 2006, \aj, 131, 1044

\bibitem[{{Drake} {et~al.}(2013{\natexlab{a}}){Drake}, {Catelan}, {Djorgovski},
  {Torrealba}, {Graham}, {Belokurov}, {Koposov}, {Mahabal}, {Prieto},
  {Donalek}, {Williams}, {Larson}, {Christensen}, \&
  {Beshore}}]{2013ApJ...763...32D}
{Drake}, A.~J., {Catelan}, M., {Djorgovski}, S.~G., {et~al.}
  2013{\natexlab{a}}, \apj, 763, 32

\bibitem[{{Drake} {et~al.}(2013{\natexlab{b}}){Drake}, {Catelan}, {Djorgovski},
  {Torrealba}, {Graham}, {Mahabal}, {Prieto}, {Donalek}, {Williams}, {Larson},
  {Christensen}, \& {Beshore}}]{2013ApJ...765..154D}
{Drake}, A.~J., {Catelan}, M., {Djorgovski}, S.~G., {et~al.}
  2013{\natexlab{b}}, \apj, 765, 154

\bibitem[{{Drake} {et~al.}(2017){Drake}, {Djorgovski}, {Catelan}, {Graham},
  {Mahabal}, {Larson}, {Christensen}, {Torrealba}, {Beshore}, {McNaught},
  {Garradd}, {Belokurov}, \& {Koposov}}]{2017MNRAS.469.3688D}
{Drake}, A.~J., {Djorgovski}, S.~G., {Catelan}, M., {et~al.} 2017, \mnras, 469,
  3688

\bibitem[{{Drake} {et~al.}(2014{\natexlab{a}}){Drake}, {G{\"a}nsicke},
  {Djorgovski}, {Wils}, {Mahabal}, {Graham}, {Yang}, {Williams}, {Catelan},
  {Prieto}, {Donalek}, {Larson}, \& {Christensen}}]{2014MNRAS.441.1186D}
{Drake}, A.~J., {G{\"a}nsicke}, B.~T., {Djorgovski}, S.~G., {et~al.}
  2014{\natexlab{a}}, \mnras, 441, 1186

\bibitem[{{Drake} {et~al.}(2014{\natexlab{b}}){Drake}, {Graham}, {Djorgovski},
  {Catelan}, {Mahabal}, {Torrealba}, {Garc{\'\i}a-{\'A}lvarez}, {Donalek},
  {Prieto}, {Williams}, {Larson}, {Christen sen}, {Belokurov}, {Koposov},
  {Beshore}, {Boattini}, {Gibbs}, {Hill}, {Kowalski}, {Johnson}, \&
  {Shelly}}]{2014ApJS..213....9D}
{Drake}, A.~J., {Graham}, M.~J., {Djorgovski}, S.~G., {et~al.}
  2014{\natexlab{b}}, \apjs, 213, 9

\bibitem[{{Dufour} {et~al.}(2011){Dufour}, {B{\'e}land}, {Fontaine}, {Chayer},
  \& {Bergeron}}]{2011ApJ...733L..19D}
{Dufour}, P., {B{\'e}land}, S., {Fontaine}, G., {Chayer}, P., \& {Bergeron}, P.
  2011, \apjl, 733, L19

\bibitem[{{Dunlap} {et~al.}(2010){Dunlap}, {Barlow}, \&
  {Clemens}}]{2010ApJ...720L.159D}
{Dunlap}, B.~H., {Barlow}, B.~N., \& {Clemens}, J.~C. 2010, \apjl, 720, L159

\bibitem[{{Eker} {et~al.}(2008){Eker}, {Ak}, {Bilir}, {Do{\v{g}}ru},
  {T{\"u}ys{\"u}z}, {Soydugan}, {Bak{\i}{\textcommabelow s}},
  {U{\v{g}}ra{\textcommabelow s}}, {Soydugan}, {Erdem}, \&
  {Demircan}}]{2008MNRAS.389.1722E}
{Eker}, Z., {Ak}, N.~F., {Bilir}, S., {et~al.} 2008, \mnras, 389, 1722

\bibitem[{{ESA}(1997)}]{1997ESASP1200.....E}
{ESA}, ed. 1997, ESA Special Publication, Vol. 1200, {The HIPPARCOS and TYCHO
  catalogues. Astrometric and photometric star catalogues derived from the ESA
  HIPPARCOS Space Astrometry Mission}

\bibitem[{{Eyer} {et~al.}(2017){Eyer}, {Mowlavi}, {Evans}, {Nienartowicz},
  {Ordonez}, {Holl}, {Lecoeur-Taibi}, {Riello}, {Clementini}, {Cuypers}, {De
  Ridder}, {Lanzafame}, {Sarro}, {Charnas}, {Guy}, {Jevardat de Fombelle},
  {Rimoldini}, {S{\"u}veges}, {Mignard}, {Busso}, {De Angeli}, {van Leeuwen},
  {Dubath}, {Beck}, {Aguado}, {Debosscher}, {Distefano}, {Fuchs}, {Koubsky},
  {Lebzelter}, {Leccia}, {Lopez}, {Moitinho}, {Regibo}, {Ripepi}, {Roelens},
  {Szabados}, {Tingley}, {Votruba}, {Zucker}, {Aerts}, {Barblan},
  {Blanco-Cuaresma}, {Grenon}, {Jan}, {Lorenz}, {Miranda}, {Morgenthaler},
  {Ordenovic}, {Palaversa}, {Prsa}, {Ruiz-Fuertes}, {Anderson}, {Delgado},
  {Dzigan}, {Hudec}, {Jonckheere}, {Klagyivik}, {Kutka}, {Moniez}, {Nicoletti},
  {Park}, {Van Hemelryck}, {Varadi}, {Kochoska}, {Lanza}, {Marconi},
  {Marschalko}, {Messina}, {Musella}, {Pagano}, {Sadowski}, \&
  {Schultheis}}]{2017arXiv170203295E}
{Eyer}, L., {Mowlavi}, N., {Evans}, D.~W., {et~al.} 2017, arXiv e-prints,
  arXiv:1702.03295

\bibitem[{{Eyer} {et~al.}(2019){Eyer}, {Rimoldini}, {Rohrbasser}, {Holl},
  {Audard}, {Evans}, {Garcia-Lario}, {Gavras}, {Clementini}, {Hodgkin},
  {Jevardat de Fombelle}, {Lanzafame}, {Lebzelter}, {Lecoeur-Taibi}, {Mowlavi},
  {Nienartowicz}, {Ripepi}, \& {Wyrzykowski}}]{2019arXiv191207659E}
{Eyer}, L., {Rimoldini}, L., {Rohrbasser}, L., {et~al.} 2019, arXiv e-prints,
  arXiv:1912.07659

\bibitem[{{Eyer et al.}(2022)}]{DR3-DPACP-162}
{Eyer et al.} 2022, \aap\ in prep.

\bibitem[{{Flesch}(2015)}]{2015PASA...32...10F}
{Flesch}, E.~W. 2015, \pasa, 32, e010

\bibitem[{{Flesch}(2019)}]{2019arXiv191205614F}
{Flesch}, E.~W. 2019, arXiv e-prints, arXiv:1912.05614

\bibitem[{{Gaia Collaboration et al.}(2022)}]{DR3-DPACP-133}
{Gaia Collaboration et al.} 2022, \aap\ in prep.

\bibitem[{{Garofalo} {et~al.}(2013){Garofalo}, {Cusano}, {Clementini},
  {Ripepi}, {Dall'Ora}, {Moretti}, {Coppola}, {Musella}, \&
  {Marconi}}]{2013ApJ...767...62G}
{Garofalo}, A., {Cusano}, F., {Clementini}, G., {et~al.} 2013, \apj, 767, 62

\bibitem[{{Gavras et al.}(2022)}]{DR3-DPACP-177}
{Gavras et al.} 2022, \aap\ in prep.

\bibitem[{{Gentile Fusillo} {et~al.}(2019){Gentile Fusillo}, {Tremblay},
  {G{\"a}nsicke}, {Manser}, {Cunningham}, {Cukanovaite}, {Hollands}, {Marsh},
  {Raddi}, {Jordan}, {Toonen}, {Geier}, {Barstow}, \&
  {Cummings}}]{2019MNRAS.482.4570G}
{Gentile Fusillo}, N.~P., {Tremblay}, P.-E., {G{\"a}nsicke}, B.~T., {et~al.}
  2019, \mnras, 482, 4570

\bibitem[{{Gianninas} {et~al.}(2005){Gianninas}, {Bergeron}, \&
  {Fontaine}}]{2005ApJ...631.1100G}
{Gianninas}, A., {Bergeron}, P., \& {Fontaine}, G. 2005, \apj, 631, 1100

\bibitem[{{Graham} {et~al.}(2019){Graham}, {Kulkarni}, {Bellm}, {Adams},
  {Barbarino}, {Blagorodnova}, {Bodewits}, {Bolin}, {Brady}, {Cenko}, {Chang},
  {Coughlin}, {De}, {Eadie}, {Farnham}, {Feindt}, {Franckowiak}, {Fremling},
  {Gezari}, {Ghosh}, {Goldstein}, {Golkhou}, {Goobar}, {Ho}, {Huppenkothen},
  {Ivezi{\'c}}, {Jones}, {Juric}, {Kaplan}, {Kasliwal}, {Kelley}, {Kupfer},
  {Lee}, {Lin}, {Lunnan}, {Mahabal}, {Miller}, {Ngeow}, {Nugent}, {Ofek},
  {Prince}, {Rauch}, {van Roestel}, {Schulze}, {Singer}, {Sollerman}, {Taddia},
  {Yan}, {Ye}, {Yu}, {Barlow}, {Bauer}, {Beck}, {Belicki}, {Biswas}, {Brinnel},
  {Brooke}, {Bue}, {Bulla}, {Burruss}, {Connolly}, {Cromer}, {Cunningham},
  {Dekany}, {Delacroix}, {Desai}, {Duev}, {Feeney}, {Flynn}, {Frederick},
  {Gal-Yam}, {Giomi}, {Groom}, {Hacopians}, {Hale}, {Helou}, {Henning},
  {Hover}, {Hillenbrand}, {Howell}, {Hung}, {Imel}, {Ip}, {Jackson}, {Kaspi},
  {Kaye}, {Kowalski}, {Kramer}, {Kuhn}, {Landry}, {Laher}, {Mao}, {Masci},
  {Monkewitz}, {Murphy}, {Nordin}, {Patterson}, {Penprase}, {Porter},
  {Rebbapragada}, {Reiley}, {Riddle}, {Rigault}, {Rodriguez}, {Rusholme}, {van
  Santen}, {Shupe}, {Smith}, {Soumagnac}, {Stein}, {Surace}, {Szkody}, {Terek},
  {Van Sistine}, {van Velzen}, {Vestrand}, {Walters}, {Ward}, {Zhang}, \&
  {Zolkower}}]{2019PASP..131g8001G}
{Graham}, M.~J., {Kulkarni}, S.~R., {Bellm}, E.~C., {et~al.} 2019, \pasp, 131,
  078001

\bibitem[{{Hamanowicz} {et~al.}(2016){Hamanowicz}, {Pietrukowicz}, {Udalski},
  {Mr{\'o}z}, {Soszy{\'n}ski}, {Szyma{\'n}ski}, {Skowron}, {Poleski},
  {Wyrzykowski}, {Koz{\l}owski}, {Pawlak}, \& {Ulaczyk}}]{2016AcA....66..197H}
{Hamanowicz}, A., {Pietrukowicz}, P., {Udalski}, A., {et~al.} 2016, \actaa, 66,
  197

\bibitem[{{Hartman} {et~al.}(2010){Hartman}, {Bakos}, {Kov{\'a}cs}, \&
  {Noyes}}]{2010MNRAS.408..475H}
{Hartman}, J.~D., {Bakos}, G.~{\'A}., {Kov{\'a}cs}, G., \& {Noyes}, R.~W. 2010,
  \mnras, 408, 475

\bibitem[{{Heber}(2016)}]{2016PASP..128h2001H}
{Heber}, U. 2016, \pasp, 128, 082001

\bibitem[{{Heinze} {et~al.}(2018){Heinze}, {Tonry}, {Denneau}, {Flewelling},
  {Stalder}, {Rest}, {Smith}, {Smartt}, \& {Weiland}}]{2018AJ....156..241H}
{Heinze}, A.~N., {Tonry}, J.~L., {Denneau}, L., {et~al.} 2018, \aj, 156, 241

\bibitem[{{Herbst} {et~al.}(1994){Herbst}, {Herbst}, {Grossman}, \&
  {Weinstein}}]{1994AJ....108.1906H}
{Herbst}, W., {Herbst}, D.~K., {Grossman}, E.~J., \& {Weinstein}, D. 1994, \aj,
  108, 1906

\bibitem[{{Herbst} \& {Shevchenko}(1999)}]{1999AJ....118.1043H}
{Herbst}, W. \& {Shevchenko}, V.~S. 1999, \aj, 118, 1043

\bibitem[{{Hermes} {et~al.}(2013{\natexlab{a}}){Hermes}, {Montgomery},
  {Gianninas}, {Winget}, {Brown}, {Harrold}, {Bell}, {Kenyon}, {Kilic}, \&
  {Castanheira}}]{2013MNRAS.436.3573H}
{Hermes}, J.~J., {Montgomery}, M.~H., {Gianninas}, A., {et~al.}
  2013{\natexlab{a}}, \mnras, 436, 3573

\bibitem[{{Hermes} {et~al.}(2013{\natexlab{b}}){Hermes}, {Montgomery},
  {Winget}, {Brown}, {Gianninas}, {Kilic}, {Kenyon}, {Bell}, \&
  {Harrold}}]{2013ApJ...765..102H}
{Hermes}, J.~J., {Montgomery}, M.~H., {Winget}, D.~E., {et~al.}
  2013{\natexlab{b}}, \apj, 765, 102

\bibitem[{{Hermes} {et~al.}(2012){Hermes}, {Montgomery}, {Winget}, {Brown},
  {Kilic}, \& {Kenyon}}]{2012ApJ...750L..28H}
{Hermes}, J.~J., {Montgomery}, M.~H., {Winget}, D.~E., {et~al.} 2012, \apjl,
  750, L28

\bibitem[{{Hey} {et~al.}(2019){Hey}, {Holdsworth}, {Bedding}, {Murphy},
  {Cunha}, {Kurtz}, {Huber}, {Fulton}, \& {Howard}}]{2019MNRAS.488...18H}
{Hey}, D.~R., {Holdsworth}, D.~L., {Bedding}, T.~R., {et~al.} 2019, \mnras,
  488, 18

\bibitem[{{Hoffman} {et~al.}(2009){Hoffman}, {Harrison}, \&
  {McNamara}}]{2009AJ....138..466H}
{Hoffman}, D.~I., {Harrison}, T.~E., \& {McNamara}, B.~J. 2009, \aj, 138, 466

\bibitem[{{Holl} {et~al.}(2018){Holl}, {Audard}, {Nienartowicz}, {Jevardat de
  Fombelle}, {Marchal}, {Mowlavi}, {Clementini}, {De Ridder}, {Evans}, {Guy},
  {Lanzafame}, {Lebzelter}, {Rimoldini}, {Roelens}, {Zucker}, {Distefano},
  {Garofalo}, {Lecoeur-Ta{\"\i}bi}, {Lopez}, {Molinaro}, {Muraveva}, {Panahi},
  {Regibo}, {Ripepi}, {Sarro}, {Aerts}, {Anderson}, {Charnas}, {Barblan},
  {Blanco-Cuaresma}, {Busso}, {Cuypers}, {De Angeli}, {Glass}, {Grenon},
  {Juh{\'a}sz}, {Kochoska}, {Koubsky}, {Lanza}, {Leccia}, {Lorenz}, {Marconi},
  {Marschalk{\'o}}, {Mazeh}, {Messina}, {Mignard}, {Moitinho}, {Moln{\'a}r},
  {Morgenthaler}, {Musella}, {Ordenovic}, {Ord{\'o}{\~n}ez}, {Pagano},
  {Palaversa}, {Pawlak}, {Plachy}, {Pr{\v{s}}a}, {Riello}, {S{\"u}veges},
  {Szabados}, {Szegedi-Elek}, {Votruba}, \& {Eyer}}]{2018A&A...618A..30H}
{Holl}, B., {Audard}, M., {Nienartowicz}, K., {et~al.} 2018, \aap, 618, A30

\bibitem[{{Holl} {et~al.}(2014){Holl}, {Mowlavi}, {Lecoeur-Ta{\"\i}bi},
  {Barblan}, {Rimoldini}, {Eyer}, {S{\"u}veges}, {Guy}, {Ordo{\~n}ez-Blanco},
  {Ruiz}, \& {Nienartowicz}}]{2014EAS....67..299H}
{Holl}, B., {Mowlavi}, N., {Lecoeur-Ta{\"\i}bi}, I., {et~al.} 2014, in EAS
  Publications Series, Vol. 67-68, EAS Publications Series, 299--303

\bibitem[{{Holl et al.}(2022)}]{DR3-DPACP-164}
{Holl et al.} 2022, \aap\ in prep.

\bibitem[{{Howell} {et~al.}(2016){Howell}, {Mason}, {Boyd}, {Smith}, \&
  {Gelino}}]{2016ApJ...831...27H}
{Howell}, S.~B., {Mason}, E., {Boyd}, P., {Smith}, K.~L., \& {Gelino}, D.~M.
  2016, \apj, 831, 27

\bibitem[{{Hubble}(1926)}]{1926ApJ....64..321H}
{Hubble}, E.~P. 1926, \apj, 64, 321

\bibitem[{{H{\"u}mmerich} {et~al.}(2018){H{\"u}mmerich}, {Mikul{\'a}{\v{s}}ek},
  {Paunzen}, {Bernhard}, {Jan{\'\i}k}, {Yakunin}, {Pribulla}, {Va{\v{n}}ko}, \&
  {Mat{\v{e}}chov{\'a}}}]{2018A&A...619A..98H}
{H{\"u}mmerich}, S., {Mikul{\'a}{\v{s}}ek}, Z., {Paunzen}, E., {et~al.} 2018,
  \aap, 619, A98

\bibitem[{{Ivezi{\'c}} {et~al.}(2007){Ivezi{\'c}}, {Smith}, {Miknaitis}, {Lin},
  {Tucker}, {Lupton}, {Gunn}, {Knapp}, {Strauss}, {Sesar}, {Doi}, {Tanaka},
  {Fukugita}, {Holtzman}, {Kent}, {Yanny}, {Schlegel}, {Finkbeiner},
  {Padmanabhan}, {Rockosi}, {Juri{\'c}}, {Bond}, {Lee}, {Stoughton}, {Jester},
  {Harris}, {Harding}, {Morrison}, {Brinkmann}, {Schneider}, \&
  {York}}]{2007AJ....134..973I}
{Ivezi{\'c}}, {\v{Z}}., {Smith}, J.~A., {Miknaitis}, G., {et~al.} 2007, \aj,
  134, 973

\bibitem[{{Jayasinghe} {et~al.}(2018){Jayasinghe}, {Kochanek}, {Stanek},
  {Shappee}, {Holoien}, {Thompson}, {Prieto}, {Dong}, {Pawlak}, {Shields},
  {Pojmanski}, {Otero}, {Britt}, \& {Will}}]{2018MNRAS.477.3145J}
{Jayasinghe}, T., {Kochanek}, C.~S., {Stanek}, K.~Z., {et~al.} 2018, \mnras,
  477, 3145

\bibitem[{{Jayasinghe} {et~al.}(2019{\natexlab{a}}){Jayasinghe}, {Stanek},
  {Kochanek}, {Shappee}, {Holoien}, {Thompson}, {Prieto}, {Dong}, {Pawlak},
  {Pejcha}, {Shields}, {Pojmanski}, {Otero}, {Britt}, \&
  {Will}}]{2019MNRAS.486.1907J}
{Jayasinghe}, T., {Stanek}, K.~Z., {Kochanek}, C.~S., {et~al.}
  2019{\natexlab{a}}, \mnras, 486, 1907

\bibitem[{{Jayasinghe} {et~al.}(2019{\natexlab{b}}){Jayasinghe}, {Stanek},
  {Kochanek}, {Shappee}, {Holoien}, {Thompson}, {Prieto}, {Dong}, {Pawlak},
  {Pejcha}, {Shields}, {Pojmanski}, {Otero}, {Hurst}, {Britt}, \&
  {Will}}]{2019MNRAS.485..961J}
{Jayasinghe}, T., {Stanek}, K.~Z., {Kochanek}, C.~S., {et~al.}
  2019{\natexlab{b}}, \mnras, 485, 961

\bibitem[{{Jiang} {et~al.}(2012){Jiang}, {Han}, {Ge}, {Yang}, \&
  {Li}}]{2012MNRAS.421.2769J}
{Jiang}, D., {Han}, Z., {Ge}, H., {Yang}, L., \& {Li}, L. 2012, \mnras, 421,
  2769

\bibitem[{{Kabath} {et~al.}(2009){Kabath}, {Erikson}, {Rauer}, {Pasternacki},
  {Csizmadia}, {Chini}, {Lemke}, {Murphy}, {Fruth}, {Titz}, \&
  {Eigm{\"u}ller}}]{2009A&A...506..569K}
{Kabath}, P., {Erikson}, A., {Rauer}, H., {et~al.} 2009, \aap, 506, 569

\bibitem[{{Kahraman Ali{\c{c}}avu{\textcommabelow s}} {et~al.}(2016){Kahraman
  Ali{\c{c}}avu{\textcommabelow s}}, {Niemczura}, {De Cat}, {Soydugan},
  {Ko{\l}aczkowski}, {Ostrowski}, {Telting}, {Uytterhoeven}, {Poretti},
  {Rainer}, {Su{\'a}rez}, {Mantegazza}, {Kilmartin}, \&
  {Pollard}}]{2016MNRAS.458.2307K}
{Kahraman Ali{\c{c}}avu{\textcommabelow s}}, F., {Niemczura}, E., {De Cat}, P.,
  {et~al.} 2016, \mnras, 458, 2307

\bibitem[{{Kepler} {et~al.}(2014){Kepler}, {Fraga}, {Winget}, {Bell},
  {C{\'o}rsico}, \& {Werner}}]{2014MNRAS.442.2278K}
{Kepler}, S.~O., {Fraga}, L., {Winget}, D.~E., {et~al.} 2014, \mnras, 442, 2278

\bibitem[{{Kim} {et~al.}(2014){Kim}, {Protopapas}, {Bailer-Jones}, {Byun},
  {Chang}, {Marquette}, \& {Shin}}]{2014A&A...566A..43K}
{Kim}, D.-W., {Protopapas}, P., {Bailer-Jones}, C. A.~L., {et~al.} 2014, \aap,
  566, A43

\bibitem[{{Kinemuchi} {et~al.}(2006){Kinemuchi}, {Smith}, {Wo{\'z}niak},
  {McKay}, \& {ROTSE Collaboration}}]{2006AJ....132.1202K}
{Kinemuchi}, K., {Smith}, H.~A., {Wo{\'z}niak}, P.~R., {McKay}, T.~A., \&
  {ROTSE Collaboration}. 2006, \aj, 132, 1202

\bibitem[{{Kirk} {et~al.}(2016){Kirk}, {Conroy}, {Pr{\v{s}}a}, {Abdul-Masih},
  {Kochoska}, {Matijevi{\v{c}}}, {Hambleton}, {Barclay}, {Bloemen}, {Boyajian},
  {Doyle}, {Fulton}, {Hoekstra}, {Jek}, {Kane}, {Kostov}, {Latham}, {Mazeh},
  {Orosz}, {Pepper}, {Quarles}, {Ragozzine}, {Shporer}, {Southworth},
  {Stassun}, {Thompson}, {Welsh}, {Agol}, {Derekas}, {Devor}, {Fischer},
  {Green}, {Gropp}, {Jacobs}, {Johnston}, {LaCourse}, {Saetre}, {Schwengeler},
  {Toczyski}, {Werner}, {Garrett}, {Gore}, {Martinez}, {Spitzer}, {Stevick},
  {Thomadis}, {Vrijmoet}, {Yenawine}, {Batalha}, \&
  {Borucki}}]{2016AJ....151...68K}
{Kirk}, B., {Conroy}, K., {Pr{\v{s}}a}, A., {et~al.} 2016, \aj, 151, 68

\bibitem[{{Koester} \& {Kepler}(2019)}]{2019A&A...628A.102K}
{Koester}, D. \& {Kepler}, S.~O. 2019, \aap, 628, A102

\bibitem[{{Koposov} \& {Bartunov}(2006)}]{2006ASPC..351..735K}
{Koposov}, S. \& {Bartunov}, O. 2006, in Astronomical Society of the Pacific
  Conference Series, Vol. 351, Astronomical Data Analysis Software and Systems
  XV, ed. C.~{Gabriel}, C.~{Arviset}, D.~{Ponz}, \& S.~{Enrique}, 735

\bibitem[{{Krone-Martins} {et~al.}(2022){Krone-Martins}, {Gavras}, {Ducourant},
  {Galluccio}, {Teixeira}, {le Campion}, {Petit}, {Guiraud}, \&
  {Managau}}]{floisvos}
{Krone-Martins}, A., {Gavras}, P., {Ducourant}, C., {et~al.} 2022, \aap\ in
  prep.

\bibitem[{{Kunkel} {et~al.}(1997){Kunkel}, {Irwin}, \&
  {Demers}}]{1997A&AS..122..463K}
{Kunkel}, W.~E., {Irwin}, M.~J., \& {Demers}, S. 1997, \aaps, 122, 463

\bibitem[{{Kurtz} {et~al.}(2013){Kurtz}, {Shibahashi}, {Dhillon}, {Marsh},
  {Littlefair}, {Copperwheat}, {G{\"a}nsicke}, \&
  {Parsons}}]{2013MNRAS.432.1632K}
{Kurtz}, D.~W., {Shibahashi}, H., {Dhillon}, V.~S., {et~al.} 2013, \mnras, 432,
  1632

\bibitem[{{Lavail} {et~al.}(2017){Lavail}, {Kochukhov}, {Hussain}, {Alecian},
  {Herczeg}, \& {Johns-Krull}}]{2017A&A...608A..77L}
{Lavail}, A., {Kochukhov}, O., {Hussain}, G.~A.~J., {et~al.} 2017, \aap, 608,
  A77

\bibitem[{{Liu} {et~al.}(2021){Liu}, {Buchner}, {Nandra}, {Merloni}, {Dwelly},
  {Sanders}, {Salvato}, {Arcodia}, {Brusa}, {Wolf}, {Georgakakis}, {Boller},
  {Krumpe}, {Lamer}, {Waddell}, {Urrutia}, {Schwope}, {Robrade}, {Wilms},
  {Dauser}, {Comparat}, {Toba}, {Ichikawa}, {Iwasawa}, {Shen}, \& {Ibarra
  Medel}}]{2021arXiv210614522L}
{Liu}, T., {Buchner}, J., {Nandra}, K., {et~al.} 2021, arXiv e-prints,
  arXiv:2106.14522

\bibitem[{{Lomb}(1976)}]{1976Ap&SS..39..447L}
{Lomb}, N.~R. 1976, \apss, 39, 447

\bibitem[{{Ma} {et~al.}(2013){Ma}, {Arias}, {Bianco}, {Boboltz}, {Bolotin},
  {Charlot}, {Engelhardt}, {Fey}, {Gaume}, {Gontier}, {Heinkelmann}, {Jacobs},
  {Kurdubov}, {Lambert}, {Malkin}, {Nothnagel}, {Petrov}, {Skurikhina},
  {Sokolova}, {Souchay}, {Sovers}, {Tesmer}, {Titov}, {Wang}, {Zharov},
  {Barache}, {Bockmann}, {Collioud}, {Gipson}, {Gordon}, {Lytvyn}, {MacMillan},
  {Ojha}, {Fey}, {Gordon}, \& {Jacobs}}]{2013yCat.1323....0M}
{Ma}, C., {Arias}, F.~E., {Bianco}, G., {et~al.} 2013, VizieR Online Data
  Catalog, I/323

\bibitem[{{Marquette} {et~al.}(2009){Marquette}, {Beaulieu}, {Buchler},
  {Szab{\'o}}, {Tisserand}, {Belghith}, {Fouqu{\'e}}, {Lesquoy}, {Milsztajn},
  {Schwarzenberg-Czerny}, {Afonso}, {Albert}, {Andersen}, {Ansari}, {Aubourg},
  {Bareyre}, {Charlot}, {Coutures}, {Ferlet}, {Glicenstein}, {Goldman},
  {Gould}, {Graff}, {Gros}, {Ha{\"\i}ssinski}, {Hamadache}, {de Kat}, {Le
  Guillou}, {Loup}, {Magneville}, {Maurice}, {Maury}, {Moniez},
  {Palanque-Delabrouille}, {Perdereau}, {Rahal}, {Rich}, {Spiro}, \&
  {Vidal-Madjar}}]{2009A&A...495..249M}
{Marquette}, J.~B., {Beaulieu}, J.~P., {Buchler}, J.~R., {et~al.} 2009, \aap,
  495, 249

\bibitem[{{Marrese} {et~al.}(2019){Marrese}, {Marinoni}, {Fabrizio}, \&
  {Altavilla}}]{2019A&A...621A.144M}
{Marrese}, P.~M., {Marinoni}, S., {Fabrizio}, M., \& {Altavilla}, G. 2019,
  \aap, 621, A144

\bibitem[{{Mart{\'\i}nez-Arn{\'a}iz} {et~al.}(2010){Mart{\'\i}nez-Arn{\'a}iz},
  {Maldonado}, {Montes}, {Eiroa}, \& {Montesinos}}]{2010A&A...520A..79M}
{Mart{\'\i}nez-Arn{\'a}iz}, R., {Maldonado}, J., {Montes}, D., {Eiroa}, C., \&
  {Montesinos}, B. 2010, \aap, 520, A79

\bibitem[{{Masci} {et~al.}(2019){Masci}, {Laher}, {Rusholme}, {Shupe}, {Groom},
  {Surace}, {Jackson}, {Monkewitz}, {Beck}, {Flynn}, {Terek}, {Landry},
  {Hacopians}, {Desai}, {Howell}, {Brooke}, {Imel}, {Wachter}, {Ye}, {Lin},
  {Cenko}, {Cunningham}, {Rebbapragada}, {Bue}, {Miller}, {Mahabal}, {Bellm},
  {Patterson}, {Juri{\'c}}, {Golkhou}, {Ofek}, {Walters}, {Graham}, {Kasliwal},
  {Dekany}, {Kupfer}, {Burdge}, {Cannella}, {Barlow}, {Van Sistine}, {Giomi},
  {Fremling}, {Blagorodnova}, {Levitan}, {Riddle}, {Smith}, {Helou}, {Prince},
  \& {Kulkarni}}]{2019PASP..131a8003M}
{Masci}, F.~J., {Laher}, R.~R., {Rusholme}, B., {et~al.} 2019, \pasp, 131,
  018003

\bibitem[{{Massaro} {et~al.}(2015){Massaro}, {Maselli}, {Leto}, {Marchegiani},
  {Perri}, {Giommi}, \& {Piranomonte}}]{2015Ap&SS.357...75M}
{Massaro}, E., {Maselli}, A., {Leto}, C., {et~al.} 2015, \apss, 357, 75

\bibitem[{{Mauron} {et~al.}(2019){Mauron}, {Maurin}, \&
  {Kendall}}]{2019A&A...626A.112M}
{Mauron}, N., {Maurin}, L.~P.~A., \& {Kendall}, T.~R. 2019, \aap, 626, A112

\bibitem[{{Medhi} {et~al.}(2007){Medhi}, {Messina}, {Parihar}, {Pagano},
  {Muneer}, \& {Duorah}}]{2007A&A...469..713M}
{Medhi}, B.~J., {Messina}, S., {Parihar}, P.~S., {et~al.} 2007, \aap, 469, 713

\bibitem[{{Mennickent} {et~al.}(2002){Mennickent}, {Pietrzy{\'n}ski}, {Gieren},
  \& {Szewczyk}}]{2002A&A...393..887M}
{Mennickent}, R.~E., {Pietrzy{\'n}ski}, G., {Gieren}, W., \& {Szewczyk}, O.
  2002, \aap, 393, 887

\bibitem[{{Messina} {et~al.}(2011){Messina}, {Desidera}, {Lanzafame},
  {Turatto}, \& {Guinan}}]{2011A&A...532A..10M}
{Messina}, S., {Desidera}, S., {Lanzafame}, A.~C., {Turatto}, M., \& {Guinan},
  E.~F. 2011, \aap, 532, A10

\bibitem[{{Messina} {et~al.}(2010){Messina}, {Desidera}, {Turatto},
  {Lanzafame}, \& {Guinan}}]{2010A&A...520A..15M}
{Messina}, S., {Desidera}, S., {Turatto}, M., {Lanzafame}, A.~C., \& {Guinan},
  E.~F. 2010, \aap, 520, A15

\bibitem[{{Mould} {et~al.}(2004){Mould}, {Saha}, \&
  {Hughes}}]{2004ApJS..154..623M}
{Mould}, J., {Saha}, A., \& {Hughes}, S. 2004, \apjs, 154, 623

\bibitem[{{Mowlavi} {et~al.}(2018){Mowlavi}, {Lecoeur-Ta{\"\i}bi}, {Lebzelter},
  {Rimoldini}, {Lorenz}, {Audard}, {De Ridder}, {Eyer}, {Guy}, {Holl},
  {Jevardat de Fombelle}, {Marchal}, {Nienartowicz}, {Regibo}, {Roelens}, \&
  {Sarro}}]{2018A&A...618A..58M}
{Mowlavi}, N., {Lecoeur-Ta{\"\i}bi}, I., {Lebzelter}, T., {et~al.} 2018, \aap,
  618, A58

\bibitem[{{Mowlavi} {et~al.}(2021){Mowlavi}, {Rimoldini}, {Evans}, {Riello},
  {De Angeli}, {Palaversa}, {Audard}, {Eyer}, {Garcia-Lario}, {Gavras}, {Holl},
  {Jevardat de Fombelle}, {Lec{\oe}ur-Ta{\"\i}bi}, \&
  {Nienartowicz}}]{2021A&A...648A..44M}
{Mowlavi}, N., {Rimoldini}, L., {Evans}, D.~W., {et~al.} 2021, \aap, 648, A44

\bibitem[{{Mr{\'o}z} {et~al.}(2015){Mr{\'o}z}, {Udalski}, {Poleski},
  {Pietrukowicz}, {Szyma{\'n}ski}, {Soszy{\'n}ski}, {Wyrzykowski}, {Ulaczyk},
  {Koz{\l}owski}, \& {Skowron}}]{2015AcA....65..313M}
{Mr{\'o}z}, P., {Udalski}, A., {Poleski}, R., {et~al.} 2015, \actaa, 65, 313

\bibitem[{{Mr{\'o}z} {et~al.}(2019){Mr{\'o}z}, {Udalski}, {Skowron},
  {Szyma{\'n}ski}, {Soszy{\'n}ski}, {Wyrzykowski}, {Pietrukowicz},
  {Koz{\l}owski}, {Poleski}, {Ulaczyk}, {Rybicki}, \&
  {Iwanek}}]{2019ApJS..244...29M}
{Mr{\'o}z}, P., {Udalski}, A., {Skowron}, J., {et~al.} 2019, \apjs, 244, 29

\bibitem[{{Musella} {et~al.}(2009){Musella}, {Ripepi}, {Clementini},
  {Dall'Ora}, {Kinemuchi}, {di Fabrizio}, {Greco}, {Marconi}, {Smith},
  {Radovich}, \& {Beers}}]{2009ApJ...695L..83M}
{Musella}, I., {Ripepi}, V., {Clementini}, G., {et~al.} 2009, \apjl, 695, L83

\bibitem[{{Musella} {et~al.}(2012){Musella}, {Ripepi}, {Marconi}, {Clementini},
  {Dall'Ora}, {Scowcroft}, {Moretti}, {Di Fabrizio}, {Greco}, {Coppola},
  {Bersier}, {Catelan}, {Grado}, {Limatola}, {Smith}, \&
  {Kinemuchi}}]{2012ApJ...756..121M}
{Musella}, I., {Ripepi}, V., {Marconi}, M., {et~al.} 2012, \apj, 756, 121

\bibitem[{{Niemczura}(2003)}]{2003A&A...404..689N}
{Niemczura}, E. 2003, \aap, 404, 689

\bibitem[{{Nitta} {et~al.}(2009){Nitta}, {Kleinman}, {Krzesinski}, {Kepler},
  {Metcalfe}, {Mukadam}, {Mullally}, {Nather}, {Sullivan}, {Thompson}, \&
  {Winget}}]{2009ApJ...690..560N}
{Nitta}, A., {Kleinman}, S.~J., {Krzesinski}, J., {et~al.} 2009, \apj, 690, 560

\bibitem[{{Palaversa} {et~al.}(2013){Palaversa}, {Ivezi{\'c}}, {Eyer},
  {Ru{\v{z}}djak}, {Sudar}, {Galin}, {Kroflin}, {Mesari{\'c}}, {Munk},
  {Vrbanec}, {Bo{\v{z}}i{\'c}}, {Loebman}, {Sesar}, {Rimoldini}, {Hunt-Walker},
  {VanderPlas}, {Westman}, {Stuart}, {Becker}, {Srdo{\v{c}}}, {Wozniak}, \&
  {Oluseyi}}]{2013AJ....146..101P}
{Palaversa}, L., {Ivezi{\'c}}, {\v{Z}}., {Eyer}, L., {et~al.} 2013, \aj, 146,
  101

\bibitem[{{Pawlak} {et~al.}(2013){Pawlak}, {Graczyk}, {Soszy{\'n}ski},
  {Pietrukowicz}, {Poleski}, {Udalski}, {Szyma{\'n}ski}, {Kubiak},
  {Pietrzy{\'n}ski}, {Wyrzykowski}, {Ulaczyk}, {Koz{\l}owski}, \&
  {Skowron}}]{2013AcA....63..323P}
{Pawlak}, M., {Graczyk}, D., {Soszy{\'n}ski}, I., {et~al.} 2013, \actaa, 63,
  323

\bibitem[{{Pawlak} {et~al.}(2016){Pawlak}, {Soszy{\'n}ski}, {Udalski},
  {Szyma{\'n}ski}, {Wyrzykowski}, {Ulaczyk}, {Poleski}, {Pietrukowicz},
  {Koz{\l}owski}, {Skowron}, {Skowron}, {Mr{\'o}z}, \&
  {Hamanowicz}}]{2016AcA....66..421P}
{Pawlak}, M., {Soszy{\'n}ski}, I., {Udalski}, A., {et~al.} 2016, \actaa, 66,
  421

\bibitem[{{Pellerin} \& {Macri}(2011)}]{2011ApJS..193...26P}
{Pellerin}, A. \& {Macri}, L.~M. 2011, \apjs, 193, 26

\bibitem[{{Pietrukowicz} {et~al.}(2017){Pietrukowicz}, {Dziembowski}, {Latour},
  {Angeloni}, {Poleski}, {di Mille}, {Soszy{\'n}ski}, {Udalski},
  {Szyma{\'n}ski}, {Wyrzykowski}, {Koz{\l}owski}, {Skowron}, {Skowron},
  {Mr{\'o}z}, {Pawlak}, \& {Ulaczyk}}]{2017NatAs...1E.166P}
{Pietrukowicz}, P., {Dziembowski}, W.~A., {Latour}, M., {et~al.} 2017, Nature
  Astronomy, 1, 0166

\bibitem[{{Pigulski} {et~al.}(2009){Pigulski}, {Pojma{\'n}ski}, {Pilecki}, \&
  {Szczygie{\l}}}]{2009AcA....59...33P}
{Pigulski}, A., {Pojma{\'n}ski}, G., {Pilecki}, B., \& {Szczygie{\l}}, D.~M.
  2009, \actaa, 59, 33

\bibitem[{{Pojmanski}(2002)}]{2002AcA....52..397P}
{Pojmanski}, G. 2002, \actaa, 52, 397

\bibitem[{{Poleski} {et~al.}(2010{\natexlab{a}}){Poleski}, {Soszy{\'n}ski},
  {Udalski}, {Szyma{\'n}ski}, {Kubiak}, {Pietrzy{\'n}ski}, {Wyrzykowski},
  {Szewczyk}, \& {Ulaczyk}}]{2010AcA....60....1P}
{Poleski}, R., {Soszy{\'n}ski}, I., {Udalski}, A., {et~al.} 2010{\natexlab{a}},
  \actaa, 60, 1

\bibitem[{{Poleski} {et~al.}(2010{\natexlab{b}}){Poleski}, {Soszy{\'n}ski},
  {Udalski}, {Szyma{\'n}ski}, {Kubiak}, {Pietrzy{\'n}ski}, {Wyrzykowski}, \&
  {Ulaczyk}}]{2010AcA....60..179P}
{Poleski}, R., {Soszy{\'n}ski}, I., {Udalski}, A., {et~al.} 2010{\natexlab{b}},
  \actaa, 60, 179

\bibitem[{{Popper}(1967)}]{1967ARA&A...5...85P}
{Popper}, D.~M. 1967, \araa, 5, 85

\bibitem[{{Pourbaix} {et~al.}(2004){Pourbaix}, {Tokovinin}, {Batten}, {Fekel},
  {Hartkopf}, {Levato}, {Morrell}, {Torres}, \& {Udry}}]{2004A&A...424..727P}
{Pourbaix}, D., {Tokovinin}, A.~A., {Batten}, A.~H., {et~al.} 2004, \aap, 424,
  727

\bibitem[{{Pritzl} {et~al.}(2002){Pritzl}, {Smith}, {Catelan}, \&
  {Sweigart}}]{2002AJ....124..949P}
{Pritzl}, B.~J., {Smith}, H.~A., {Catelan}, M., \& {Sweigart}, A.~V. 2002, \aj,
  124, 949

\bibitem[{{Pritzl} {et~al.}(2003){Pritzl}, {Smith}, {Stetson}, {Catelan},
  {Sweigart}, {Layden}, \& {Rich}}]{2003AJ....126.1381P}
{Pritzl}, B.~J., {Smith}, H.~A., {Stetson}, P.~B., {et~al.} 2003, \aj, 126,
  1381

\bibitem[{{Quirion} {et~al.}(2007){Quirion}, {Fontaine}, \&
  {Brassard}}]{2007ApJS..171..219Q}
{Quirion}, P.~O., {Fontaine}, G., \& {Brassard}, P. 2007, \apjs, 171, 219

\bibitem[{{Reinhold} \& {Gizon}(2015)}]{2015A&A...583A..65R}
{Reinhold}, T. \& {Gizon}, L. 2015, \aap, 583, A65

\bibitem[{{Renson} \& {Manfroid}(2009)}]{2009A&A...498..961R}
{Renson}, P. \& {Manfroid}, J. 2009, \aap, 498, 961

\bibitem[{{Richards} {et~al.}(2012){Richards}, {Starr}, {Miller}, {Bloom},
  {Butler}, {Brink}, \& {Crellin-Quick}}]{2012ApJS..203...32R}
{Richards}, J.~W., {Starr}, D.~L., {Miller}, A.~A., {et~al.} 2012, \apjs, 203,
  32

\bibitem[{{Ricker} {et~al.}(2015){Ricker}, {Winn}, {Vanderspek}, {Latham},
  {Bakos}, {Bean}, {Berta-Thompson}, {Brown}, {Buchhave}, {Butler}, {Butler},
  {Chaplin}, {Charbonneau}, {Christensen-Dalsgaard}, {Clampin}, {Deming},
  {Doty}, {De Lee}, {Dressing}, {Dunham}, {Endl}, {Fressin}, {Ge}, {Henning},
  {Holman}, {Howard}, {Ida}, {Jenkins}, {Jernigan}, {Johnson}, {Kaltenegger},
  {Kawai}, {Kjeldsen}, {Laughlin}, {Levine}, {Lin}, {Lissauer}, {MacQueen},
  {Marcy}, {McCullough}, {Morton}, {Narita}, {Paegert}, {Palle}, {Pepe},
  {Pepper}, {Quirrenbach}, {Rinehart}, {Sasselov}, {Sato}, {Seager},
  {Sozzetti}, {Stassun}, {Sullivan}, {Szentgyorgyi}, {Torres}, {Udry}, \&
  {Villasenor}}]{2015JATIS...1a4003R}
{Ricker}, G.~R., {Winn}, J.~N., {Vanderspek}, R., {et~al.} 2015, Journal of
  Astronomical Telescopes, Instruments, and Systems, 1, 014003

\bibitem[{{Rimoldini} {et~al.}(2019{\natexlab{a}}){Rimoldini}, {Holl},
  {Audard}, {Mowlavi}, {Nienartowicz}, {Evans}, {Guy}, {Lecoeur-Ta{\"\i}bi},
  {Jevardat de Fombelle}, {Marchal}, {Roelens}, {De Ridder}, {Sarro}, {Regibo},
  {Lopez}, {Clementini}, {Ripepi}, {Molinaro}, {Garofalo}, {Moln{\'a}r},
  {Plachy}, {Juh{\'a}sz}, {Szabados}, {Lebzelter}, {Teyssier}, \&
  {Eyer}}]{2019A&A...625A..97R}
{Rimoldini}, L., {Holl}, B., {Audard}, M., {et~al.} 2019{\natexlab{a}}, \aap,
  625, A97

\bibitem[{{Rimoldini} {et~al.}(2019{\natexlab{b}}){Rimoldini}, {Nienartowicz},
  {S{\"u}veges}, {Charnas}, {Guy}, {Jevardat de Fombelle}, {Holl},
  {Lecoeur-Ta{\"\i}bi}, {Mowlavi}, {Ord{\'o}{\~n}ez-Blanco}, \&
  {Eyer}}]{2019ASPC..521..307R}
{Rimoldini}, L., {Nienartowicz}, K., {S{\"u}veges}, M., {et~al.}
  2019{\natexlab{b}}, in Astronomical Society of the Pacific Conference Series,
  Vol. 521, Astronomical Data Analysis Software and Systems XXVI, ed.
  M.~{Molinaro}, K.~{Shortridge}, \& F.~{Pasian}, 307

\bibitem[{{Rimoldini et al.}(2022)}]{DR3-DPACP-165}
{Rimoldini et al.} 2022, \aap\ in prep.

\bibitem[{{Ripepi} {et~al.}(2019){Ripepi}, {Molinaro}, {Musella}, {Marconi},
  {Leccia}, \& {Eyer}}]{2019A&A...625A..14R}
{Ripepi}, V., {Molinaro}, R., {Musella}, I., {et~al.} 2019, \aap, 625, A14

\bibitem[{{Ritter} \& {Kolb}(2003)}]{2003A&A...404..301R}
{Ritter}, H. \& {Kolb}, U. 2003, \aap, 404, 301

\bibitem[{{Romero} {et~al.}(2019){Romero}, {Amaral}, {Klippel}, {Sanmartim},
  {Fraga}, {Ourique}, {Pelisoli}, {Lauffer}, {Kepler}, \&
  {Koester}}]{2019MNRAS.490.1803R}
{Romero}, A.~D., {Amaral}, L.~A., {Klippel}, T., {et~al.} 2019, \mnras, 490,
  1803

\bibitem[{{Rowan} {et~al.}(2019){Rowan}, {Tucker}, {Shappee}, \&
  {Hermes}}]{2019MNRAS.486.4574R}
{Rowan}, D.~M., {Tucker}, M.~A., {Shappee}, B.~J., \& {Hermes}, J.~J. 2019,
  \mnras, 486, 4574

\bibitem[{{Ruf}(2019)}]{lomb}
{Ruf}, T. 2019, {lomb: Lomb-Scargle Periodogram}, r package version 3.5.2

\bibitem[{{Sabogal} {et~al.}(2008){Sabogal}, {Mennickent}, {Pietrzy{\'n}ski},
  {Garc{\'\i}a}, {Gieren}, \& {Kolaczkowski}}]{2008A&A...478..659S}
{Sabogal}, B.~E., {Mennickent}, R.~E., {Pietrzy{\'n}ski}, G., {et~al.} 2008,
  \aap, 478, 659

\bibitem[{{Sabogal} {et~al.}(2005){Sabogal}, {Mennickent}, {Pietrzy{\'n}ski},
  \& {Gieren}}]{2005MNRAS.361.1055S}
{Sabogal}, B.~E., {Mennickent}, R.~E., {Pietrzy{\'n}ski}, G., \& {Gieren}, W.
  2005, \mnras, 361, 1055

\bibitem[{{Salvato} {et~al.}(2021){Salvato}, {Wolf}, {Dwelly}, {Georgakakis},
  {Brusa}, {Merloni}, {Liu}, {Toba}, {Nandra}, {Lamer}, {Buchner}, {Schneider},
  {Freund}, {Rau}, {Schwope}, {Nishizawa}, {Klein}, {Arcodia}, {Comparat},
  {Musiimenta}, {Nagao}, {Brunner}, {Malyali}, {Finoguenov}, {Anderson},
  {Shen}, {Ibarra-Mendel}, {Trump}, {Brandt}, {Urry}, {Rivera}, {Krumpe},
  {Urrutia}, {Miyaji}, {Ichikawa}, {Schneider}, {Fresco}, {Wilms}, {Boller},
  {Haase}, {Brownstein}, {Lane}, {Bizyaev}, \&
  {Nitschelm}}]{2021arXiv210614520S}
{Salvato}, M., {Wolf}, J., {Dwelly}, T., {et~al.} 2021, arXiv e-prints,
  arXiv:2106.14520

\bibitem[{{Samus'} {et~al.}(2017){Samus'}, {Kazarovets}, {Durlevich},
  {Kireeva}, \& {Pastukhova}}]{2017ARep...61...80S}
{Samus'}, N.~N., {Kazarovets}, E.~V., {Durlevich}, O.~V., {Kireeva}, N.~N., \&
  {Pastukhova}, E.~N. 2017, Astronomy Reports, 61, 80

\bibitem[{{Sarro} {et~al.}(2013){Sarro}, {Debosscher}, {Neiner},
  {Bello-Garc{\'\i}a}, {Gonz{\'a}lez-Marcos}, {Prendes-Gero}, {Ordieres},
  {Le{\'o}n}, {Aerts}, \& {de Batz}}]{2013A&A...550A.120S}
{Sarro}, L.~M., {Debosscher}, J., {Neiner}, C., {et~al.} 2013, \aap, 550, A120

\bibitem[{{Scargle}(1982)}]{1982ApJ...263..835S}
{Scargle}, J.~D. 1982, \apj, 263, 835

\bibitem[{{Sesar} {et~al.}(2014){Sesar}, {Banholzer}, {Cohen}, {Martin},
  {Grillmair}, {Levitan}, {Laher}, {Ofek}, {Surace}, {Kulkarni}, {Prince}, \&
  {Rix}}]{2014ApJ...793..135S}
{Sesar}, B., {Banholzer}, S.~R., {Cohen}, J.~G., {et~al.} 2014, \apj, 793, 135

\bibitem[{{Sesar} {et~al.}(2017){Sesar}, {Hernitschek}, {Mitrovi{\'c}},
  {Ivezi{\'c}}, {Rix}, {Cohen}, {Bernard}, {Grebel}, {Martin}, {Schlafly},
  {Burgett}, {Draper}, {Flewelling}, {Kaiser}, {Kudritzki}, {Magnier},
  {Metcalfe}, {Tonry}, \& {Waters}}]{2017AJ....153..204S}
{Sesar}, B., {Hernitschek}, N., {Mitrovi{\'c}}, S., {et~al.} 2017, \aj, 153,
  204

\bibitem[{{Shappee} {et~al.}(2014){Shappee}, {Prieto}, {Grupe}, {Kochanek},
  {Stanek}, {De Rosa}, {Mathur}, {Zu}, {Peterson}, {Pogge}, {Komossa}, {Im},
  {Jencson}, {Holoien}, {Basu}, {Beacom}, {Szczygie{\l}}, {Brimacombe},
  {Adams}, {Campillay}, {Choi}, {Contreras}, {Dietrich}, {Dubberley},
  {Elphick}, {Foale}, {Giustini}, {Gonzalez}, {Hawkins}, {Howell}, {Hsiao},
  {Koss}, {Leighly}, {Morrell}, {Mudd}, {Mullins}, {Nugent}, {Parrent},
  {Phillips}, {Pojmanski}, {Rosing}, {Ross}, {Sand}, {Terndrup}, {Valenti},
  {Walker}, \& {Yoon}}]{2014ApJ...788...48S}
{Shappee}, B.~J., {Prieto}, J.~L., {Grupe}, D., {et~al.} 2014, \apj, 788, 48

\bibitem[{{Shibayama} {et~al.}(2013){Shibayama}, {Maehara}, {Notsu}, {Notsu},
  {Nagao}, {Honda}, {Ishii}, {Nogami}, \& {Shibata}}]{2013ApJS..209....5S}
{Shibayama}, T., {Maehara}, H., {Notsu}, S., {et~al.} 2013, \apjs, 209, 5

\bibitem[{{Siegel}(2006)}]{2006ApJ...649L..83S}
{Siegel}, M.~H. 2006, \apjl, 649, L83

\bibitem[{{Sikora} {et~al.}(2019){Sikora}, {David-Uraz}, {Chowdhury}, {Bowman},
  {Wade}, {Khalack}, {Kobzar}, {Kochukhov}, {Neiner}, \&
  {Paunzen}}]{2019MNRAS.487.4695S}
{Sikora}, J., {David-Uraz}, A., {Chowdhury}, S., {et~al.} 2019, \mnras, 487,
  4695

\bibitem[{{Skottfelt} {et~al.}(2015){Skottfelt}, {Bramich}, {Figuera Jaimes},
  {J{\o}rgensen}, {Kains}, {Arellano Ferro}, {Alsubai}, {Bozza}, {Calchi
  Novati}, {Ciceri}, {D'Ago}, {Dominik}, {Galianni}, {Gu}, {Harps{\o}e},
  {Haugb{\o}lle}, {Hinse}, {Hundertmark}, {Juncher}, {Korhonen}, {Liebig},
  {Mancini}, {Popovas}, {Rabus}, {Rahvar}, {Scarpetta}, {Schmidt}, {Snodgrass},
  {Southworth}, {Starkey}, {Street}, {Surdej}, {Wang}, {Wertz}, \& {Mindstep
  Consortium}}]{2015A&A...573A.103S}
{Skottfelt}, J., {Bramich}, D.~M., {Figuera Jaimes}, R., {et~al.} 2015, \aap,
  573, A103

\bibitem[{{Slawson} {et~al.}(2011){Slawson}, {Pr{\v{s}}a}, {Welsh}, {Orosz},
  {Rucker}, {Batalha}, {Doyle}, {Engle}, {Conroy}, {Coughlin}, {Gregg},
  {Fetherolf}, {Short}, {Windmiller}, {Fabrycky}, {Howell}, {Jenkins}, {Uddin},
  {Mullally}, {Seader}, {Thompson}, {Sand erfer}, {Borucki}, \&
  {Koch}}]{2011AJ....142..160S}
{Slawson}, R.~W., {Pr{\v{s}}a}, A., {Welsh}, W.~F., {et~al.} 2011, \aj, 142,
  160

\bibitem[{{Soszy{\'n}ski}(2007)}]{2007ApJ...660.1486S}
{Soszy{\'n}ski}, I. 2007, \apj, 660, 1486

\bibitem[{{Soszy{\'n}ski} {et~al.}(2011{\natexlab{a}}){Soszy{\'n}ski},
  {Dziembowski}, {Udalski}, {Poleski}, {Szyma{\'n}ski}, {Kubiak},
  {Pietrzy{\'n}ski}, {Wyrzykowski}, {Ulaczyk}, {Koz{\l}owski}, \&
  {Pietrukowicz}}]{2011AcA....61....1S}
{Soszy{\'n}ski}, I., {Dziembowski}, W.~A., {Udalski}, A., {et~al.}
  2011{\natexlab{a}}, \actaa, 61, 1

\bibitem[{{Soszy{\'n}ski} {et~al.}(2016{\natexlab{a}}){Soszy{\'n}ski},
  {Pawlak}, {Pietrukowicz}, {Udalski}, {Szyma{\'n}ski}, {Wyrzykowski},
  {Ulaczyk}, {Poleski}, {Koz{\l}owski}, {Skowron}, {Skowron}, {Mr{\'o}z}, \&
  {Hamanowicz}}]{2016AcA....66..405S}
{Soszy{\'n}ski}, I., {Pawlak}, M., {Pietrukowicz}, P., {et~al.}
  2016{\natexlab{a}}, \actaa, 66, 405

\bibitem[{{Soszynski} {et~al.}(2008){Soszynski}, {Poleski}, {Udalski},
  {Szymanski}, {Kubiak}, {Pietrzynski}, {Wyrzykowski}, {Szewczyk}, \&
  {Ulaczyk}}]{2008AcA....58..163S}
{Soszynski}, I., {Poleski}, R., {Udalski}, A., {et~al.} 2008, \actaa, 58, 163

\bibitem[{{Soszy{\'n}ski} {et~al.}(2010{\natexlab{a}}){Soszy{\'n}ski},
  {Poleski}, {Udalski}, {Szyma{\'n}ski}, {Kubiak}, {Pietrzy{\'n}ski},
  {Wyrzykowski}, {Szewczyk}, \& {Ulaczyk}}]{2010AcA....60...17S}
{Soszy{\'n}ski}, I., {Poleski}, R., {Udalski}, A., {et~al.} 2010{\natexlab{a}},
  \actaa, 60, 17

\bibitem[{{Soszy{\'n}ski} {et~al.}(2015{\natexlab{a}}){Soszy{\'n}ski},
  {St{\k{e}}pie{\'n}}, {Pilecki}, {Mr{\'o}z}, {Udalski}, {Szyma{\'n}ski},
  {Pietrzy{\'n}ski}, {Wyrzykowski}, {Ulaczyk}, {Poleski}, {Koz{\l}owski},
  {Pietrukowicz}, {Skowron}, \& {Pawlak}}]{2015AcA....65...39S}
{Soszy{\'n}ski}, I., {St{\k{e}}pie{\'n}}, K., {Pilecki}, B., {et~al.}
  2015{\natexlab{a}}, \actaa, 65, 39

\bibitem[{{Soszy{\'n}ski} {et~al.}(2011{\natexlab{b}}){Soszy{\'n}ski},
  {Udalski}, {Pietrukowicz}, {Szyma{\'n}ski}, {Kubiak}, {Pietrzy{\'n}ski},
  {Wyrzykowski}, {Ulaczyk}, {Poleski}, \& {Koz{\l}owski}}]{2011AcA....61..285S}
{Soszy{\'n}ski}, I., {Udalski}, A., {Pietrukowicz}, P., {et~al.}
  2011{\natexlab{b}}, \actaa, 61, 285

\bibitem[{{Soszy{\'n}ski} {et~al.}(2012){Soszy{\'n}ski}, {Udalski}, {Poleski},
  {Koz{\l}owski}, {Wyrzykowski}, {Pietrukowicz}, {Szyma{\'n}ski}, {Kubiak},
  {Pietrzy{\'n}ski}, {Ulaczyk}, \& {Skowron}}]{2012AcA....62..219S}
{Soszy{\'n}ski}, I., {Udalski}, A., {Poleski}, R., {et~al.} 2012, \actaa, 62,
  219

\bibitem[{{Soszy{\'n}ski} {et~al.}(2010{\natexlab{b}}){Soszy{\'n}ski},
  {Udalski}, {Szyma{\'n}ski}, {Kubiak}, {Pietrzy{\'n}ski}, {Wyrzykowski},
  {Ulaczyk}, \& {Poleski}}]{2010AcA....60..165S}
{Soszy{\'n}ski}, I., {Udalski}, A., {Szyma{\'n}ski}, M.~K., {et~al.}
  2010{\natexlab{b}}, \actaa, 60, 165

\bibitem[{{Soszy{\'n}ski} {et~al.}(2008){Soszy{\'n}ski}, {Udalski},
  {Szyma{\'n}ski}, {Kubiak}, {Pietrzy{\'n}ski}, {Wyrzykowski}, {Szewczyk},
  {Ulaczyk}, \& {Poleski}}]{2008AcA....58..293S}
{Soszy{\'n}ski}, I., {Udalski}, A., {Szyma{\'n}ski}, M.~K., {et~al.} 2008,
  \actaa, 58, 293

\bibitem[{{Soszy{\'n}ski} {et~al.}(2009{\natexlab{a}}){Soszy{\'n}ski},
  {Udalski}, {Szyma{\'n}ski}, {Kubiak}, {Pietrzy{\'n}ski}, {Wyrzykowski},
  {Szewczyk}, {Ulaczyk}, \& {Poleski}}]{2009AcA....59....1S}
{Soszy{\'n}ski}, I., {Udalski}, A., {Szyma{\'n}ski}, M.~K., {et~al.}
  2009{\natexlab{a}}, \actaa, 59, 1

\bibitem[{{Soszy{\'n}ski} {et~al.}(2009{\natexlab{b}}){Soszy{\'n}ski},
  {Udalski}, {Szyma{\'n}ski}, {Kubiak}, {Pietrzy{\'n}ski}, {Wyrzykowski},
  {Szewczyk}, {Ulaczyk}, \& {Poleski}}]{2009AcA....59..239S}
{Soszy{\'n}ski}, I., {Udalski}, A., {Szyma{\'n}ski}, M.~K., {et~al.}
  2009{\natexlab{b}}, \actaa, 59, 239

\bibitem[{{Soszy{\'n}ski} {et~al.}(2009{\natexlab{c}}){Soszy{\'n}ski},
  {Udalski}, {Szyma{\'n}ski}, {Kubiak}, {Pietrzy{\'n}ski}, {Wyrzykowski},
  {Szewczyk}, {Ulaczyk}, \& {Poleski}}]{2009AcA....59..335S}
{Soszy{\'n}ski}, I., {Udalski}, A., {Szyma{\'n}ski}, M.~K., {et~al.}
  2009{\natexlab{c}}, \actaa, 59, 335

\bibitem[{{Soszy{\'n}ski} {et~al.}(2010{\natexlab{c}}){Soszy{\'n}ski},
  {Udalski}, {Szyma{\'n}ski}, {Kubiak}, {Pietrzy{\'n}ski}, {Wyrzykowski},
  {Ulaczyk}, \& {Poleski}}]{2010AcA....60...91S}
{Soszy{\'n}ski}, I., {Udalski}, A., {Szyma{\'n}ski}, M.~K., {et~al.}
  2010{\natexlab{c}}, \actaa, 60, 91

\bibitem[{{Soszy{\'n}ski} {et~al.}(2011{\natexlab{c}}){Soszy{\'n}ski},
  {Udalski}, {Szyma{\'n}ski}, {Kubiak}, {Pietrzy{\'n}ski}, {Wyrzykowski},
  {Ulaczyk}, {Poleski}, {Koz{\l}owski}, \&
  {Pietrukowicz}}]{2011AcA....61..217S}
{Soszy{\'n}ski}, I., {Udalski}, A., {Szyma{\'n}ski}, M.~K., {et~al.}
  2011{\natexlab{c}}, \actaa, 61, 217

\bibitem[{{Soszy{\'n}ski} {et~al.}(2013){Soszy{\'n}ski}, {Udalski},
  {Szyma{\'n}ski}, {Kubiak}, {Pietrzy{\'n}ski}, {Wyrzykowski}, {Ulaczyk},
  {Poleski}, {Koz{\l}owski}, {Pietrukowicz}, \&
  {Skowron}}]{2013AcA....63...21S}
{Soszy{\'n}ski}, I., {Udalski}, A., {Szyma{\'n}ski}, M.~K., {et~al.} 2013,
  \actaa, 63, 21

\bibitem[{{Soszy{\'n}ski} {et~al.}(2014){Soszy{\'n}ski}, {Udalski},
  {Szyma{\'n}ski}, {Pietrukowicz}, {Mr{\'o}z}, {Skowron}, {Koz{\l}owski},
  {Poleski}, {Skowron}, {Pietrzy{\'n}ski}, {Wyrzykowski}, {Ulaczyk}, \&
  {Kubiak}}]{2014AcA....64..177S}
{Soszy{\'n}ski}, I., {Udalski}, A., {Szyma{\'n}ski}, M.~K., {et~al.} 2014,
  \actaa, 64, 177

\bibitem[{{Soszy{\'n}ski} {et~al.}(2020){Soszy{\'n}ski}, {Udalski},
  {Szyma{\'n}ski}, {Pietrukowicz}, {Skowron}, {Skowron}, {Poleski},
  {Koz{\l}owski}, {Mr{\'o}z}, {Ulaczyk}, {Rybicki}, {Iwanek}, {Wrona}, \&
  {Gromadzki}}]{2020AcA....70..101S}
{Soszy{\'n}ski}, I., {Udalski}, A., {Szyma{\'n}ski}, M.~K., {et~al.} 2020,
  \actaa, 70, 101

\bibitem[{{Soszy{\'n}ski} {et~al.}(2015{\natexlab{b}}){Soszy{\'n}ski},
  {Udalski}, {Szyma{\'n}ski}, {Skowron}, {Pietrzy{\'n}ski}, {Poleski},
  {Pietrukowicz}, {Skowron}, {Mr{\'o}z}, {Koz{\l}owski}, {Wyrzykowski},
  {Ulaczyk}, \& {Pawlak}}]{2015AcA....65..297S}
{Soszy{\'n}ski}, I., {Udalski}, A., {Szyma{\'n}ski}, M.~K., {et~al.}
  2015{\natexlab{b}}, \actaa, 65, 297

\bibitem[{{Soszy{\'n}ski} {et~al.}(2016{\natexlab{b}}){Soszy{\'n}ski},
  {Udalski}, {Szyma{\'n}ski}, {Wyrzykowski}, {Ulaczyk}, {Poleski},
  {Pietrukowicz}, {Koz{\l}owski}, {Skowron}, {Skowron}, {Mr{\'o}z}, \&
  {Pawlak}}]{2016AcA....66..131S}
{Soszy{\'n}ski}, I., {Udalski}, A., {Szyma{\'n}ski}, M.~K., {et~al.}
  2016{\natexlab{b}}, \actaa, 66, 131

\bibitem[{{Soszy{\'n}ski} {et~al.}(2017){Soszy{\'n}ski}, {Udalski},
  {Szyma{\'n}ski}, {Wyrzykowski}, {Ulaczyk}, {Poleski}, {Pietrukowicz},
  {Koz{\l}owski}, {Skowron}, {Skowron}, {Mr{\'o}z}, {Pawlak}, {Rybicki}, \&
  {Jacyszyn-Dobrzeniecka}}]{2017AcA....67..297S}
{Soszy{\'n}ski}, I., {Udalski}, A., {Szyma{\'n}ski}, M.~K., {et~al.} 2017,
  \actaa, 67, 297

\bibitem[{{Soszy{\'n}ski} {et~al.}(2019){Soszy{\'n}ski}, {Udalski}, {Wrona},
  {Szyma{\'n}ski}, {Pietrukowicz}, {Skowron}, {Skowron}, {Poleski},
  {Koz{\l}owski}, {Mr{\'o}z}, {Ulaczyk}, {Rybicki}, {Iwanek}, \&
  {Gromadzki}}]{2019AcA....69..321S}
{Soszy{\'n}ski}, I., {Udalski}, A., {Wrona}, M., {et~al.} 2019, \actaa, 69, 321

\bibitem[{{Southworth}(2011)}]{2011MNRAS.417.2166S}
{Southworth}, J. 2011, \mnras, 417, 2166

\bibitem[{{Southworth} {et~al.}(2021){Southworth}, {Bowman}, \&
  {Pavlovski}}]{2021MNRAS.501L..65S}
{Southworth}, J., {Bowman}, D.~M., \& {Pavlovski}, K. 2021, \mnras, 501, L65

\bibitem[{{Spano} {et~al.}(2011){Spano}, {Mowlavi}, {Eyer}, {Burki},
  {Marquette}, {Lecoeur-Ta{\"\i}bi}, \& {Tisserand}}]{2011A&A...536A..60S}
{Spano}, M., {Mowlavi}, N., {Eyer}, L., {et~al.} 2011, \aap, 536, A60

\bibitem[{{Stankov} \& {Handler}(2005)}]{2005ApJS..158..193S}
{Stankov}, A. \& {Handler}, G. 2005, \apjs, 158, 193

\bibitem[{{Suh} \& {Hong}(2017)}]{2017JKAS...50..131S}
{Suh}, K.-W. \& {Hong}, J. 2017, Journal of Korean Astronomical Society, 50,
  131

\bibitem[{{S{\"u}veges} {et~al.}(2012){S{\"u}veges}, {Sesar}, {V{\'a}radi},
  {Mowlavi}, {Becker}, {Ivezi{\'c}}, {Beck}, {Nienartowicz}, {Rimoldini},
  {Dubath}, {Bartholdi}, \& {Eyer}}]{2012MNRAS.424.2528S}
{S{\"u}veges}, M., {Sesar}, B., {V{\'a}radi}, M., {et~al.} 2012, \mnras, 424,
  2528

\bibitem[{{Szkody} {et~al.}(2011){Szkody}, {Anderson}, {Brooks},
  {G{\"a}nsicke}, {Kronberg}, {Riecken}, {Ross}, {Schmidt}, {Schneider},
  {Ag{\"u}eros}, {Gomez-Moran}, {Knapp}, {Schreiber}, \&
  {Schwope}}]{2011AJ....142..181S}
{Szkody}, P., {Anderson}, S.~F., {Brooks}, K., {et~al.} 2011, \aj, 142, 181

\bibitem[{{Szkody} {et~al.}(2020){Szkody}, {Dicenzo}, {Ho}, {Hillenbrand },
  {van Roestel}, {Ridder}, {DeJesus Lima}, {Graham}, {Bellm}, {Burdge},
  {Kupfer}, {Prince}, {Masci}, {Mr{\'o}z}, {Golkhou}, {Coughlin}, {Cunningham},
  {Dekany}, {Graham}, {Hale}, {Kaplan}, {Kasliwal}, {Miller}, {Neill},
  {Patterson}, {Riddle}, {Smith}, \& {Soumagnac}}]{2020AJ....159..198S}
{Szkody}, P., {Dicenzo}, B., {Ho}, A. Y.~Q., {et~al.} 2020, \aj, 159, 198

\bibitem[{{Taylor}(2005)}]{2005ASPC..347...29T}
{Taylor}, M.~B. 2005, in Astronomical Society of the Pacific Conference Series,
  Vol. 347, Astronomical Data Analysis Software and Systems XIV, ed.
  P.~{Shopbell}, M.~{Britton}, \& R.~{Ebert}, 29

\bibitem[{{Tian} {et~al.}(2020){Tian}, {Liu}, {Yuan}, {Fang}, {Chen}, {Xiang},
  {Huang}, {Bi}, {Yang}, {Wu}, {Wang}, {Zhang}, {Huo}, {Yang}, {Liu}, {Guo}, \&
  {Zhang}}]{2020ApJS..249...22T}
{Tian}, Z., {Liu}, X., {Yuan}, H., {et~al.} 2020, \apjs, 249, 22

\bibitem[{{Torrealba} {et~al.}(2015){Torrealba}, {Catelan}, {Drake},
  {Djorgovski}, {McNaught}, {Belokurov}, {Koposov}, {Graham}, {Mahabal},
  {Larson}, \& {Christensen}}]{2015MNRAS.446.2251T}
{Torrealba}, G., {Catelan}, M., {Drake}, A.~J., {et~al.} 2015, \mnras, 446,
  2251

\bibitem[{{Udalski} {et~al.}(2018){Udalski}, {Soszy{\'n}ski}, {Pietrukowicz},
  {Szyma{\'n}ski}, {Skowron}, {Skowron}, {Mr{\'o}z}, {Poleski}, {Koz{\l}owski},
  {Ulaczyk}, {Rybicki}, {Iwanek}, \& {Wrona}}]{2018AcA....68..315U}
{Udalski}, A., {Soszy{\'n}ski}, I., {Pietrukowicz}, P., {et~al.} 2018, \actaa,
  68, 315

\bibitem[{{Udalski} {et~al.}(2015){Udalski}, {Szyma{\'n}ski}, \&
  {Szyma{\'n}ski}}]{2015AcA....65....1U}
{Udalski}, A., {Szyma{\'n}ski}, M.~K., \& {Szyma{\'n}ski}, G. 2015, \actaa, 65,
  1

\bibitem[{{Uytterhoeven} {et~al.}(2011){Uytterhoeven}, {Moya},
  {Grigahc{\`e}ne}, {Guzik}, {Guti{\'e}rrez-Soto}, {Smalley}, {Hand ler},
  {Balona}, {Niemczura}, {Fox Machado}, {Benatti}, {Chapellier}, {Tkachenko},
  {Szab{\'o}}, {Su{\'a}rez}, {Ripepi}, {Pascual}, {Mathias},
  {Mart{\'\i}n-Ru{\'\i}z}, {Lehmann}, {Jackiewicz}, {Hekker}, {Gruberbauer},
  {Garc{\'\i}a}, {Dumusque}, {D{\'\i}az-Fraile}, {Bradley}, {Antoci}, {Roth},
  {Leroy}, {Murphy}, {De Cat}, {Cuypers}, {Kjeldsen}, {Christensen-Dalsgaard},
  {Breger}, {Pigulski}, {Kiss}, {Still}, {Thompson}, \& {van
  Cleve}}]{2011A&A...534A.125U}
{Uytterhoeven}, K., {Moya}, A., {Grigahc{\`e}ne}, A., {et~al.} 2011, \aap, 534,
  A125

\bibitem[{{Van Reeth} {et~al.}(2015){Van Reeth}, {Tkachenko}, {Aerts},
  {P{\'a}pics}, {Triana}, {Zwintz}, {Degroote}, {Debosscher}, {Bloemen},
  {Schmid}, {De Smedt}, {Fremat}, {Fuentes}, {Homan}, {Hrudkova},
  {Karjalainen}, {Lombaert}, {Nemeth}, {{\O}stensen}, {Van De Steene}, {Vos},
  {Raskin}, \& {Van Winckel}}]{2015ApJS..218...27V}
{Van Reeth}, T., {Tkachenko}, A., {Aerts}, C., {et~al.} 2015, \apjs, 218, 27

\bibitem[{{Varga-Vereb{\'e}lyi} {et~al.}(2020){Varga-Vereb{\'e}lyi}, {Kun},
  {Szegedi-Elek}, {{\'A}brah{\'a}m}, {Varga}, {Kiss}, {K{\'o}sp{\'a}l},
  {Marton}, \& {Szabados}}]{2020IAUS..345..378V}
{Varga-Vereb{\'e}lyi}, E., {Kun}, M., {Szegedi-Elek}, E., {et~al.} 2020, in IAU
  Symposium, Vol. 345, IAU Symposium, ed. B.~G. {Elmegreen}, L.~V. {T{\'o}th},
  \& M.~{G{\"u}del}, 378--379

\bibitem[{{Vivas} {et~al.}(2020){Vivas}, {Walker}, {Mart{\'\i}nez-V{\'a}zquez},
  {Monelli}, {Bono}, {Dorta}, {Nidever}, {Fiorentino}, {Gallart}, {Andreuzzi},
  {Braga}, {Dall'Ora}, {Olsen}, \& {Stetson}}]{2020MNRAS.492.1061V}
{Vivas}, A.~K., {Walker}, A.~R., {Mart{\'\i}nez-V{\'a}zquez}, C.~E., {et~al.}
  2020, \mnras, 492, 1061

\bibitem[{{{\v{Z}}erjal} {et~al.}(2017){{\v{Z}}erjal}, {Zwitter},
  {Matijevi{\v{c}}}, {Grebel}, {Kordopatis}, {Munari}, {Seabroke}, {Steinmetz},
  {Wojno}, {Bienaym{\'e}}, {Bland-Hawthorn}, {Conrad}, {Freeman}, {Gibson},
  {Gilmore}, {Kunder}, {Navarro}, {Parker}, {Reid}, {Siviero}, {Watson}, \&
  {Wyse}}]{2017ApJ...835...61Z}
{{\v{Z}}erjal}, M., {Zwitter}, T., {Matijevi{\v{c}}}, G., {et~al.} 2017, \apj,
  835, 61

\bibitem[{{Walkowicz} {et~al.}(2011){Walkowicz}, {Basri}, {Batalha},
  {Gilliland}, {Jenkins}, {Borucki}, {Koch}, {Caldwell}, {Dupree}, {Latham},
  {Meibom}, {Howell}, {Brown}, \& {Bryson}}]{2011AJ....141...50W}
{Walkowicz}, L.~M., {Basri}, G., {Batalha}, N., {et~al.} 2011, \aj, 141, 50

\bibitem[{{Watkins} {et~al.}(2009){Watkins}, {Evans}, {Belokurov}, {Smith},
  {Hewett}, {Bramich}, {Gilmore}, {Irwin}, {Vidrih}, {Wyrzykowski}, \&
  {Zucker}}]{2009MNRAS.398.1757W}
{Watkins}, L.~L., {Evans}, N.~W., {Belokurov}, V., {et~al.} 2009, \mnras, 398,
  1757

\bibitem[{{Watson} {et~al.}(2006){Watson}, {Henden}, \&
  {Price}}]{2006SASS...25...47W}
{Watson}, C.~L., {Henden}, A.~A., \& {Price}, A. 2006, Society for Astronomical
  Sciences Annual Symposium, 25, 47

\bibitem[{{Wesselink}(1946)}]{1946BAN....10...91W}
{Wesselink}, A.~J. 1946, \bain, 10, 91

\bibitem[{{Williams} {et~al.}(2016){Williams}, {Montgomery}, {Winget},
  {Falcon}, \& {Bierwagen}}]{2016ApJ...817...27W}
{Williams}, K.~A., {Montgomery}, M.~H., {Winget}, D.~E., {Falcon}, R.~E., \&
  {Bierwagen}, M. 2016, \apj, 817, 27

\bibitem[{{Wood} {et~al.}(1999){Wood}, {Alcock}, {Allsman}, {Alves}, {Axelrod},
  {Becker}, {Bennett}, {Cook}, {Drake}, {Freeman}, {Griest}, {King}, {Lehner},
  {Marshall}, {Minniti}, {Peterson}, {Pratt}, {Quinn}, {Stubbs}, {Sutherland},
  {Tomaney}, {Vandehei}, \& {Welch}}]{1999IAUS..191..151W}
{Wood}, P.~R., {Alcock}, C., {Allsman}, R.~A., {et~al.} 1999, in Asymptotic
  Giant Branch Stars, ed. T.~{Le Bertre}, A.~{Lebre}, \& C.~{Waelkens}, Vol.
  191, 151

\bibitem[{{Wo{\'z}niak} {et~al.}(2004){Wo{\'z}niak}, {Williams}, {Vestrand}, \&
  {Gupta}}]{2004AJ....128.2965W}
{Wo{\'z}niak}, P.~R., {Williams}, S.~J., {Vestrand}, W.~T., \& {Gupta}, V.
  2004, \aj, 128, 2965

\bibitem[{{Wraight} {et~al.}(2012){Wraight}, {Fossati}, {Netopil}, {Paunzen},
  {Rode-Paunzen}, {Bewsher}, {Norton}, \& {White}}]{2012MNRAS.420..757W}
{Wraight}, K.~T., {Fossati}, L., {Netopil}, M., {et~al.} 2012, \mnras, 420, 757

\bibitem[{{Wu} {et~al.}(2015){Wu}, {Ip}, \& {Huang}}]{2015ApJ...798...92W}
{Wu}, C.-J., {Ip}, W.-H., \& {Huang}, L.-C. 2015, \apj, 798, 92

\bibitem[{{Zwintz} \& {Weiss}(2006)}]{2006A&A...457..237Z}
{Zwintz}, K. \& {Weiss}, W.~W. 2006, \aap, 457, 237

\end{thebibliography}
